%% file: effNote.tex
\pdfoutput=1
\documentclass[twocolumn,epjc3]{svjour3}  

\usepackage{preprintcover}
\PreprintCoverPaperTitle{Electron reconstruction and identification efficiency
  measurements with the ATLAS detector using the 2011 LHC
  proton--proton collision data}  
\PreprintIdNumber{CERN-PH-EP-2014-040}  
\PreprintCoverAbstract{Many of the interesting physics processes to be measured at the LHC have a signature involving one or more isolated electrons.
The electron reconstruction and identification efficiencies
  of the ATLAS detector at the LHC have been evaluated using
  proton--proton collision data collected in 2011 at $\rts = 7$~TeV
  and corresponding to an integrated luminosity of 4.7~\ifb.
  Tag-and-probe methods using events with leptonic decays of \Wboson~and \Zboson\ bosons
  and \jpsi\ mesons are employed to benchmark these performance
  parameters. The combination of all measurements results in
  identification efficiencies determined with an accuracy at the few
  per mil level for electron transverse energy greater than 30~GeV. 
}  
\PreprintJournalName{EPJC}  

\RequirePackage{graphicx}
\usepackage{atlasphysics}
\usepackage{subfigure}
\usepackage{mathrsfs}
\usepackage{rotating}
\usepackage{amsmath}
\usepackage{amssymb}
\usepackage[hyperindex,breaklinks]{hyperref} 
\hypersetup{
  colorlinks   = true, 
  urlcolor     = blue, 
  linkcolor    = blue, 
  citecolor   = blue 
}
\journalname{Eur. Phys. J. C}

\begin{document}

\title{\boldmath Electron reconstruction and identification efficiency
  measurements with the ATLAS detector using the 2011 LHC
  proton--proton collision data}

\author{The ATLAS Collaboration\thanksref{CERN}}
\institute{CERN \label{CERN}}

\newcommand{\tnp}{tag-and-probe}
\newcommand{\Pythia}{\textsc{Pythia}}
\newcommand{\Powheg}{\textsc{Powheg}}
\newcommand{\Photos}{\textsc{Photos}}
\newcommand{\antikt}{\ensuremath{\mathrm{anti-}k_t}}
\newcommand{\metiso}{\ensuremath{\Delta \phi_{\met\mathrm{-jet}}}}
\newcommand{\SF}{\ensuremath{\mathrm{SF}}}
\newcommand{\epscluster}{\ensuremath{\epsilon_\mathrm{cluster}}}
\newcommand{\epsreco}{\ensuremath{\epsilon_\mathrm{reco}}}
\newcommand{\epsid}{\ensuremath{\epsilon_\mathrm{id}}}
\newcommand{\epstrig}{\ensuremath{\epsilon_\mathrm{trig}}}
\newcommand{\epsother}{\ensuremath{\epsilon_\mathrm{other}}}
\newcommand{\epscharge}{\ensuremath{\epsilon_\mathrm{charge}}}
\newcommand{\epsloose}{\ensuremath{\epsilon_\mathrm{loose}}}
\newcommand{\epsmedium}{\ensuremath{\epsilon_\mathrm{medium}}}
\newcommand{\epstight}{\ensuremath{\epsilon_\mathrm{tight}}}
\sloppy
\date{Received: 16.4.2014}

\maketitle

\abstract{Many of the interesting physics processes to be measured at the LHC have a signature involving one or more isolated electrons.
The electron reconstruction and identification efficiencies
  of the ATLAS detector at the LHC have been evaluated using
  proton--proton collision data collected in 2011 at $\rts = 7$~TeV
  and corresponding to an integrated luminosity of 4.7~\ifb.
  Tag-and-probe methods using events with leptonic decays of \Wboson~and \Zboson\ bosons 
  and \jpsi\ mesons are employed to benchmark these performance
  parameters. The combination of all measurements results in
  identification efficiencies determined with an accuracy at the few
  per mil level for electron transverse energy greater than 30~GeV. 
}

\section{Introduction}

The good performance of electron\footnote{Throughout this paper, the term ``electron'' usually indicates both
electrons and positrons.}
 reconstruction and identification in
the ATLAS experiment at the Large Hadron Collider (LHC) based at the
CERN Laboratory has been an essential ingredient to its successful
scientific programme. It has played a critical role in several
analyses, as for instance in Standard Model
measurements~\cite{phistar,ATLAS:2012mec,ATLAS:ZZxs,Aad:2014xca}, the
discovery of a Higgs boson~\cite{Aad:2012tfa}, and the searches for
new physics beyond the Standard Model~\cite{Aad:2012hf}. 
Isolated electrons produced in many interesting physics processes can be subject to large backgrounds from
misidentified hadrons, electrons from photon conversions, and
non-isolated electrons originating from heavy-flavour decays. For this
reason, it is important to efficiently reconstruct and identify
electrons over the full acceptance of the detector, while at the same
time to have a significant background rejection. In ATLAS, this is
accomplished using a combination of powerful detector technologies:
silicon detectors and a transition radiation tracker to identify the
track of the electron and a longitudinally layered electromagnetic
calorimeter system with fine lateral segmentation to measure the
electron's energy deposition, followed by hadronic calorimeters used
to veto particles giving rise to significant hadronic activity.

During the 2011 data-taking period at $\sqrt s= 7$~TeV, the LHC
steadily increased the instantaneous luminosity from $5 \times
10^{32}$~cm$^{-2}$~s$^{-1}$ to $3.7 \times
10^{33}$~cm$^{-2}$~s$^{-1}$, with an average superposition
(``pile-up'') of approximately nine proton--proton interactions per
beam crossing. In contrast to the electron performance goals for the
2010 period~\cite{2011mk}, which focused on robustness for the first
LHC running, the goals for the 2011 period aimed at substantially
increasing the background rejection power in this much busier
environment to keep the online output rate of events triggered by 
electron signatures within its allocated budget while at the same time
preserving high reconstruction and identification efficiencies for electrons.
During this period, ATLAS collected large samples of isolated
electrons from \Wen, \Zee, and \Jpsiee~events, allowing precise
measurements of the electron reconstruction and identification
efficiencies over the range of transverse energies, \et, from 7 to
50~GeV. This paper reports on the methods used to perform these
measurements, describes the improvements with respect to previous
results~\cite{2011mk}, and benchmarks the performance of the 2011
electron reconstruction and identification used in various analyses
performed with proton--proton collisions.

The structure of the paper is as follows. Section~\ref{sec:ATLAS}
provides a brief summary of the main components of the ATLAS detector.
The electron trigger design, the algorithm for electron reconstruction
and the electron identification criteria are described in
Section~\ref{sec:RecoIDDescription}. Section~\ref{sec:method} focuses
on the method used to compute the various efficiencies. The data and
simulation samples used in this work are given in
Section~\ref{sec:data} together with the main triggers that enabled
the event collection. Section~\ref{sec:IDmeas} reports on the
identification efficiency measurement, presenting the background
evaluation and the results obtained with the tag-and-probe technique.
A similar methodology, but using a subset of the samples available for
the identification efficiency measurement, is used to extract the
efficiency of the electron reconstruction described in
Section~\ref{sec:RecoEff}. The study of the probability to mismeasure
the charge of an electron is presented in
Section~\ref{Sec:chargemisID}. The summary of the work is given in
Section~\ref{sec:Conclusion}.

\section{The ATLAS detector}
\label{sec:ATLAS}

The ATLAS detector is designed to observe particles produced in
high-energy proton--proton and heavy-ion collisions. It is composed of
an inner tracking detector (ID) immersed in a 2~T axial magnetic field
produced by a thin superconducting solenoid, electromagnetic (EM) and
hadronic calorimeters outside the solenoid, and air-core-toroid muon
spectrometers. A three-level triggering system reduces the total
data-taking rate from a bunch-crossing frequency of approximately
20~MHz to several hundred Hz. A detailed description of the detector
is provided elsewhere~\cite{ATLASdetector}. In the following, only an
overview of the main systems relevant to the results reported in this
paper is provided.

The inner tracking detector provides precise reconstruction of tracks
within a pseudorapidity range\footnote{ATLAS uses a right-handed
  coordinate system with its origin at the nominal interaction point
  (IP) in the centre of the detector and the $z$-axis along the
  beam-pipe. The $x$-axis points from the IP to the centre of the LHC
  ring, and the $y$-axis points upward. Cylindrical coordinates
  $(r,\phi)$ are used in the transverse plane, $\phi$ being the
  azimuthal angle around the beam-pipe. The pseudorapidity is defined
  in terms of the polar angle $\theta$ as $\eta=-\ln\tan(\theta/2)$.
  Transverse momenta and energies are defined as $\pT=p\sin\theta$ and
  $\et=E\sin\theta$, respectively.} $|\eta| \lesssim 2.5$. The
innermost part of the ID consists of a silicon pixel detector
providing typically three measurement points for charged particles originating
in the beam-interaction region. The closest layer to the beam-pipe
(referred to as the b-layer) contributes significantly to precision
vertexing and provides discrimination against photon conversions. A
SemiConductor Tracker (SCT) consisting of modules with two layers of silicon micro-strip sensors
surrounds the pixel detector, providing typically eight
hits per track at intermediate radii. The outermost region of the ID
is covered by a Transition Radiation Tracker (TRT) consisting of straw
drift tubes filled with a Xenon mixture, interleaved with polypropylene/polyethylene transition radiators. For charged particles with transverse momentum
$\pT>0.5$ GeV within its pseudorapidity coverage ($|\eta| \lesssim
2$), the TRT provides typically 35 hits per track. The TRT offers
additional electron identification capability via the detection of
transition-radiation photons generated by the radiators.

The ATLAS calorimeter system has both electromagnetic and hadronic
components and covers the pseudorapidity range $|\eta|<4.9$, with
finer granularity over the region matched to the inner detector. The
central EM calorimeters are of an accordion-geometry design made from
lead/liquid-argon (LAr) detectors, providing a full $\phi$ coverage.
These detectors are divided into two half-barrels ($-1.475<\eta<0$ and
$0<\eta<1.475$) and two endcap (EMEC) components ($1.375<|\eta|<3.2$),
with a transition region between the barrel and the endcaps
($1.37<\abseta<1.52$) which contains a relatively large amount of
inactive material. Over the region devoted to precision measurements
($|\eta|<2.47$, excluding the transition regions), the EM calorimeter
is segmented into longitudinal (depth) compartments called
\emph{front} (also known as \emph{strips}), \emph{middle}, and
\emph{back}. The front layer consists of strips finely grained in the
$\eta$ direction, offering excellent discrimination between photons
and $\pi^{0}\to\gamma\gamma$. At high electron or photon energy, most of the energy is
collected in the middle layer, which has a lateral granularity of
$0.025 \times 0.025$ in $(\eta,\phi)$ space, while the back layer
provides measurements of energy deposited in the tails of the shower.
The hadronic calorimeters, which surround the EM detectors, provide
additional discrimination through further energy measurements of
possible shower tails. The central EM calorimeter is complemented by
two presampler detectors in the region $|\eta|<1.52$ (barrel) and
$1.5<|\eta|<1.8$ (endcaps), made of a thin LAr layer, providing a
sampling for particles that start showering in front of the EM
calorimeters. The forward calorimeter (FCal), a copper--tungsten/LAr
detector, provides coverage at high pseudorapidity ($3.1<|\eta|<4.9$)
with EM-shower identification capability given by its lateral
granularity and longitudinal segmentation into three layers; this
calorimeter plays an important role in extending the pseudorapidity
range where electrons from $Z$-boson decays can be identified.

\begin{figure}
  \begin{center}
    \includegraphics[width=0.5\textwidth]{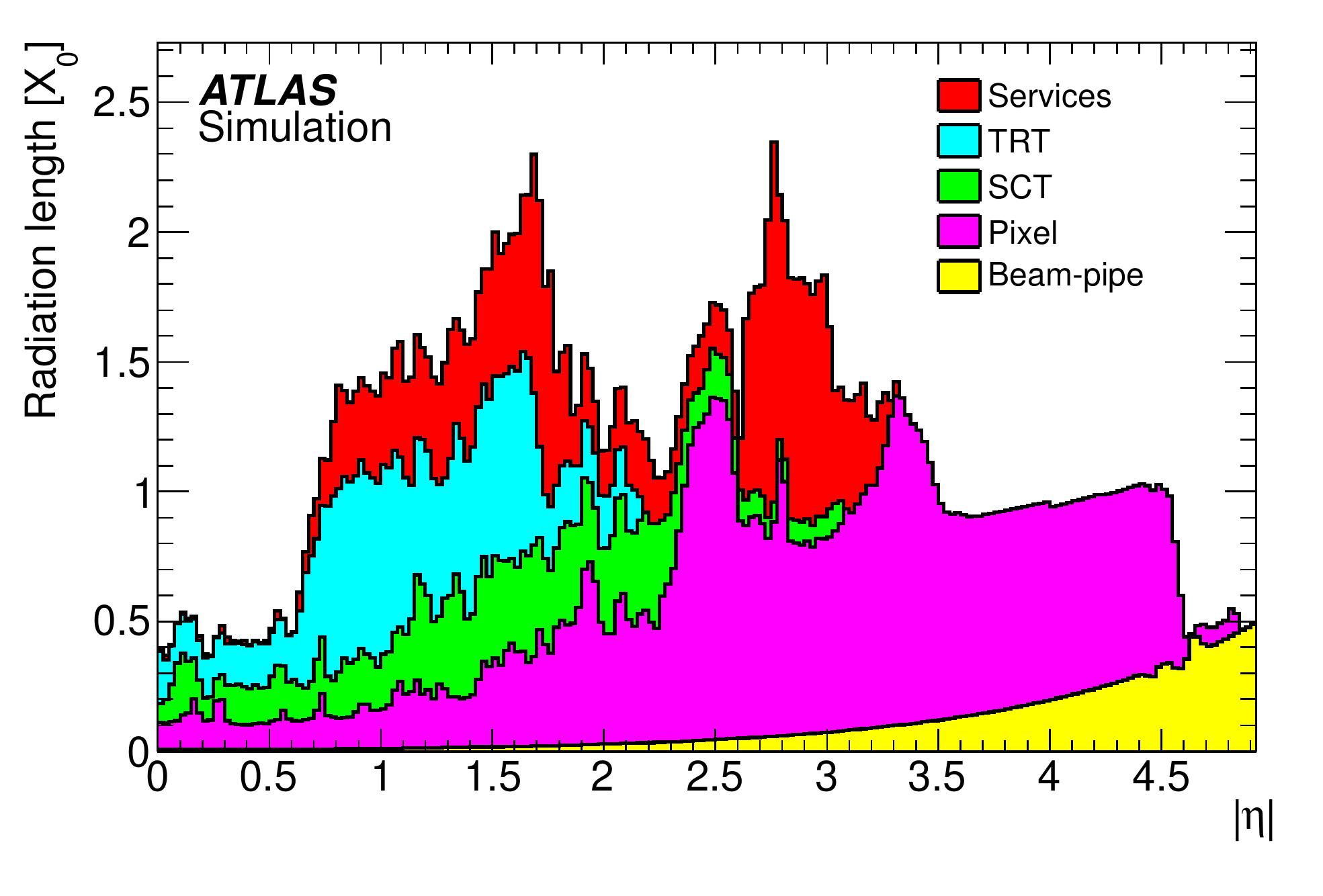}
  \end{center}
  \caption{Amount of material in front of the cryostat, housing the solenoid 
    and the EM calorimeters, in units of
    radiation length $X_{0}$, traversed by a particle as a function of
    $|\eta|$. The contributions of the different detector elements,
    including the services, are shown separately by filled colour
    areas. }
  \label{fig:material}
\end{figure}

The inner detectors, including their services, as well as the cryostat
containing the LAr calorimeter system correspond to a significant
pseudorapidity-dependent amount of material located in front of the EM
calorimeters and can impact the electron reconstruction and
identification performance. Figure~\ref{fig:material} shows the
distribution of the material in front of the cryostat in terms of radiation lengths as a function of pseudorapidity. The observed material variations
suggest a pseudorapidity-dependent optimisation of the selection criteria.

\section{Electron trigger, reconstruction, and identification}
\label{sec:RecoIDDescription}

\subsection{Trigger}

The trigger system in ATLAS~\cite{ATLASdetector,ATLAStrigger}
comprises a hardware-based Level-1 trigger (L1) and software-based
High-Level Triggers (HLT), composed of the Level-2 trigger (L2) and
the Event Filter (EF). Inside the L1, the transverse energy \et\ of
electromagnetic showers collected in the calorimeters is computed
within a granularity of $\Delta \eta \times \Delta \phi \approx 0.1
\times 0.1$. The selected objects must satisfy an \et\ threshold and
are used to seed the L2 reconstruction, which combines calorimetric
and track information using fast algorithms. In the EF, offline-like
algorithms are deployed for the reconstruction of the calorimetric
quantities while an adapted version of the offline software is used to
treat the information of the inner detector. During the 2011 run, the
L1 output rate was kept below 60~kHz, the L2 rate below 5~kHz and the
EF rate was approximately 400~Hz, averaged over the LHC fills.

\subsection{Reconstruction}
\label{sec:tworeco}

\subsubsection{Central electrons}
\label{sec:centralreco}

The electron-reconstruction algorithm used in the central region of
the detector equipped with the ID ($|\eta|<2.5$) identifies energy
deposits in the EM calorimeter and associates these clusters of energy
with reconstructed tracks in the inner detector. The three-step
process is as follows.

\emph{Cluster reconstruction:} EM clusters are seeded from energy
deposits with total transverse energy above 2.5\\GeV by using a
sliding-window algorithm with window size $3 \times 5$ in units of
$0.025 \times 0.025$ in \etaphi\ space. From Monte Carlo (MC)
simulations of \Wboson\ and \Zboson\ leptonic decays, the efficiency
of the initial cluster reconstruction is expected to be approximately
97\% at $\ET=7\GeV$ and almost 100\% for electrons with $\ET>20\GeV$.

\emph{Track association with the cluster:} Within the tracking volume,
tracks with $\pt>0.5$~GeV are extrapolated from their last measured
point to the middle layer of the EM calorimeter. The extrapolated
$\eta$ and $\phi$ coordinates of the impact point are compared to a
corresponding seed cluster position in that layer. A track and a
cluster are considered to be successfully matched if the distance
between the track impact point and the EM cluster barycentre is
$|\Delta\eta| < 0.05$. To account for the effect of bremsstrahlung
losses on the azimuthal distance, the size of the $\Delta\phi$
track--cluster matching window is 0.1 on the side where the
extrapolated track bends as it traverses the solenoidal magnetic
field. An electron candidate is considered to be reconstructed if at
least one track is matched to the seed cluster. In the case where more
than one track is matched to a cluster, tracks with hits in the pixel
detector or the SCT are given priority, and the match with the
smallest $\Delta R=\sqrt{(\Delta\eta)^2+(\Delta \phi)^2}$ distance is
chosen. In the absence of a matching track, the cluster is classified
as an unconverted photon candidate. Electrons are distinguished from
converted photons by investigating the presence of pairs of close-by
tracks originating from a vertex displaced from the interaction point
and by verifying the location of the first hits along the path of the
single tracks \cite{electronphoton}.

\emph{Reconstructed electron candidate:} After a successful
track--cluster matching, the cluster sizes are optimised to take into
account the overall energy distributions in the different regions of
the calorimeter. In the EM barrel region, the energy of the electron
cluster is collected by enlarging its size to $3 \times 7$ in units of
$0.025 \times 0.025$ in ($\eta$,$\phi$) space, while in the EM endcaps
the size is increased to $5 \times 5$. The total reconstructed
electron-candidate energy is determined from the sum of four
contributions~\cite{cscbook}: the estimated energy deposit in the
material in front of the EM calorimeter; the measured energy deposit
in the cluster, corrected for the estimated fraction of energy
measured by the sampling calorimeter; the estimated energy deposit
outside the cluster (lateral leakage); and the estimated energy
deposit beyond the EM calorimeter (longitudinal leakage). The
correction for the material is aided by the measured presampler
signal, while the other three corrections are derived from MC
simulations. The $(\eta,\phi)$ spatial coordinates of the electron
candidate are taken from the parameters of the matched track at the
interaction vertex. The absolute energy scale and the intercalibration
of the different parts of the EM calorimeter are determined using
tightly selected electrons from \Zee, \Jpsiee~and \Wen~ decays
\cite{2011mk}.

\begin{figure*}
  \begin{center}
\subfigure[]{    \includegraphics[width=0.47\textwidth]{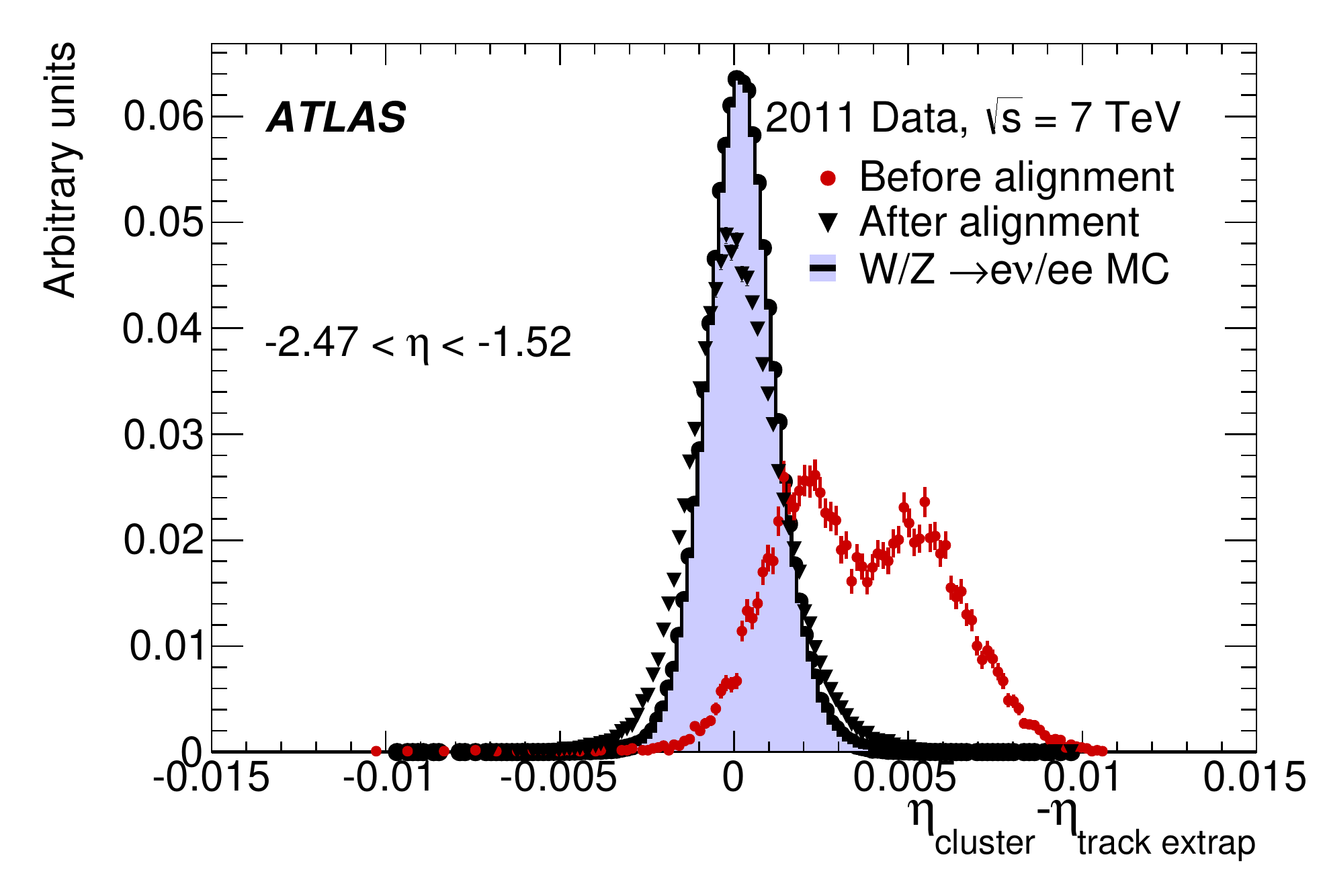}}
\subfigure[]{    \includegraphics[width=0.47\textwidth]{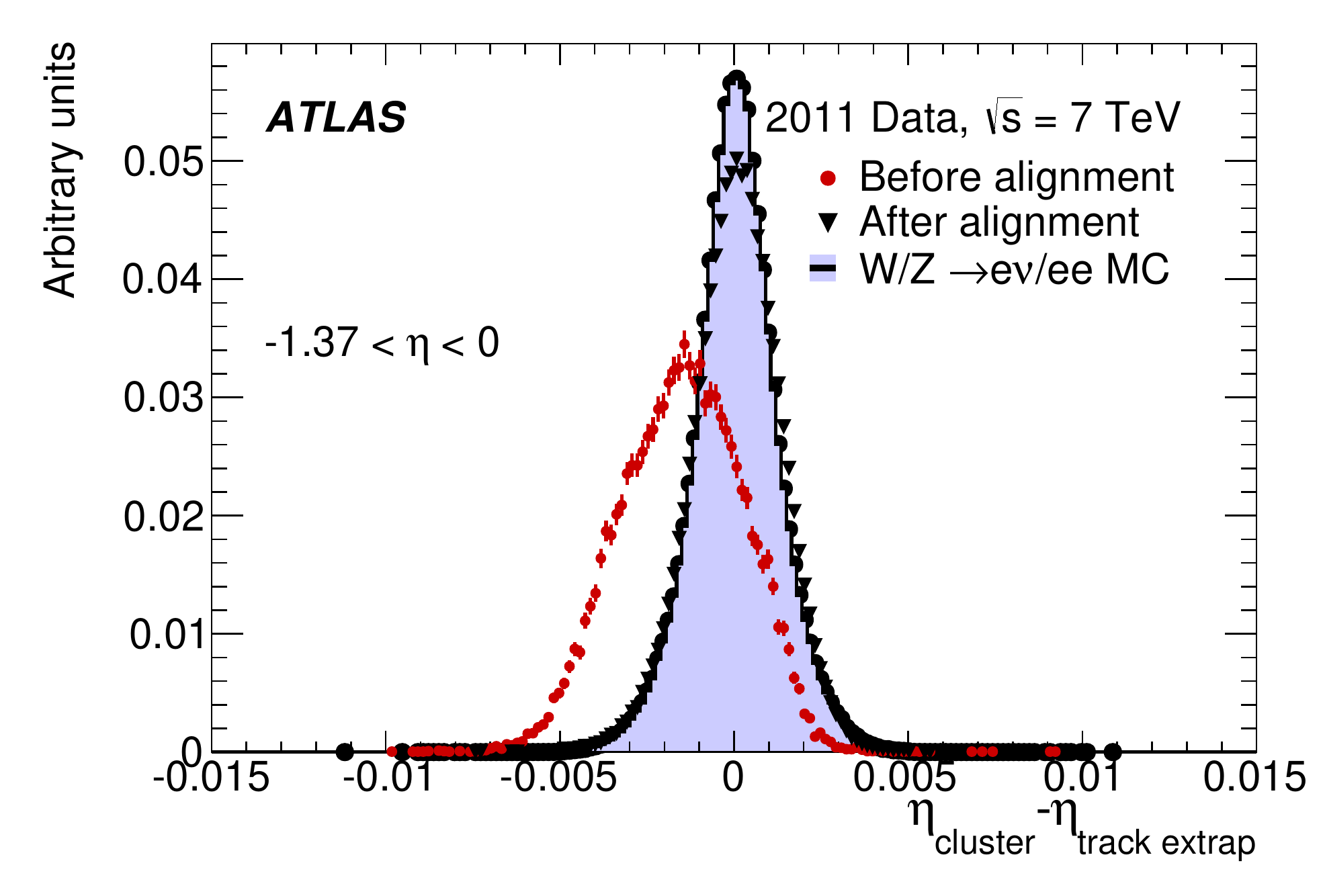}}
\subfigure[]{    \includegraphics[width=0.47\textwidth]{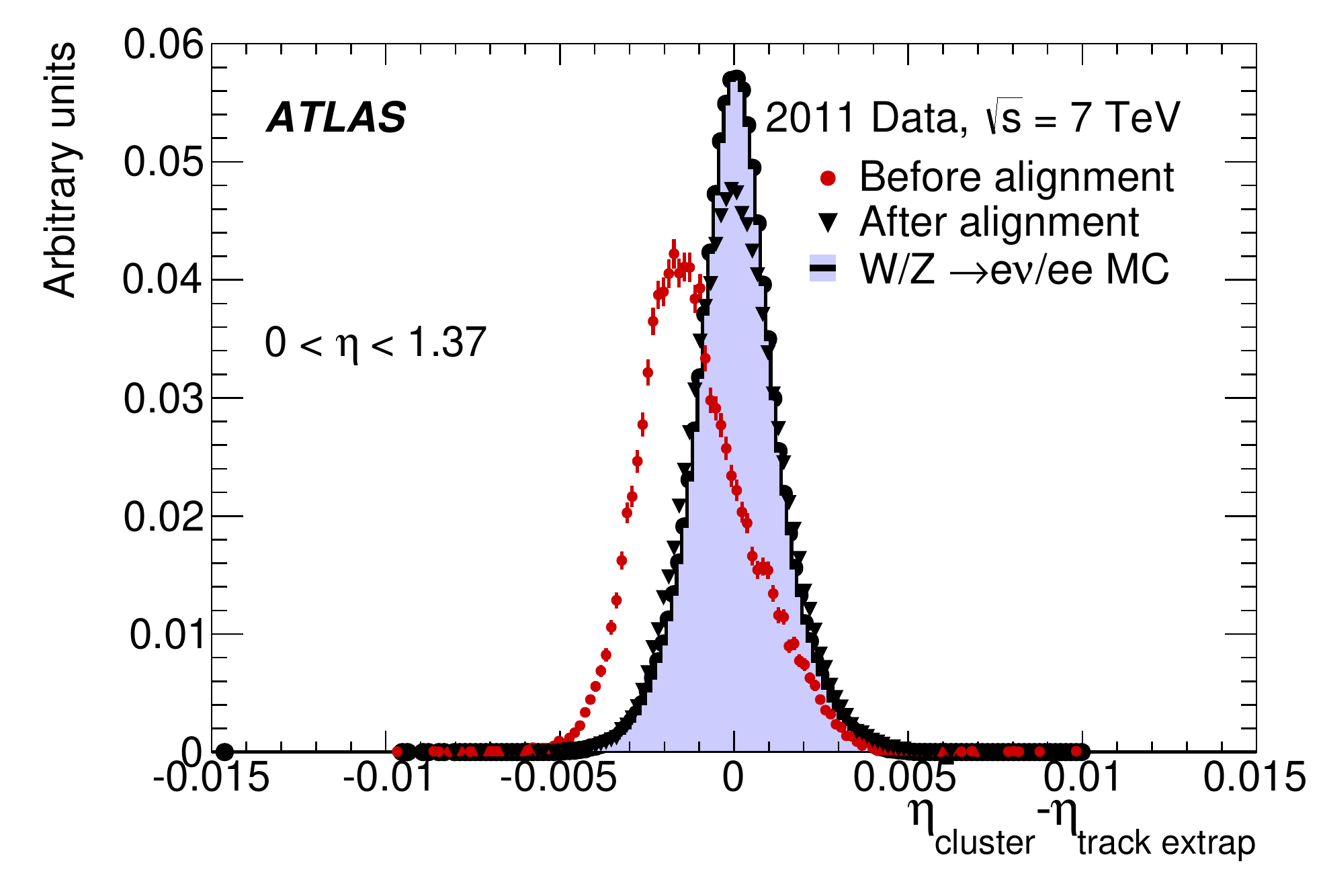}}
\subfigure[]{    \includegraphics[width=0.47\textwidth]{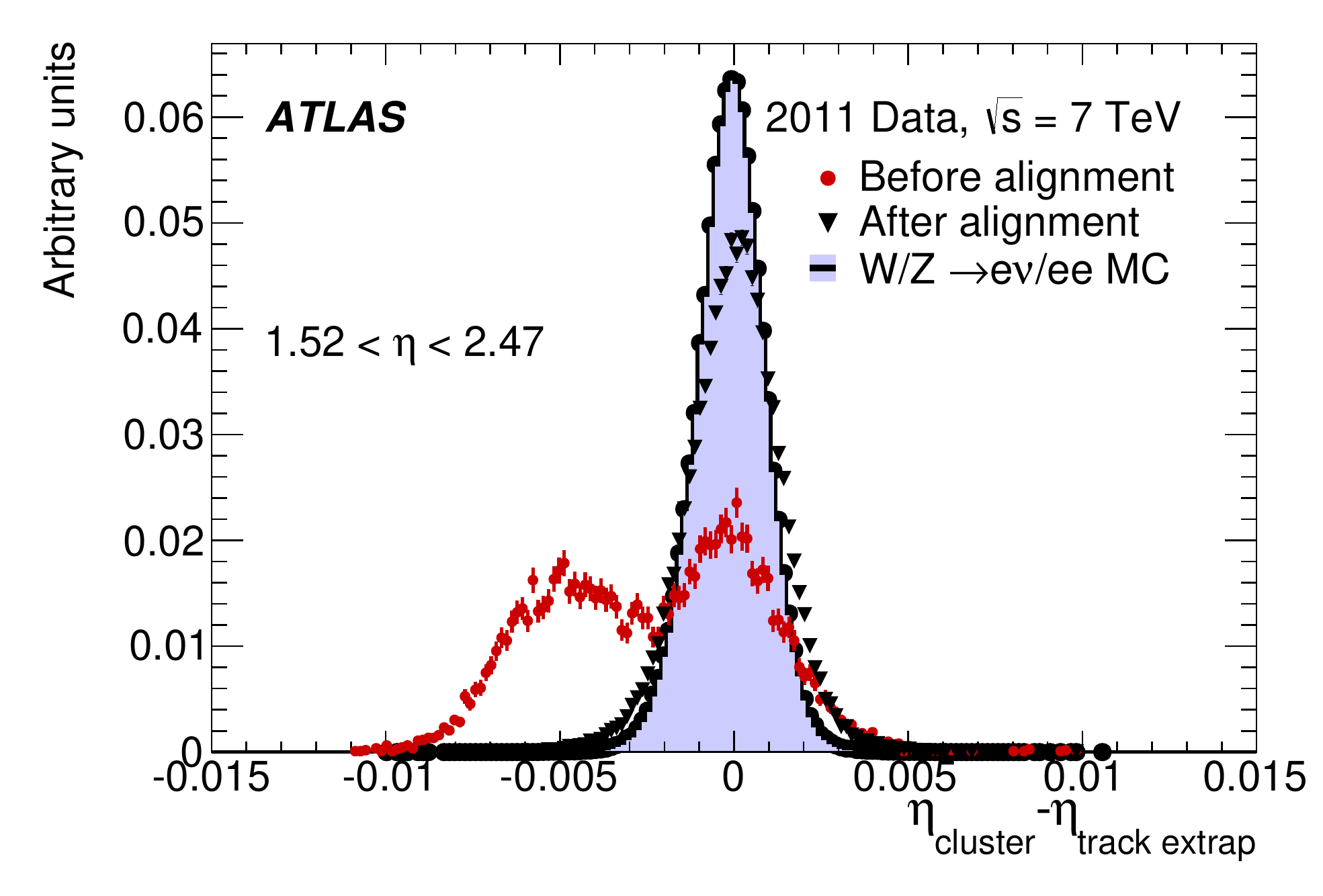}}
  \end{center}
  \caption{Distributions of the difference between the cluster
    \eta~position determined from the first layer of the EM
    calorimeter, and the \eta~position of the ID track extrapolated to
    the entrance of that layer. Before the alignment procedure, the
    estimated detector positions were based on the best knowledge from
    survey and construction. The distribution is shown before (red
    points) and after (black triangles) the alignment corrections.
    Monte Carlo distributions using a perfect tracker--calorimeter
    alignment are also shown as a coloured histogram. The four figures
    correspond to two half-barrels ($-1.37<\eta<0$ in (b) and
    $0<\eta<1.37$ in (c))
    and the two endcaps ($-2.47<\eta<-1.52$ in (a) and
    $1.52<\eta<2.47$ n (d)). The
    two-peak structure visible in the endcap plots (a) and (d) before alignment is
    due to an endcap transverse displacement of 5~mm with respect to
    the beam-line.}
  \label{fig:align}
\end{figure*}

The relative alignment of the calorimeter components with respect to
the inner detector has been measured using electron candidates with
transverse energy $\et>20$~GeV selected with strict identification
criteria, similar to those used for the energy calibration, and compatible with coming from the decay of \Wboson\ or
\Zboson\ bosons.
The difference between the electron cluster position
and the impact point of the track extrapolation to the calorimeter
indicates the size of possible relative displacements between the two
detectors. The derived alignment constants are applied to correct both
the \eta\ (as shown in Figure~\ref{fig:align}) and $\phi$ electron
cluster coordinates.

\subsubsection{Forward electrons}

In the forward region ($2.5<|\eta|<4.9$), which is not equipped with
tracking detectors, the electron reconstruction uses only the
information from the EMEC and forward calorimeters and therefore no
distinction is possible between electrons and photons. Due to the reduced detector 
information in this region, the use of forward electrons in physics analyses
is restricted to the range $\et>20$~GeV. In contrast to
the fixed-size sliding-window clustering used in the central region,
the forward region uses a topological clustering
algorithm~\cite{topoNote}: cells with deposited energy significantly
above the noise level are grouped in three dimensions in an iterative
procedure, starting from seed cells. The number of cells in the
cluster is not fixed and the sum of their energies defines the energy
of the cluster, with corrections made to account for energy losses in
the passive material in front of the calorimeters. As determined from
simulation, the efficiency of the cluster reconstruction is better
than 99\% for $\et>20$~GeV. An electron candidate in the forward
region is reconstructed if it has a transverse energy of $\et>5$~GeV
and has only a small energy component in the hadronic calorimeters.
The direction of the forward-electron candidates is defined by the
barycentre of the cells belonging to the cluster.

\begin{table*}
  \renewcommand{\arraystretch}{1.4}
  \footnotesize
  \caption{Variables used in the \lpp,
    \mpp, and \tpp\ electron identification criteria in the central
    region of the detector ($|\eta|<2.47$).}
  \label{tab:IDcuts}
  \begin{center}
    \begin{tabular}{p{0.25\textwidth}p{0.6\textwidth}c}
      \hline \hline
      Category & Description & Variable \\ 
      \hline
      \multicolumn{3}{c}{\bf{\lpp}} \\
      \hline      
      Acceptance & $|\eta|<2.47$ &       \\ 
      Hadronic leakage 
      & 
      In $|\eta| <0.8$ and  $|\eta| >1.37$: ratio of \et\ in the first layer of the hadronic calorimeter to \et\ of the EM cluster & $R_{\rm{had,1}}$\\
      & In $0.8<|\eta|< 1.37$: ratio of \et\ in whole hadronic calorimeter to \et\ of the EM cluster  &   $R_{\rm{had}}$ 
      \\
      Middle layer of the EM
      & 
      Ratio of energies in $3 \times 7$ cells over $7 \times 7$ cells & $R_{\eta}$ 
      \\
      &
      Lateral width of the shower & $w_{\rm \eta2}$ 
      \\
      Front layer of the EM        
      & 
      Total shower width  &  $w_{\rm stot}$ 
      \\  
      & 
      Energy difference of
      the largest and second largest energy deposits in the cluster 
      divided by their sum                        &   $E_{\rm{ratio}}$ 
      \\
      Track quality and track--cluster matching       
      & Number of hits in the pixel detector ($>0$) &
      \\
      & Number of hits in the silicon detectors ($\geq 7$) &
      \\
      & $|\Delta\eta|$ between the cluster position in the first layer and the extrapolated track ($< 0.015$) &  $\Delta\eta_1$
      \\      
      \hline
      \multicolumn{3}{c}{\bf{\mpp} (includes \lpp\ with tighter requirements on shower shapes)}\\
      \hline
      Track quality and track--cluster matching 
      & Number of hits in the b-layer $>0$ for $|\eta|<2.01$ & \\
      & Number of hits in the pixel detector $>1$ for $|\eta|>2.01$ & \\
     & Transverse impact parameter $|d_{0}|<5$~mm & $d_0$ \\      
      & Tighter $|\Delta\eta_1|$ cut ($< 0.005$) &  \\
      TRT       
      & Loose cut on TRT high-threshold fraction & \\ 
      \hline
      \multicolumn{3}{c}{\bf{\tpp} (includes \mpp)}\\
      \hline
      Track quality and track--cluster matching 
      & Tighter transverse impact parameter cut ($|d_0|<1$~mm) & \\      
      & Asymmetric cut on $\Delta\phi$ between the cluster position in the middle layer and the extrapolated track & $\Delta \phi$ \\
      & Ratio of the cluster energy to the track momentum & $E/p $ \\
      TRT                 
      & Total number of hits in the  TRT & \\
      & Tighter cut on the TRT high-threshold fraction & \\
      Conversions         
      & Reject electron candidates matched to reconstructed photon conversions & \\
      \hline
      \hline
\\
    \end{tabular}
  \end{center}
\end{table*}

\begin{table*}
  \renewcommand{\arraystretch}{1.4}
  \footnotesize
  \caption{Variables used to identify electrons in the forward region of the detector ($2.5<|\eta|<4.9$). }
  \label{tab:IDfwdcuts}
	\begin{center}
	\begin{tabular}{p{0.23\textwidth}p{0.55\textwidth}c}
	\hline\hline
	Category & Description & Variable \\ \hline
        Acceptance & $2.5<|\eta|<4.9$ &       \\  
	Shower depth & Distance of the shower barycentre from the calorimeter front face measured along the shower axis 
        & $\lambda_{\rm{centre}}$ \\
        Maximum cell energy &
        Fraction of cluster energy in the most energetic cell & $f_{\rm{max}}$ \\
	Longitudinal second moment & Second moment of the distance of each cell to the shower centre in the longitudinal direction ($\lambda_i$)
	& $\langle \lambda^2 \rangle$ \\
	Transverse second moment & Second moment of the distance of each cell to the shower centre in the transverse direction ($r_i$)
	& $\langle r^2 \rangle$ \\
	Normalised lateral moment & $w_2$ and $w_{\rm{max}}$ are second moments of $r_i$ for different weights per cell &  $\frac{w_2}{w_2+w_{\rm{max}}}$ \\
	Normalised longitudinal moment & $l_2$ and  $l_{\rm{max}}$ are the second moments of $\lambda_i$ for different weights per cell & $\frac{l_2}{l_2+l_{\rm{max}}}$ \\
	\hline\hline
	\end{tabular}
\end{center}
\end{table*}

\subsection{Electron identification}
\label{sec:Id}

\subsubsection{Central electrons}

The identification criteria for central-electron candidates are
implemented based on sequential cuts on calorimeter, on tracking, and
on combined track--cluster variables. These requirements are optimised
in 10 cluster-$\eta$ bins, motivated by the structure of the detector, and 11 \et\ bins (from 5
to 80~GeV), in order to provide good separation between signal
(isolated) electrons and background from hadrons misidentified as
electrons, non-isolated electrons (e.g. from semileptonic decays of
heavy-flavour particles), and electrons from photon conversions.

Three sets of reference selection criteria, labelled \lpp, \mpp~and
\tpp, are designed for use in analyses. These three sets 
were revisited with respect to those described in
Ref.~\cite{2011mk}, which were designed mostly for robustness at
the startup of the LHC machine with low-luminosity conditions. These
criteria are designed in a hierarchical way so as to provide
increasing background-rejection power at some cost to the
identification efficiency. The increased background-rejection power
was obtained both by adding discriminating variables at each step and
by tightening the requirements on the original variables. The
different selections used for central-electron identification are
detailed in Table~\ref{tab:IDcuts} and described below.
 
\emph{Loose:} The \lpp~selection uses shower-shape variables in both
the first and second layers of the EM calorimeter, in contrast to the
original selection~\cite{2011mk}, which did not use the former. As
before, hadronic leakage information is used. Additional requirements
on the quality of the electron track and track--cluster matching
improve the rejection of hadronic backgrounds by a factor of $\sim 5$
in the \et~range 30 to 40~GeV while maintaining a high identification
efficiency.

\emph{Medium:} The \mpp~selection adds to the \lpp~discriminating
variables by requiring the presence of a measured hit in the innermost
layer of the pixel detector (to reject electrons from photon
conversions), applying a loose selection requirement on the transverse impact parameter $|d_{0}|$, and identifying transition radiation in the TRT (to
reject charged-hadron background), when available. The requirements on
the discriminating variables in common with the \lpp~selection are
also tightened, allowing the background-rejection power to increase by
approximately an order of magnitude with respect to \lpp.
 
\emph{Tight:} The \tpp~selection makes full use of the
particle-identification tools available for electron identification.
In addition to the generally tighter requirements on \mpp~selection
discriminating variables, stricter requirements on track quality in
the presence of a track extension in the TRT detector, on the ratio of
the EM cluster energy to the track momentum, and a veto on
reconstructed photon conversion vertices associated with the cluster~\cite{electronphoton}
are applied. Overall, a rejection power higher by a factor of two is
achieved with respect to the \mpp~selection.

The \lpp, \mpp, and \tpp\ identification criteria naturally exclude a large fraction of candidates with additional close-by activity, such as electrons within jets. It is important to note that 
none of the electron identification criteria explicitly apply requirements 
on the presence of other particles (additional tracks or energy deposits outside the EM cluster) close to the identified electrons. The optimisation of such
dedicated requirements (so-called
\emph{isolation} requirements), is strongly dependent on the physics process and is
performed separately in each analysis.

\subsubsection{Forward electrons}
\label{sec:fwdid}

Electron identification in the forward region also is based on
sequential cuts on discriminating variables; however, these variables
are mostly based on topological cluster moments\footnote{The cluster
  moment of degree $n$ for a variable $x$ is defined as:
  \begin{equation*}
    \langle x^n\rangle=\left[\sum_i{E_i\,x_i^n}\right]/\left[\sum_i{E_i}\right],
  \end{equation*}
  where $i$ is the cell index within the cluster.}, as defined in
Table~\ref{tab:IDfwdcuts}. As for the central region, three
reference sets of selection criteria, labelled \lpp, \mpp, and \tpp,
are defined. To compensate for the absence of tracking information in the forward region, variables describing both the lateral and longitudinal shower development are employed. In addition, due to the significantly harsher pile-up
conditions at high pseudorapidity with respect to those described in
Ref.~\cite{2011mk}, the identification criteria for forward electrons
were redesigned and optimised directly with data in nine
cluster-$\eta$ bins: six in the EMEC calorimeter ($2.5<|\eta|<3.16$)
and three in the FCal ($3.35<|\eta|<4.90$). The transition region
between the two calorimeters ($3.16<|\eta|<3.35$) is excluded from the
study. No explicit dependence on cluster \et\ or isolation energy
is introduced in the forward-electron identification
criteria. However, in contrast to the central electrons, the
identification criteria are also optimised in four bins of the number
of primary vertices reconstructed in the event $N_{\mathrm{PV}}$
(1--3, 4--6, 7--10, $>$10), allowing for similar
electron-identification efficiency for different pile-up conditions.
These three reference sets use the same variables in each set, but
with increasing background rejection power coming from tightened
requirements, with the \tpp~identification providing a rejection
factor approximately two to three times higher than the
\lpp~selection.

\subsection{Bremsstrahlung-mitigation algorithms}

An electron can lose a significant amount of its energy due to
bremsstrahlung when interacting with the material it traverses.
Because of the electron's small mass, radiative losses can be
substantial, resulting in alterations of the curvature of the
electron's trajectory when it propagates through a magnetic field and
hence of the reconstructed electron track. The electron-reconstruction
scheme described in Section~\ref{sec:centralreco} employs the same tracking
algorithm for all charged particles, with all tracks fitted using a
pion mass hypothesis to estimate the material effects. The lack of
special treatment for bremsstrahlung effects results in inefficiencies
in reconstructing the electron trajectory. It also results in the
degradation of the estimated track parameters, increasing with the
amount of material encountered. The effect is strongly dependent on
the electron pseudorapidity, as shown in Figure~\ref{fig:material}. By
taking into account possible bremsstrahlung losses (and the resulting
alteration of the track curvature), the estimated electron track
parameters can be improved. In 2011--2012, a two-step programme was
underway in ATLAS to improve electron reconstruction: first to correct
all track parameters associated with electron candidates by performing
a bremsstrahlung refitting procedure prior to the matching with the
electron cluster, and then performing bremsstrahlung recovery at the
initial step of the electron trajectory formation, to allow more
efficient track reconstruction. By the end of the 2011 data-taking
period, the first step~\cite{GSFNote} was made available to analyses,
improving the track-related electron identification variables. The
second step was implemented in time for the 2012 data-taking period,
increasing the electron reconstruction efficiency by several percent, especially at
low \et. Results presented in this paper do not use the
bremsstrahlung-mitigation algorithms.

\section{Methodology for efficiency measurements}
\label{sec:method}

Isolated electrons are important ingredients in Standard Model
measurements and searches for physics beyond the Standard Model.
However, the experimentally determined electron spectra must be
corrected for instrumentation inefficiencies, such as those related to
trigger, reconstruction, and identification, before absolute
measurements can be made. These inefficiencies may be directly
estimated from data using so-called tag-and-probe
methods~\cite{2011mk}. These methods are used to select, from known
resonances such as \Zee, unbiased samples of electrons (\emph{probes})
by using strict selection requirements on the second object produced
from the particle's decay (\emph{tags}). The efficiency of a
requirement can then be determined by applying it directly to the
probe sample after accounting for residual background contamination.
The efficiency factor relating a true single-electron spectrum to one
determined experimentally may be factorised as a product of different
efficiency terms:
\begin{equation*}
  \epsilon_\mathrm{e} = \epscluster \cdot \epsreco \cdot
  \epsid \cdot \epstrig \cdot \epsother\,,
\end{equation*}
where $\epscluster$ is the efficiency to reconstruct an
electromagnetic cluster, $\epsreco$ is the electron reconstruction
algorithm efficiency given the presence of the cluster
(Section~\ref{sec:tworeco}), and $\epsid$ is the efficiency of
identification criteria with respect to the reconstructed electron
candidates (Section~\ref{sec:Id}). The variable $\epstrig$ denotes the
trigger efficiency with respect to reconstructed electron candidates
passing the identification criteria. The variable $\epsother$ is the
efficiency of any extra selection requirements applied to the
electrons satisfying the identification criteria, such as isolation of
the electron cluster and/or track, or selections on the significance
of the impact parameter of the fitted electron track (both are used in
many analyses). This paper reports on the measurement of the
reconstruction efficiency $\epsreco$ and the identification efficiency
$\epsid$ as determined from data and compared with expectations from
simulated events. The term $\epscluster$ is determined from simulation
to be close to unity, with typical values in the central and forward
regions provided in Section~\ref{sec:tworeco}. Measurements of the
trigger efficiency $\epstrig$ can be found in
Ref.~\cite{TriggerEfficiencyNote}. The term $\epsother$ is largely
process-dependent and so must be measured separately in each analysis.
Section~\ref{Sec:chargemisID} presents a measurement of the efficiency
to correctly identify the charge of an electron, $\epscharge$, with
respect to the reconstructed electron candidates satisfying the
various identification criteria.

Tag-and-probe-based measurements based on samples of \Zee, \Wen, and
\Jpsiee\ events are presented. The combination of the three samples
allows efficiency measurements over a significant \et~range, from 7 to
50~GeV, while still providing overlapping measurements between the
samples\footnote{Results in the high transverse energy region
  $\et>50$~GeV are discussed in Ref.~\cite{Aad:2013iua}.}. In the case
of \Zee~and \Jpsiee~decays, events are selected on the basis of the
electron-positron invariant mass and strict identification criteria
applied to the tag electron. Electron identification efficiencies are
also extracted from \Wen~decays, tagging on the presence of missing
transverse momentum in the event; this channel contributes
significantly to the overall efficiency determination due to its high
statistical power. At the LHC, \Jpsi\ mesons are produced directly and
in $b$-hadron decays. Prompt \Jpsi\ decays occur
in the vicinity of the primary event vertex while many of the
non-prompt \Jpsi\ particles have displaced decay vertices due to the
relatively long lifetime of their $b$-hadron parent. The \Jpsi\
candidates come from a mixture of these two processes; however, their
ability to extend the reach of efficiency measurements to low \et\
makes them nonetheless very attractive, in spite of this added
complication.

The shower profiles of electrons in the calorimeters depend on both
the energy of the electrons and the amount of material traversed by
the electrons before reaching the calorimeter. For this reason,
electron efficiency measurements in the central region ($|\eta|<2.47$)
are made binned in two dimensions, both transverse energy and
pseudorapidity, in contrast to the previous results~\cite{2011mk}
whose statistical precision could only provide one-dimensional binning
in either variable. Eight bins of 5~GeV in transverse energy are used
in the range from 10 to 50~GeV, with an additional bin covering the
low \et~range from 7 to 10~GeV. Depending on the available statistics
in each \et\ bin, efficiencies are measured in three different,
largely detector-motivated, $\eta$~granularities:
\begin{itemize}
\item \emph{coarse:} 11 bins in $\eta$  with limits  $-2.47$, $-2.01$, $-1.52$, $-1.37$, $-0.8$, $-0.1$, $0.1$, $0.8$, $1.37$, $1.52$, $2.01$, $2.47$
\item \emph{middle:} 20 bins in $\eta$ with \abseta~limits $0.0$, $0.1$, $0.6$, $0.8$, $1.15$, $1.37$, $1.52$, $1.81$, $2.01$, $2.37$, $2.47$
\item \emph{fine:} 50 bins in $\eta$ with a typical granularity of 0.1
  covering the full pseudorapidity range ($|\eta|<2.47$)
\end{itemize}

In the forward region the measurements are performed binned only in 
absolute electron pseudorapidity:
\begin{itemize}
\item \emph{forward:} 9 bins in $|\eta|$  with limits  $2.5$, $2.6$, $2.7,
  2.8$, $2.9$, $3.0$, $3.16$, $3.35$, $3.6$, $4.0$, $4.9$
\end{itemize}

The efficiency is defined as the fraction of electrons passing a
particular selection in a given (\et,\eta) bin. For the case of
$\epsreco$, the electron reconstruction efficiency is calculated with
respect to the sample satisfying the cluster-building step. Hence,
clusters associated with reconstructed photons are also included in the
denominator of the measured reconstruction efficiency, provided that
they are separated by $\Delta R>0.4$ from any other cluster associated
with a reconstructed electron. As no reconstructed charge is available
for clusters without an associated track, no requirement on the charge
of the tag and the probe is applied. For the case of $\epsid$, the
efficiency to identify an electron as \lpp, \mpp, or \tpp\ is
calculated with respect to a reconstructed electron candidate,
resulting in three ratios: $\epsloose$, $\epsmedium$, and $\epstight$,
respectively. For the case of $\epscharge$, the efficiency to
correctly identify the charge of an electron is calculated by
comparing the ensemble of di-electron pairs without any requirement on
the sign of the charge of the track to that of the yield of opposite-sign pairs
consistent with the decay of a $Z$ boson. The statistical uncertainty
of these efficiencies is computed assuming a binomial distribution. If
the evaluation of the number of events (before or after the selection
under investigation) is the result of a background subtraction, the
corresponding uncertainties are also included in the statistical
uncertainty.

\section{The 2011 data and simulation samples}
\label{sec:data}

The data recorded during the 2011 proton--proton collision run at
7~TeV are subdivided into several periods corresponding to the
changing conditions of the detector, including the energy thresholds
of the primary triggers, as well as the instantaneous luminosity of
the LHC. Monte Carlo samples are generated to mimic the same period
granularity. In order to reproduce the pile-up effects observed in the
data, additional inelastic proton--proton interactions in the form of
simulated \Pythia~\cite{Sjostrand:2006za} minimum-bias events are
included in the Monte Carlo simulation.

\subsection{Samples}

All data collected by the ATLAS detector undergo careful scrutiny to
ensure the quality of the recorded information. In particular, data
used for the efficiency measurements are filtered requiring that all
detector subsystems needed in the analysis (calorimeters and tracking
detectors) are operating nominally. Several detector defects had minor
impacts on the quality of the 2011 data set. The total integrated
luminosity used for the measurement presented in this paper is
$\mathcal{L}=4.7$~fb$^{-1}$~\cite{Aad:2013ucp}.

Samples of simulated \Zee, \Wen, and \Jpsiee~decays are used to
benchmark the expected electron reconstruction and identification
performance. The primary \Zee\ and \Wen\ MC samples are generated with
\Powheg\ version
r1556~\cite{Nason:2004rx,Frixione:2007vw,Alioli:2010xd,Alioli:2008gx}
and parton showering is accomplished using \Pythia\ version 6.425. The
\jpsi\ samples are generated using the same version of \Pythia. All
generators are interfaced to \Photos\ version
3.0~\cite{Golonka:2005pn} to simulate the effect of final-state QED
radiation. The generated event samples are passed through a detailed
ATLAS detector simulation~\cite{ATLAS:2010wqa} using
GEANT4~\cite{Agostinelli:2002hh}. The MC events are reconstructed
using the same software suite as used for the data. Because
background subtraction is not performed on the MC signal samples when
assessing the expected electron efficiency, generator-level
information is used to select electrons originating only from \Zee,
\Wen, or \Jpsiee~ decays. Correction factors are applied to the
simulation to account for known discrepancies with the data. These
include corrections in the form of event weights applied to the
simulated events to match the average interaction rate per bunch
crossing and the width of the beam-spot in the $z$-direction,
both as measured in the 2011 data set. Both corrections are important
for the measurements presented in this paper since the identification
efficiency depends on the instantaneous luminosity and the position of
the primary interaction.

\begin{figure*}
  \begin{center}
\subfigure[]{    \includegraphics[width=0.47\textwidth]{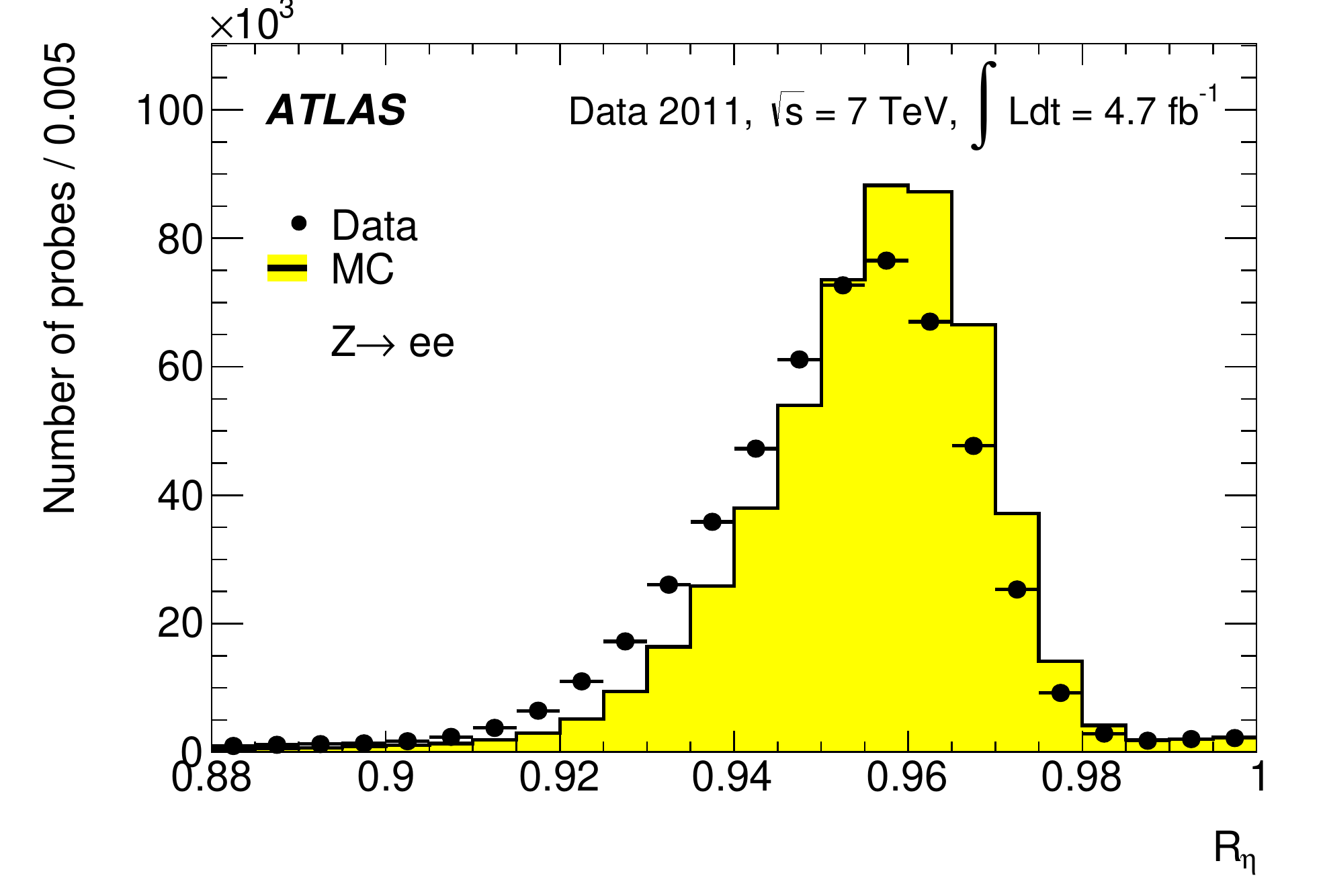}}
\subfigure[]{    \includegraphics[width=0.47\textwidth]{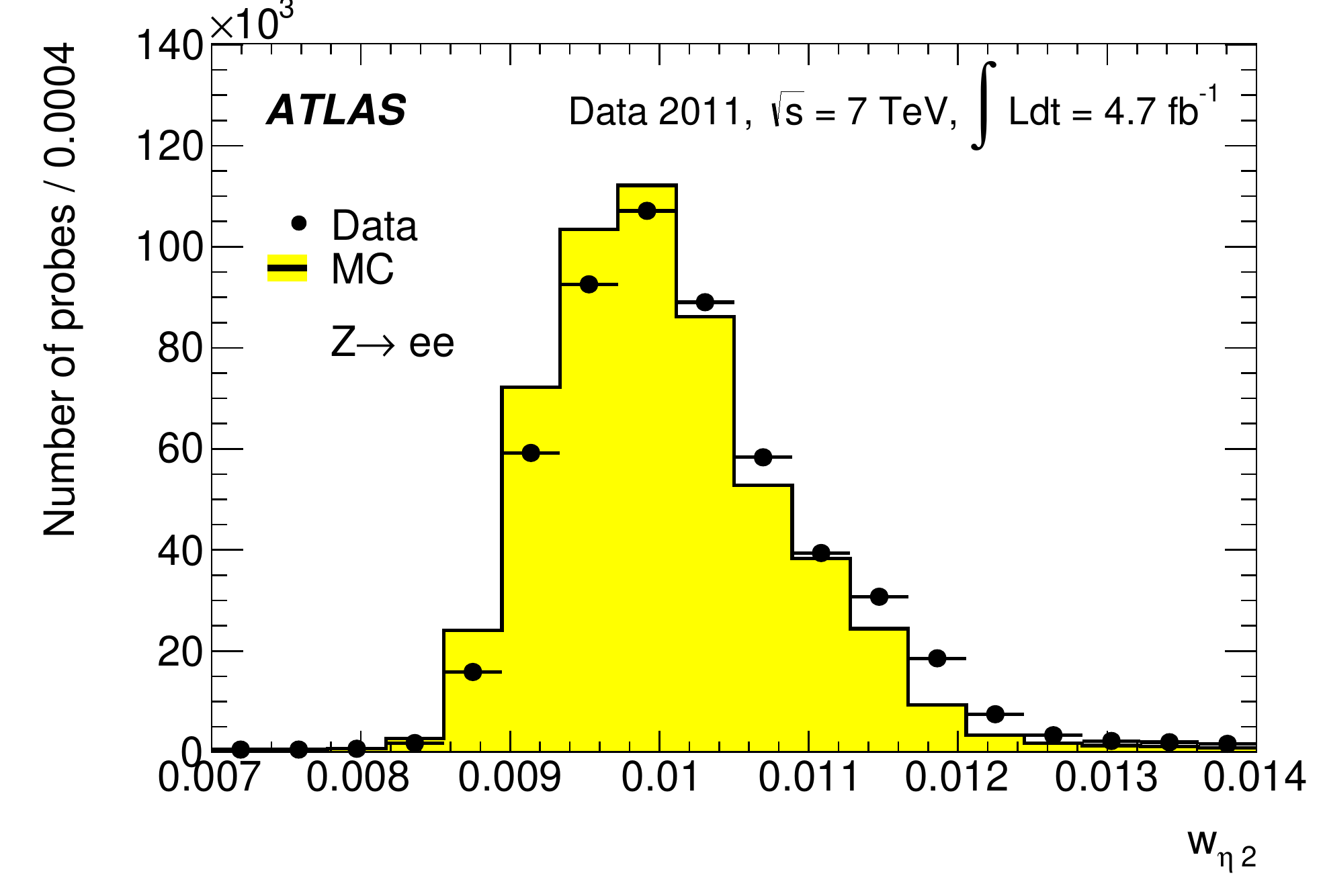}}
  \end{center}
  \caption{Comparison of the shapes in data and MC simulation for two
    variables related to the lateral shower extension in the second
    layer of the EM calorimeter (see Table~\protect\ref{tab:IDcuts}):
    $R_{\eta}$ in (a) and $w_{\eta 2}$ in (b). Electrons with \et~ in
    the range 40--45~GeV from \Zee~ decays are used to extract these
    shapes.}
  \label{fig:showerShapes}
\end{figure*}

Important improvements to the ATLAS GEANT4 simulation were made as a
consequence of observed Monte Carlo--data discrepancies in 2010
related to the transverse shower shapes of electrons in the EM
calorimeter~\cite{2011mk}. The implementation of a new GEANT4 version
(4.9.3), combined with a change of the ATLAS geometry description
resulted in a significant improvement in the 2011 MC simulation
samples. The residual differences that are still observed when
comparing data and MC for some variables, as shown in
Figure~\ref{fig:showerShapes}, have to be taken into account in the
analyses by applying appropriate data-to-MC efficiency corrections as
presented in this paper.

\subsection{Triggers}

The samples used in these measurements were selected by the primary
electron triggers as well as by specifically designed supporting
triggers. In order to keep the trigger rates to an acceptable level
with the increase of the instantaneous luminosity in 2011, the primary
single-electron trigger selection had to be adjusted several times by
raising the minimum transverse energy threshold and tightening the
selection criteria. These same trigger conditions are also implemented
in the Monte Carlo simulations.

\begin{itemize}
\item{\Zee~events were collected using the unprescaled single-electron triggers, requiring the candidates to pass a minimum  \et\ threshold. These events were also required to satisfy strict quality criteria; initially, the so-called \emph{medium} and later \emph{medium1} criteria introduced to tighten the requirements on the shower shapes and track properties, limitations on the amount of energy deposited in the hadronic calorimeter, and $\eta$-dependent \et\ thresholds (indicated in the trigger name by ``vh'') at L1. These triggers are summarised in Table~\ref{tab:periods}~\cite{TriggerEfficiencyNote}.

\begin{table}[t]

\caption{\label{tab:periods} Single-electron trigger evolution during the 2011 data taking, with their respective \et~thresholds at EF level.}
\begin{center}
\begin{tabular}{ccc}
\hline\hline
Single-electron  & Luminosity & \et~threshold\\
Triggers         & $[\mathrm{cm}^{-2}\mathrm{s}^{-1}]$ & [GeV]\\
\hline
e20\_medium & up to 2$\times$10$^{33}$ & 20 \\
e22\_medium & 2--2.3 $\times$ 10$^{33}$ & 22 \\
e22vh\_medium1 & $>$2.3$\times$10$^{33}$ & 22 \\
\hline\hline
\end{tabular}
\end{center}
\end{table}
} 
\item{\Wen~events were collected with specialised triggers based on
    the missing transverse momentum\footnote{In a collider event, the missing transverse momentum is defined as the momentum imbalance in the plane transverse to the beam axis and is obtained from the negative vector sum of the momenta of all particles detected in the event.} \etmis~significance
    $x_{s}=\etmis/ ( \alpha ( \sqrt{\sum{\et}-c}) )$, where the sum runs over all energy deposits and the
    constants $\alpha$ and $c$ are optimised such that the denominator
    represents the \etmis~resolution. The $x_{s}$ variable offers the ability to
    suppress the background significantly, allowing the triggers to run
    unprescaled at any pile-up rate. An $x_{s}$ selection requirement was used in combination with
    an electron \et~cluster threshold of 10 or 13~GeV. During the
    2011 run, additional track-quality requirements were applied to
    the probe electron candidates. The \etmis~vector was required
    to be separated by at least $\Delta \phi=0.7$ from any jet with $\pt>10$~GeV, where the jets were reconstructed with the anti-$k_t$ algorithm~\cite{antikt} with distance parameter $R=0.4$. }
\item{\Jpsiee~events were collected with five dedicated prescaled di-electron triggers, mainly enabled towards the end of LHC fills, by requiring a candidate with \tpp\ identification criteria exceeding a minimum \et~threshold for the tag electron,  an electromagnetic cluster exceeding a minimum \et~threshold for the probe electron, and a tag--probe invariant mass between 1 and 6~GeV. These triggers are summarised in Table~\ref{tab:JpsiT}.
\begin{table}
\caption{\label{tab:JpsiT} Di-electron triggers used for collecting \Jpsiee~events. The first part of each trigger name indicates the threshold of the tight tag electron, while the second corresponds to the loosely selected probe one.  The di-electron mass is required to be in the 1--6~GeV mass range.}
\begin{center}
\begin{tabular}{ccc}
\hline\hline
Di-electron  & Tag electron & Probe electron \\
triggers                     & \et~threshold [GeV] & \et~threshold [GeV]\\ 
\hline
e5e4                 & 5 & 4 \\
e5e9                 & 5 & 9 \\
e5e14                & 5 & 14 \\
e9e4                 & 9 & 4 \\
e14e4                & 14 & 4 \\
\hline\hline
\end{tabular}
\end{center}
\end{table}
}

\end{itemize}
While the triggers used for the collection of \Wen\ and \Jpsiee\ events do apply some requirements on probe electrons and on the event topology, these are chosen to be
looser than the offline selection and thus do not impact the efficiency measurement. In the case of \Zee\ collection, it is ensured that the tag electron was sufficient to trigger the event, thus avoiding any bias on the probe properties.

\section{Identification efficiency measurement}

\label{sec:IDmeas} \subsection{Central-electron identification efficiency}

Events from \Wen, \Zee, and \Jpsiee\ samples are used to measure the
central-electron identification efficiencies for various
identification criteria, in the transverse energy range from 7 to
50~GeV and pseudorapidity range \abseta$<2.47$.

\subsubsection{Selection requirements and sample sizes} 
\label{sec:eventSel}

A common set of requirements is applied to all triggered events to
ensure good data quality and suppress contamination from background
events. All electron candidates, whether they be tag or probe
electrons, must be reconstructed within \abseta$<2.47$ with at least
six hits in the SCT and one in the pixel detector. The effect of these requirements is accounted for in
the reconstruction efficiency; see Section \ref{sec:RecoEff}. Tight
selection criteria are applied to the tagging object that triggered
the event, that is, to one of the two electrons in \Zee\ and \Jpsiee\
events or to \etmis\ in the case of \Wen\ events. For the case of
\Wen\ and \Zee\ candidates, the probe electrons must also satisfy a
requirement limiting the amount of leakage of the shower into the
hadronic calorimeter (also accounted for in the reconstruction efficiency;
see Section~\ref{sec:RecoEff}). Further criteria are imposed in each
channel to improve the separation between signal and background
events.

\emph{\Wen~channel:} A range of requirements is applied to the minimum
value of the transverse
mass\footnote{$m_\mathrm{T}=\sqrt{2\et\etmis(1-\cos\Delta\phi)}$ where
  $\Delta\phi$ is the azimuthal separation between the directions of
  the electron and missing transverse momentum.} $m_\mathrm{T}$ (40 to
50~GeV), and on the missing transverse momentum, \etmis\ (25 to
40~GeV), of the event in order to obtain event samples with differing
background fractions. A minimum transverse-energy requirement of
$\et>15$~GeV is applied to the probe electrons and the entire event is
discarded if more than one probe candidate in a given event satisfies
the \mpp\ criteria. Two additional requirements are imposed in order
to reduce contributions from hadrons misidentified as electrons. The
probe electron candidate is required to be separated from any $R=0.4$ anti-$k_t$ jet
with $\pt>25$~GeV found within a cone of
radius $\Delta R = 0.4$. Similarly, the \etmis\ vector must be separated
from jets with $\pt > 25$~GeV by at least an angular distance of
$\Delta \phi=0.7$. After the final selection, a sample of 6.8 million
\Wen\ candidate events was collected when requiring $\etmis > 25$~GeV
and $m_\mathrm{T}> 40$~GeV.

\emph{\Zee~channel:} The tag electron is required to have $\et>20$~GeV
and to lie outside the calorimeter transition region
($1.37<\abseta<1.52$). The probe electron must have $\et>15$~GeV and
be separated from any jet with $\pt>20$~GeV found within a cone of $\Delta
R = 0.4$. For each pair, the tag and the probe electrons are required to have opposite reconstructed charges.
A typical di-electron invariant mass range used in this
analysis is 80 to 100~GeV, although this range is varied in systematic
studies. After the final selection, a sample of 2.1 million probes
from \Zee\ candidate events with opposite-charge electrons is
extracted from the 2011 data set.

\begin{figure}
  \centering
  \includegraphics[width=.5\textwidth] {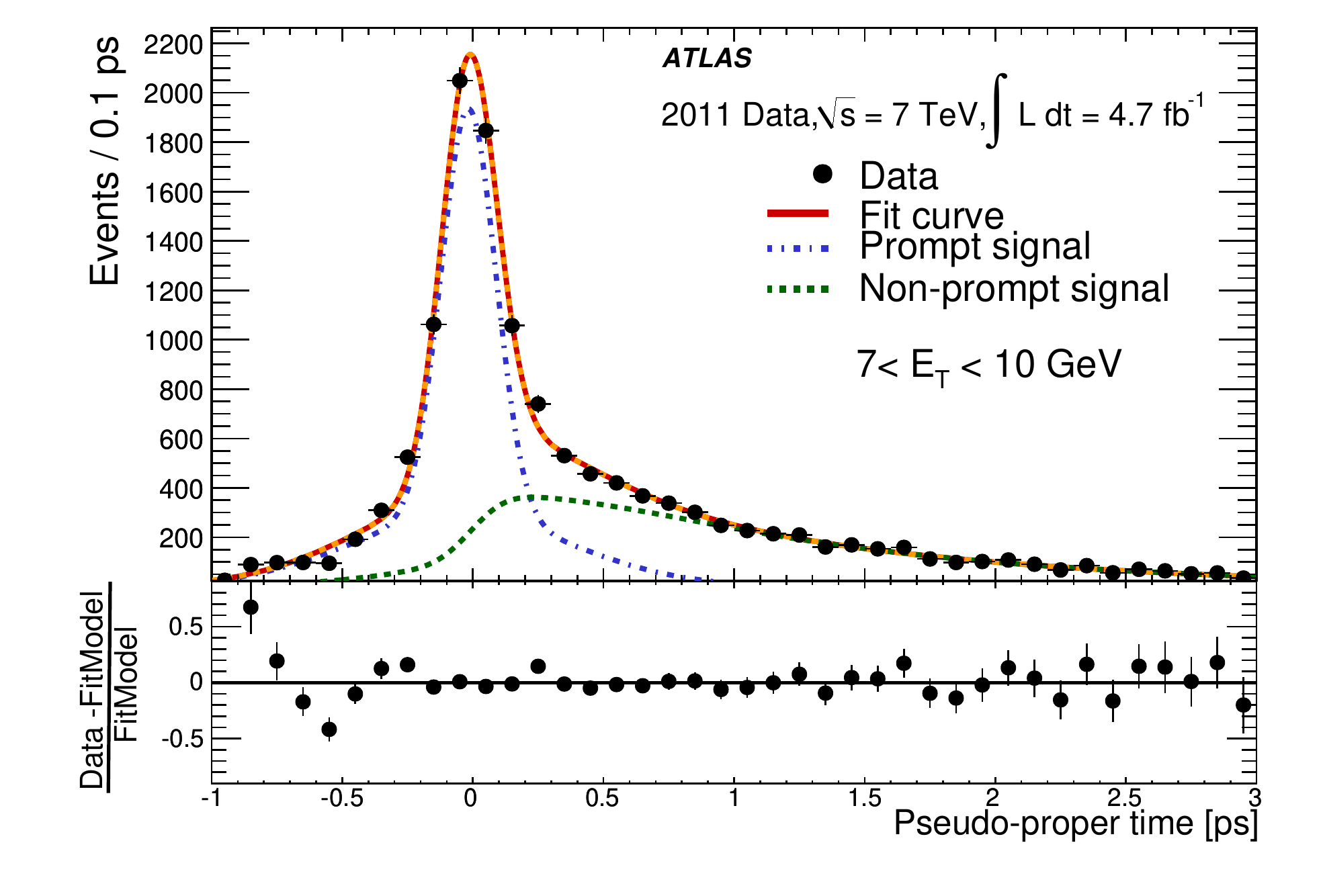}
  \caption{Pseudo-proper time fit of \Jpsiee\ candidate events for all
    selected probes within the \et\ range 7--10~GeV and integrated
    over \eta. The prompt contribution is modelled by two Gaussian
    functions, while the non-prompt component uses an exponential
    function convolved with two Gaussians. Points with error bars
    represent the data sample after background subtraction. The blue
    dashed line shows the prompt signal component while the non-prompt
    component is drawn with a dashed green line. The red curve is the
    sum of the fitted prompt and non-prompt components.}
  \label{fig:lifetimeFit}
\end{figure}

\emph{\Jpsiee~channel:} The \Jpsiee\ events come from a mixture of
both the prompt and non-prompt decays, with their relative fraction
depending both on the triggers used to collect the data and also on
the \et\ of the probe electrons. Given the difficulties associated
with the fact that electrons from non-prompt decays are often
surrounded by hadronic activity, two methods have been developed to
measure the efficiency for isolated electrons at low \et, both
exploiting the pseudo-proper time variable\footnote{The pseudo-proper time is defined
  as $t_{0}=L_\mathrm{xy}\cdot m^{\jpsi}_{\rm
    PDG}/p_\mathrm{T}^{\jpsi}$, where $L_\mathrm{xy}$ is the
  displacement of the \jpsi~ vertex with respect to the primary vertex
  projected onto the flight direction of the \jpsi\ in the transverse
  plane, $m^{\jpsi}_{\rm PDG}$ is the nominal
  \jpsi~mass~\cite{Beringer:1900zz} and $p_\mathrm{T}^{\jpsi}$ is the
  \jpsi~reconstructed transverse momentum.}. The first method, the
so-called ``{\it{short-lifetime method}}'' uses \Jpsiee\ decays
measured within very small values of the pseudo-proper time where the
prompt component is enhanced, thereby limiting the non-prompt
contribution ($f_\mathrm{NP}$) to 8--20\% of the yield. The second
method, the so-called ``{\it{lifetime-fit method}}'', uses the full
\Jpsiee\ candidate sample, corrected for the non-prompt fraction,
which is obtained by performing a fit of the pseudo-proper time distribution at
each identification stage. An example of this pseudo-proper time fit
is shown in Figure~\ref{fig:lifetimeFit}. For both \Jpsiee\ methods,
the main challenge is the suppression of the large background present
in the low electron \et~region. In order to reduce this background,
tighter requirements are imposed on the quantities measured with the
TRT hits associated with the tag electron, and the probe electron is
required to be isolated from surrounding energy deposits\footnote{Tighter TRT and isolation requirements are applied on the probe samples entering in both the numerator and denominator of the efficiency ratio; both criteria were verified in simulation not to affect the measured identification efficiency.}. Moreover,
both the tag and probe tracks are required to originate from the same
primary vertex and to be within 0.2~mm of each other in the
$z$-direction at the vertex $(x,y)$-position. The probe electron must
have $\et>5$~GeV. Both the tag and probe are permitted to point toward
the calorimeter transition region. After the final selection, a sample
of 120,000 \Jpsiee\ candidate events with opposite-charge electrons is
collected in the invariant mass range 2.8--3.3~GeV.

\begin{figure*}
  \centering
\subfigure[]{  \includegraphics[width=.45\textwidth] {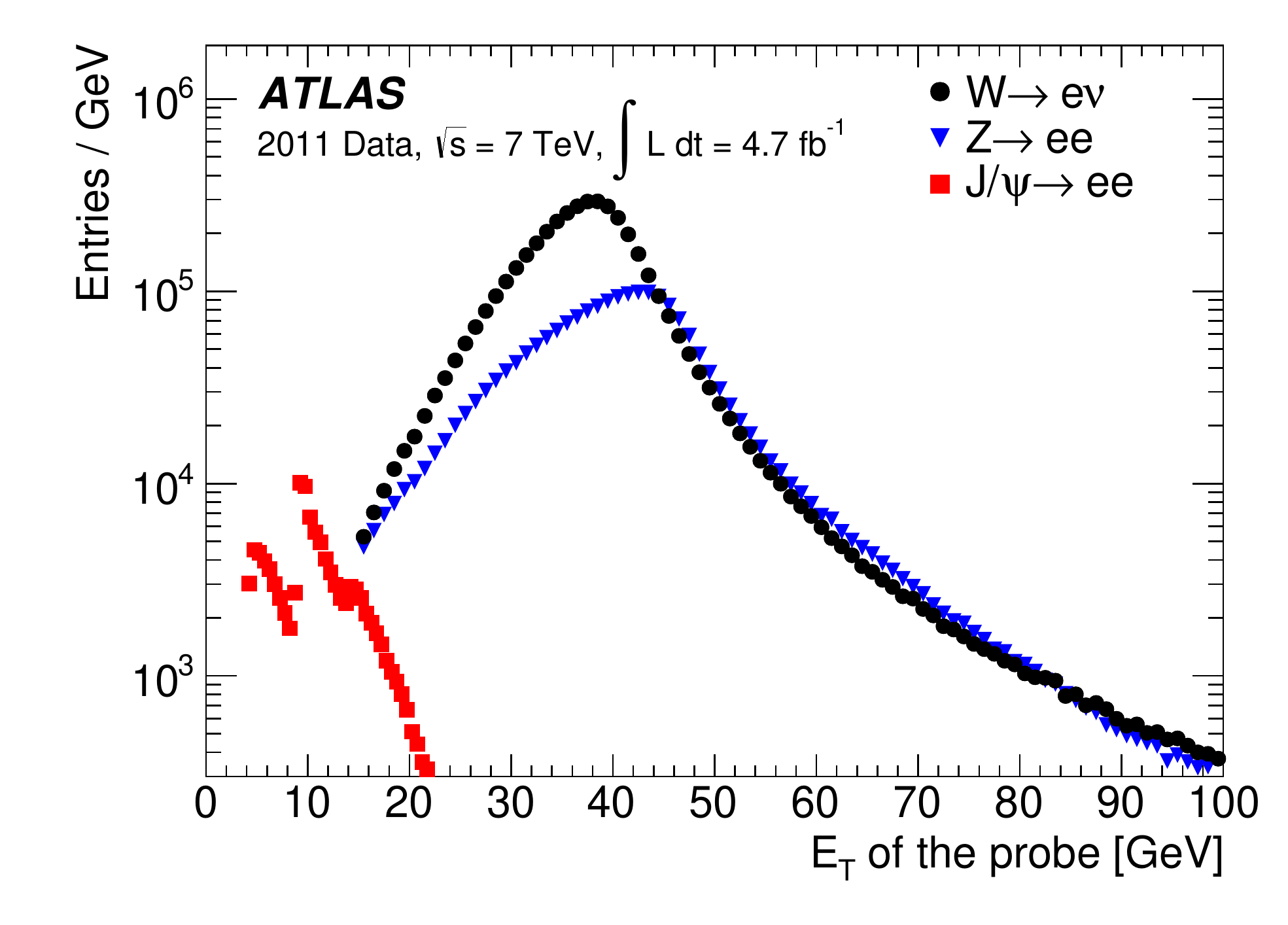}}
\subfigure[]{  \includegraphics[width=.45\textwidth] {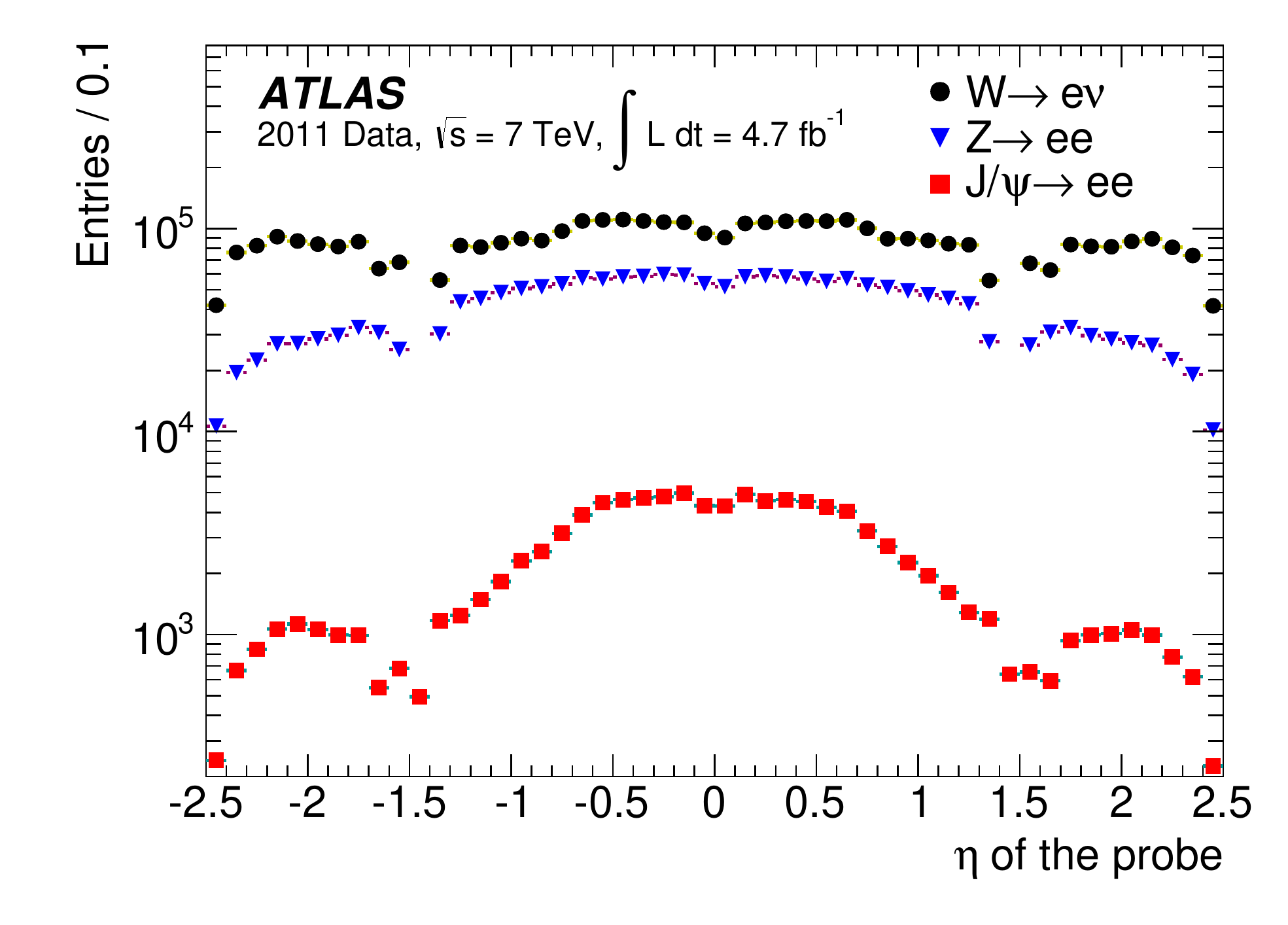}}
  \caption{Distributions of probe \et~in (a) and \eta~in (b) for the
    three samples of probes satisfying \tpp\ identification criteria.
    The non-continuous \et\ spectrum of the \Jpsiee\ sample is due to
    the different \et\ thresholds of the triggers utilised to collect
    this sample. }
  \label{fig:EtaEtsamples}
\end{figure*}

The \et\ and \eta\ distributions of \tpp\ electron probes for the
three \tandp\ samples are shown in Figure~\ref{fig:EtaEtsamples}.

\subsubsection{Background evaluation}
\label{sec:backgComp}

After the selections described in Section~\ref{sec:eventSel} are
applied to the data, the three samples still contain background
originating from hadrons misidentified as electrons as well as from
true electrons from photon conversions and non-isolated electrons
originating from heavy-flavour decays. For each sample and in each
(\et,\eta) bin, the level of background is evaluated by the use of
sensitive discriminating variables to build templates able to provide
some separation between signal and background events. These templates
are then either fitted or normalised to data to evaluate and subtract
the estimated background component in the signal sample.

\begin{figure*}
  \centering
  \subfigure[]{  \includegraphics[width=.45\textwidth] {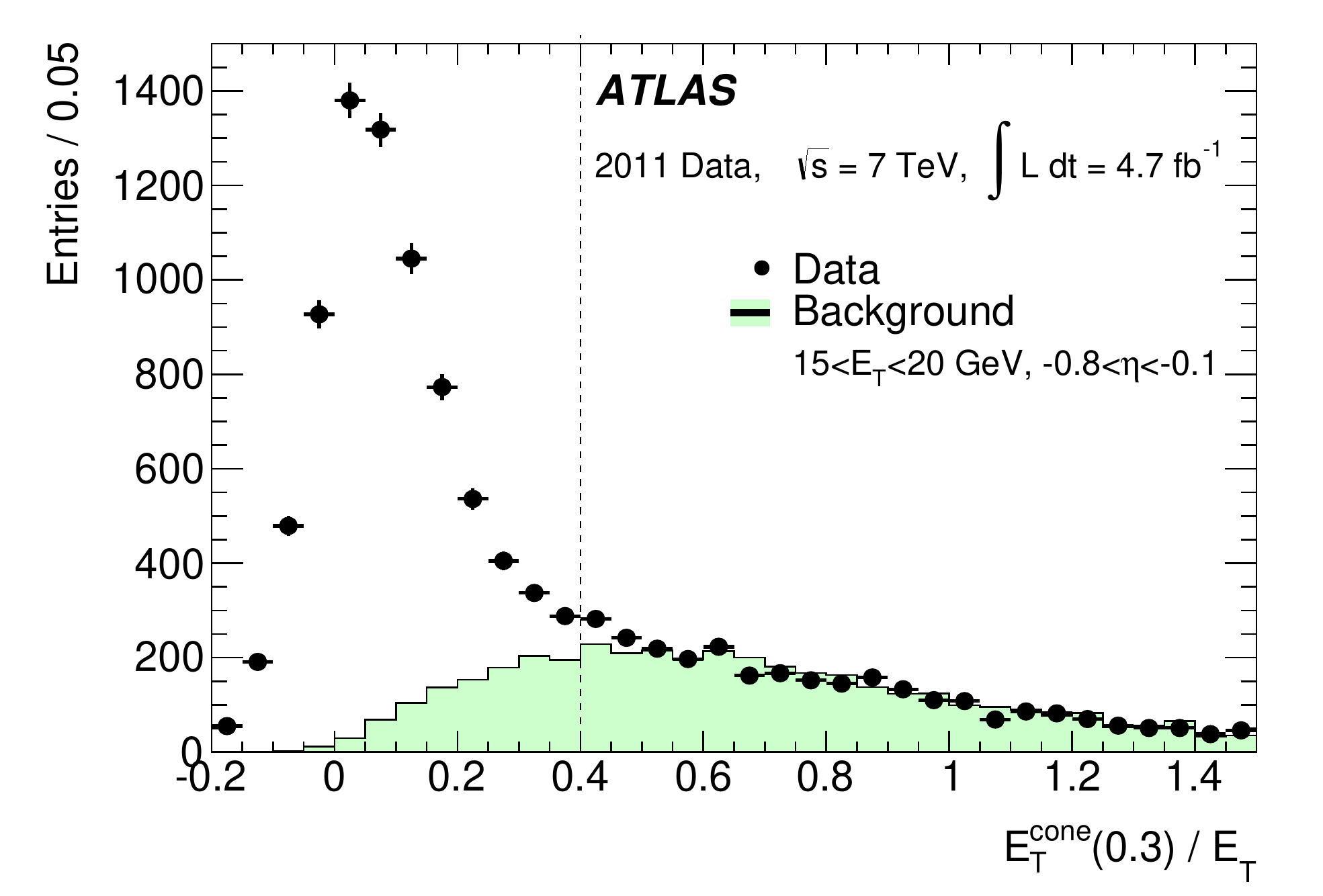}}
  \subfigure[]{  \includegraphics[width=.45\textwidth] {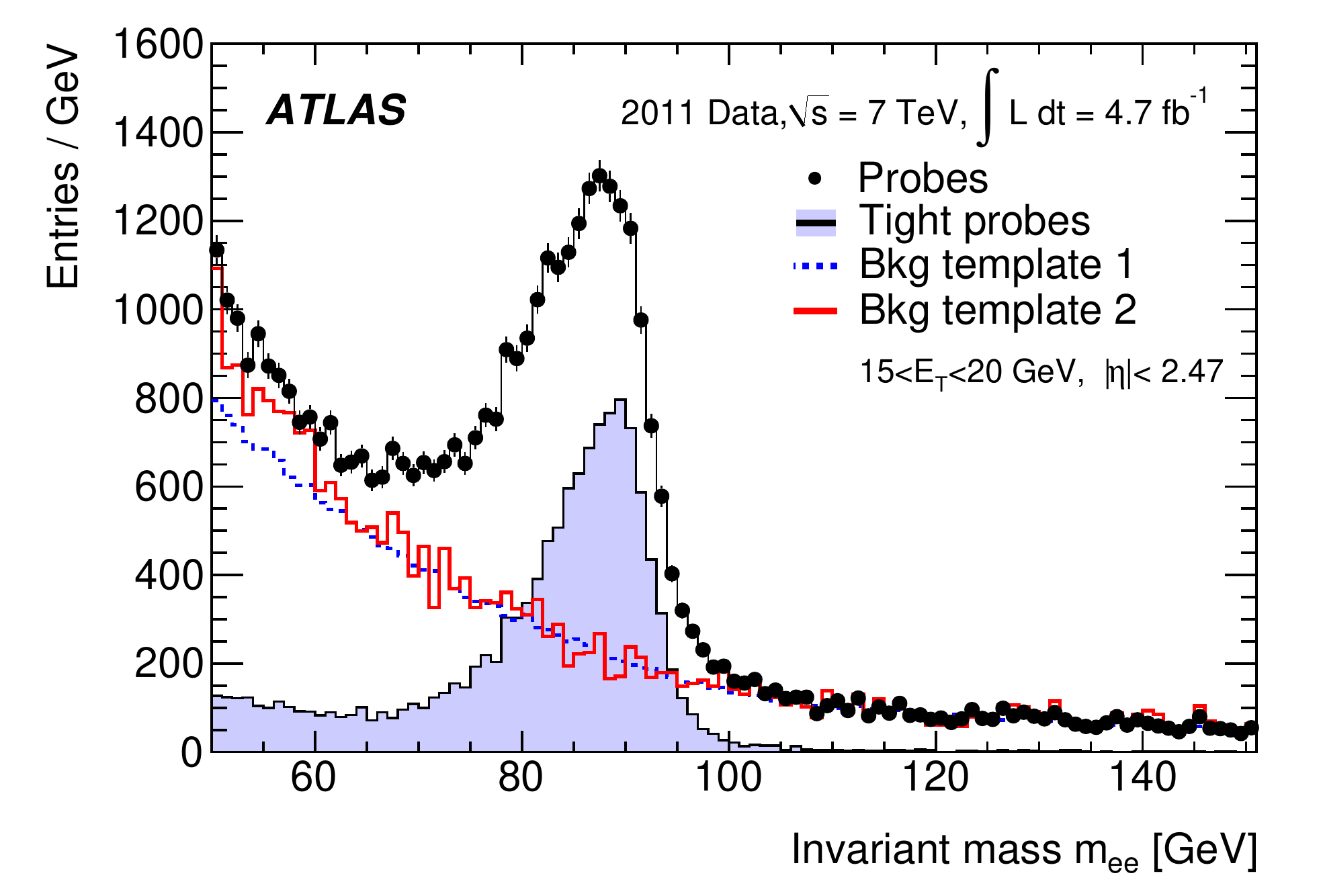}}
  \subfigure[]{  \includegraphics[width=.45\textwidth] {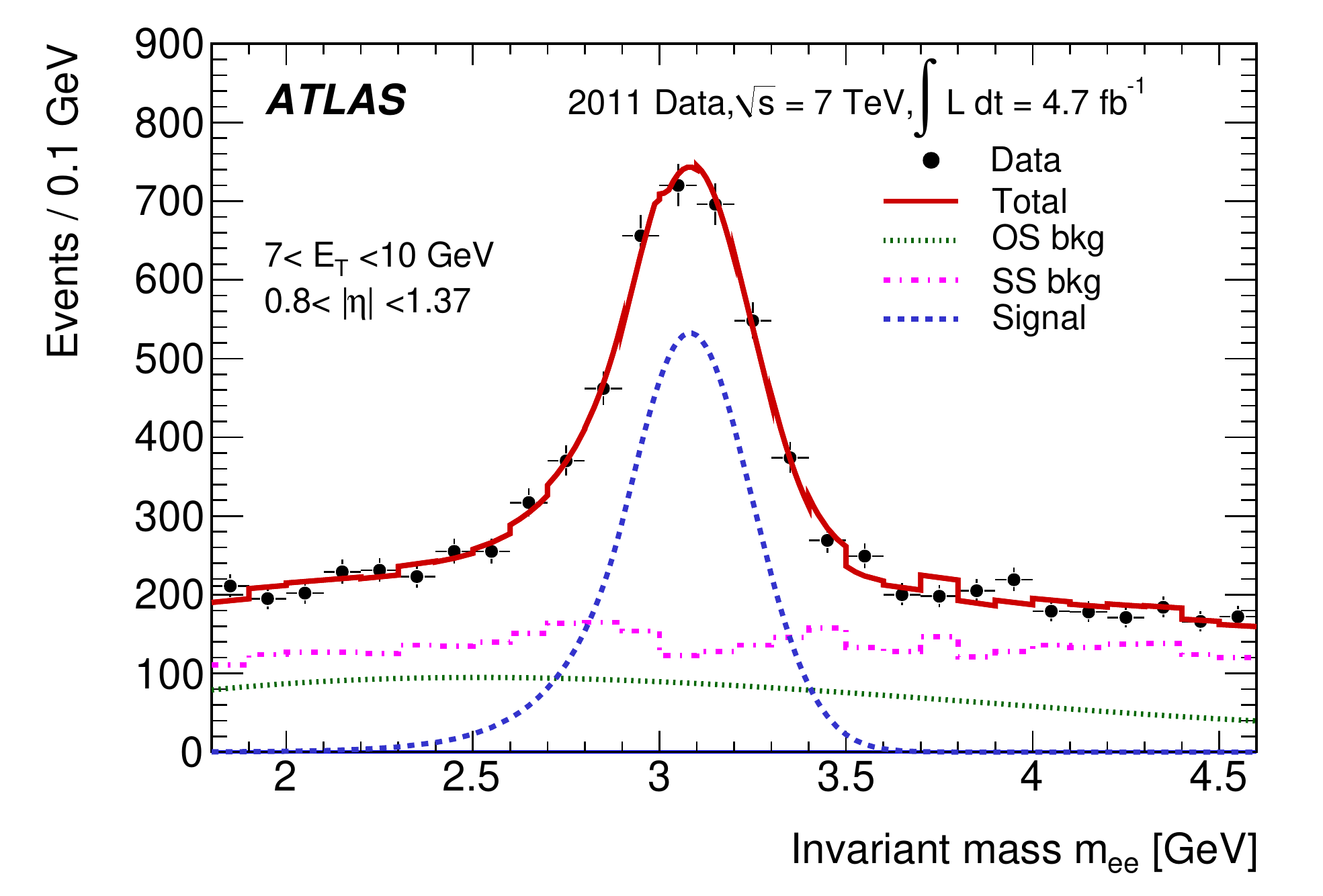}}
  \subfigure[] {\includegraphics[width=.45\textwidth] {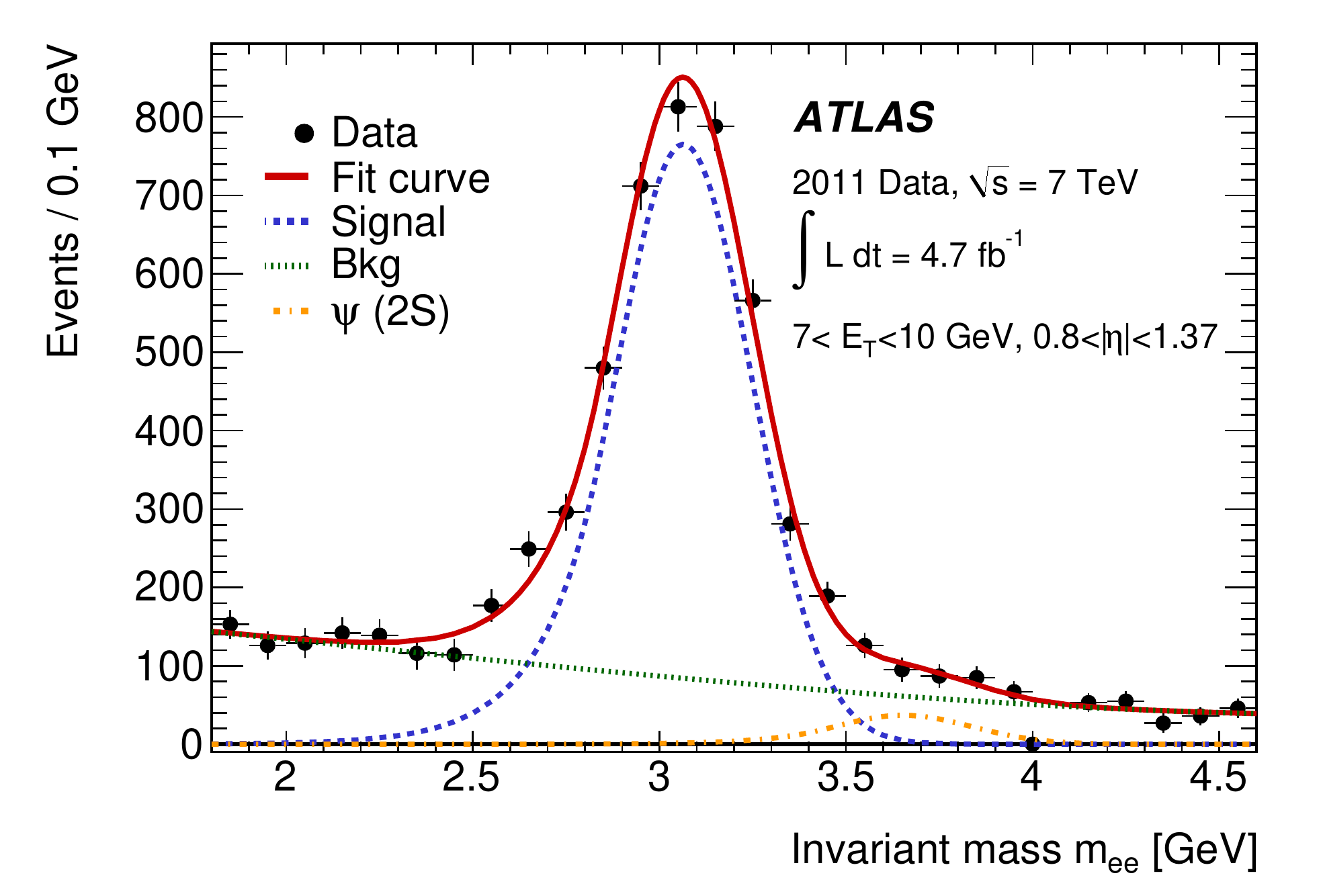}}
  \caption{Examples of discriminating variables and
    background-subtraction techniques for illustrative (\et,\eta)
    bins. (a) The \etcontr\ distribution of probes in the \Wen\ sample
    superimposed with the normalised background template. The black
    dashed line indicates the threshold chosen to delineate the signal
    and background regions. The \etcontr\ variable may take negative values due to the applied average corrections for electronic noise and pileup. (b) Invariant mass distribution in the
    \Zee\ sample. The normalised shapes of two different background
    templates are also shown (see text for details). The invariant mass for pairs where the probe satisfies the \tpp\ criteria is also shown. (c) Invariant
    mass distribution for the \Jpsiee\ sample in the short-lifetime
    range. The purple curve corresponds to the measured background
    with same-sign (SS) pairs, the dashed green line shows the
    opposite-sign (OS) background, the blue curve indicates the
    extracted signal and the red line is the fit to data taking into
    account signal, background, and $\psi(2S)$ (not shown in the
    figure) contributions. For presentational purposes the red line
    has been smoothed. (d) Invariant mass distribution for the
    \Jpsiee\ sample using the lifetime-fit method. Points with error bars
    represent the number of opposite-sign minus the number of
    same-sign data pairs, the fitted signal is drawn by the dashed
    blue line, and the $\psi(2S)$ resonance by the dashed orange line.
    The residual opposite-sign background is represented by the dashed
    green curve. }
  \label{fig:discrimPlot}
\end{figure*}

\emph{\Wen~channel:} Electron
isolation~\cite{Isoref1}
is used as the discriminating variable. Templates are built from the
sum of the transverse energies in the electromagnetic and hadronic
calorimeters contained in a cone of size $\Delta R= X$ around the
probe, excluding the probe's contribution. The size $X$ of the cone is
typically 0.3 or 0.4. This isolation variable is corrected on an
event-by-event basis for pile-up and underlying event
contributions~\cite{directphoton} and then normalised to the probe's
transverse energy. The resulting quantity is referred to as \etconX.
The background template is constructed from the probe selection by
reversing two of the electron identification criteria, namely the
total shower width $w_{\mathrm{stot}}$ and the ratio of high-threshold
hits to all TRT hits (see Table~\ref{tab:IDcuts}). To ensure adequate
statistics in each bin, the background templates are constructed in
(\et,\abseta) bins, assuming similar background at positive and
negative pseudorapidity values. In the outermost \abseta\ bins where
no information from the TRT is available, the template from the last
bin with TRT information is employed. A threshold requirement is
applied to the \etconX\ variable to separate the signal-dominated and
background-dominated regions located below and above this threshold,
respectively. The \etconX\ spectrum is normalised to the data in this
latter region and then used to estimate the background fraction in the
signal-dominated region located below the threshold.
Figure~\ref{fig:discrimPlot}(a) shows a typical \etcontr\ distribution
together with the normalised template shape. The signal-to-background
ratio S/B typically varies from 6 to 60 for probes with \et\ in the
ranges of 15--20~GeV to 35--40~GeV, respectively. After performing
this background subtraction, 5.2 million events remain in the signal
region. As part of the systematic uncertainties studies, templates are
also built by applying an additional reverse requirement on $R_{\phi}$
\footnote{$R_{\phi}$ is the ratio of the energy contained in
  3$\times$3 in ($\eta \times \phi$) cells, to the energy in 3$\times$7
  cells, computed in the middle layer of the EM calorimeter.} to the
original template selection. Both sets of templates adequately describe the
high \etconX\ tail while offering differing shapes close to the signal
region.

\emph{\Zee~channel:} Two discriminating variables are used to evaluate
the background yield in this channel. The first variable is the
invariant mass distribution $m_{ee}$ of the tag--probe pair. In this
case, the background template is constructed from events failing at
least two \lpp\ identification requirements and having a significant
energy deposit in a cone around the probe (see ``Bkg template 1'' in
Figure~\ref{fig:discrimPlot}(b)). This template is normalised to the
invariant mass distribution of reconstructed events in the high-mass
region of $m_{ee}>120$~GeV and then used to evaluate the background
fraction in the signal region (typically defined as $80<
m_{ee}<100$~GeV). A small correction of $\leq$1\% is performed to
account for $Z/\gamma^* \to ee$ signal contribution in the high-mass
tail. This is estimated from signal MC normalized to data in the peak
region after tight identification cuts. In comparison to using a
functional fit to describe the background shape, this method has the
advantage of providing reliable results over the entire (\et,\eta)
kinematic range. The second variable employed is the \etconX\ value of
the probe, as used in the \Wen\ channel and following the same
background subtraction techniques. A typical invariant mass
distribution is shown in Figure~\ref{fig:discrimPlot}(b). The S/B
ratio typically varies from 5 to 160 for probes with \et\ in the
ranges of 15--20~GeV and 35--40~GeV, respectively. After performing
this background subtraction, two million probes remain in the signal
region.

\emph{\Jpsiee~channel:} As for the \Zee\ channel, the discriminating
variable is the tag--probe invariant mass distribution. The $m_{ee}$
spectrum of opposite-sign pairs is fitted, typically in the range of
1.8 to 4.6~GeV, considering four distinct components. Two Crystal Ball
functions \cite{CBfunction} separately model the signal shape and that
of the $\psi(2S)$ resonance (the latter function is centred on the
nominal PDG~\cite{Beringer:1900zz} value). The background contribution
in the signal region is largely modelled by same-sign pairs as
measured in data, with an additional Chebychev polynomial used to
model the remaining background from opposite-sign pairs. For the
short-lifetime method, these contributions are fitted to the $m_{ee}$
spectrum as measured in data to evaluate the background contribution
in the signal region (see Figure~\ref{fig:discrimPlot}(c)). For the
lifetime-fit method, an unbinned maximum likelihood fit is performed,
where same (opposite)-sign pairs are considered with a negative
(positive) weight (see Figure~\ref{fig:discrimPlot}(d)). The \Jpsiee\
sample suffers from a higher background contamination than the other
two channels such that the S/B ratio in the typical signal extraction
range of $2.8<m_{ee}<3$~GeV varies between $\sim$0.5 and $\sim$3.
After performing the background subtraction, 88,000 (66,000) events
remain in the signal region in the full (short) pseudo-proper time
range.

\subsubsection{Identification efficiency measurement systematics}
\vspace{-0.1cm}
For all three channels, the dominant systematic uncertainties are
related to the evaluation of the background contribution to the signal
region. Possible biases affecting the efficiency measurement are
investigated by varying the selection of events such that the
signal-to-background ratio is modified substantially or by
re-evaluating the efficiencies with alternative templates or
background models. Each analysis is repeated with a large set of
variations and the spread of the corresponding results is used to
quantify the systematic uncertainties. These variations are designed
to allow a reasonable modification of the S/B ratio depending on the
background level affecting each mode.

\emph{\Wen~channel:} The baseline sample of \Wen\ events is varied by
using alternative \etmis\ and $m_{\mathrm{T}}$ selection requirements,
and by changing the isolation discriminating variable (\etconqu\ and
\etcontr) as well as its associated threshold requirement used to
delineate the signal and background regions. For each variation, both
sets of background templates are used to normalise the isolation distributions
above the thresholds. Within the 80 variations used, the S/B ratio
distribution in the signal region exhibits an RMS (Root Mean Square)
of $\sim 30$\% at low \et\ (15--20~GeV) and $\sim 25$\% at high \et\
(35--40~GeV). The combined effect of the charge misidentification and
the different $W^+$ and $W^-$ production cross-sections at the LHC
leads to an up to 5\% difference in efficiency using the \tpp\
criteria between $e^+$ and $e^-$ in the calorimeter endcap bins for
probes with $25<$ \et~$<30$~GeV. This difference is very well modelled
in the MC efficiency, leading to a negligible uncertainty for most
analyses.
 
\begin{figure*}
  \centering
  \subfigure[]{  \includegraphics[width=.45\textwidth] {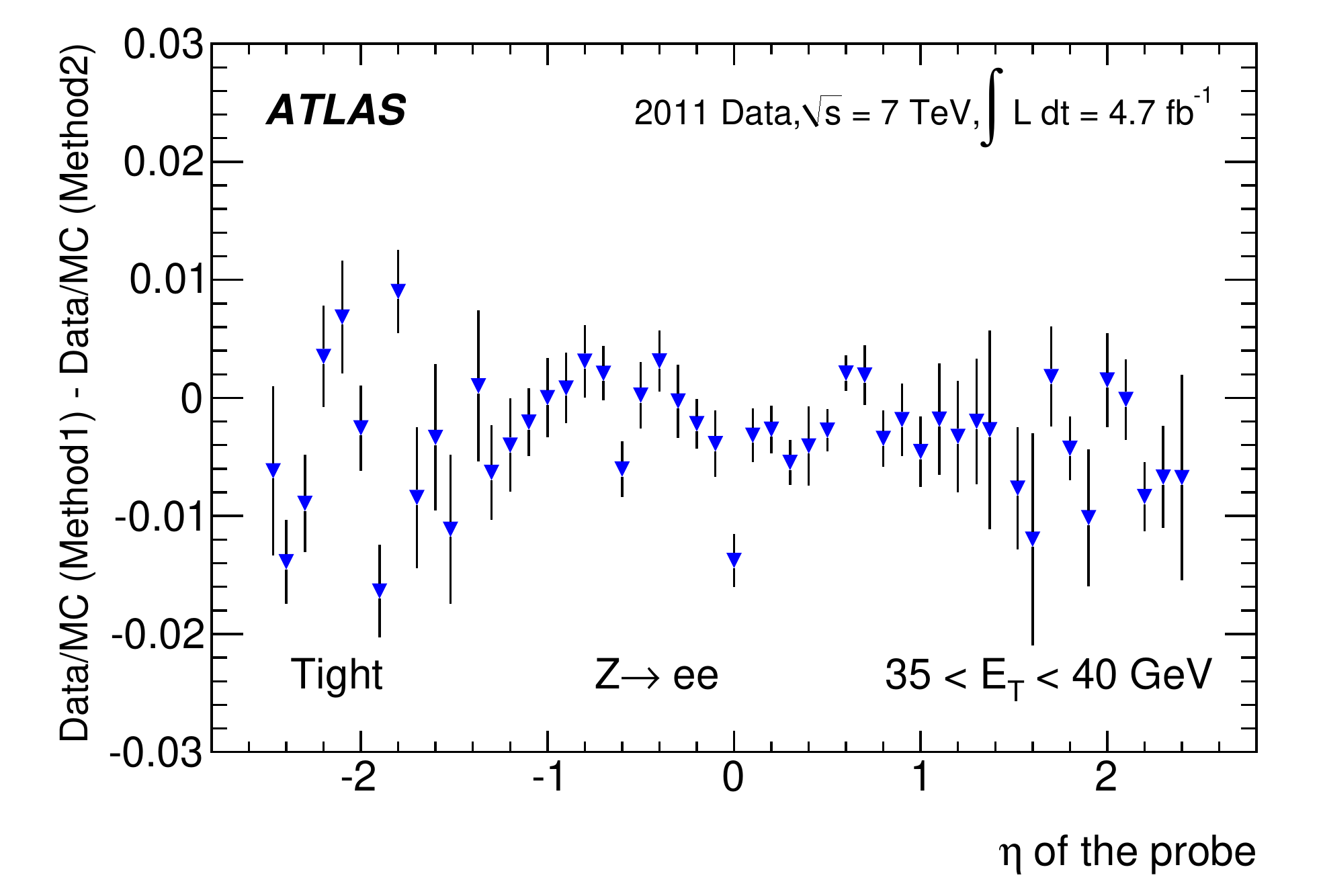} }
  \subfigure[]{  \includegraphics[width=.45\textwidth] {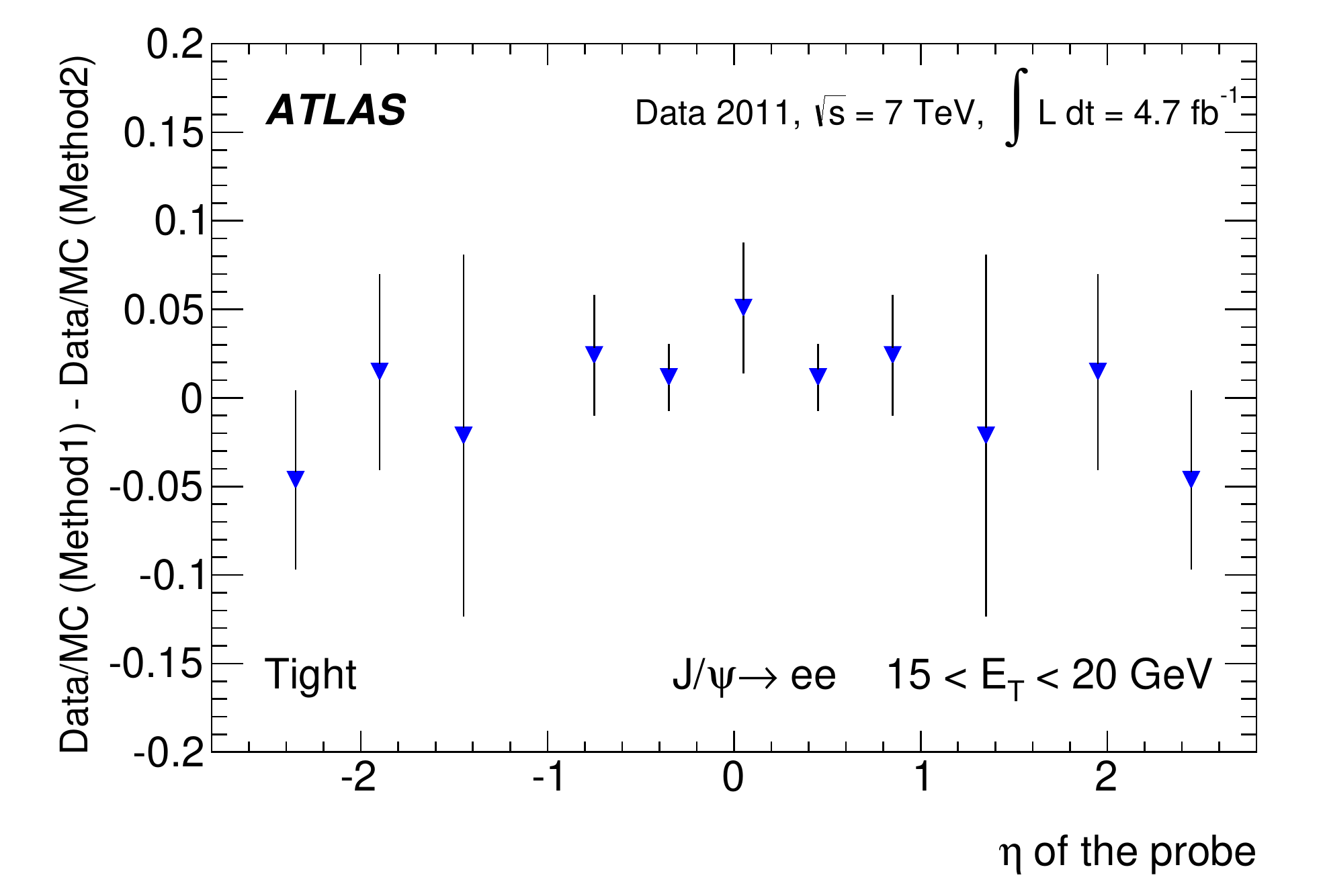}}
  \caption{(a) Data-to-MC efficiency ratio difference between the two
    methods to estimate background (Method 1: invariant mass, Method 2: isolation) used in
    the \Zee\ analysis for central electrons, for the \tpp\ criteria and for probes in the
    35--40~GeV \et\ bin. (b) The same difference for the lifetime-fit
    (Method 1) and short-lifetime (Method 2) methods used for the
    \Jpsiee\ analysis for \tpp~criteria and for probes in 15--20~GeV
    bin. In both figures, the error bars represent only the systematic
    uncertainties associated with the individual methods.}
  \label{fig:Methodcomp}
\end{figure*}

\emph{\Zee~channel:} The baseline sample of \Zee\ events is modified
by using alternative selection criteria defining the tag electrons.
Three $m_{ee}$ windows (80--100, 70--100 and 75--105~GeV) are used to
extract the signal events. Moreover, the size and composition of the
background are varied by modifying the reverse requirements used to
generate the templates. As an example, the curves ``Bkg template 1 and
2'' in Figure~\ref{fig:discrimPlot}(b) are similar in that the events
used to build these templates are required to fail some of the \lpp\
identification requirements (template 1 fails at least two
requirements while template 2 fails three) and have a significant
energy deposit in a cone around the probe. However, in contrast to
template 1, template 2 is also built from events passing additional
track-quality requirements and having little hadronic activity
associated with the candidate. In the case where the invariant mass is
the discriminating variable, an isolation condition
($E_{\mathrm{T}}^{\rm{cone}} (0.4) <5$~GeV) is optionally applied to
the tag requirement. A total of 36 variations are performed, for which
the S/B ratio distribution exhibits an RMS of $\sim 10$\%. In the case
where the isolation of the probe electron plays the role of
discriminating variable, the radius of the isolation cone and its
associated threshold are also varied, giving in total 120 variations.
The method employing the invariant mass as the discriminating variable
is used as the primary efficiency measurement. However, the
efficiencies computed using either variable agree well with each other
within the systematic uncertainties. Figure~\ref{fig:Methodcomp}(a)
shows the differences of the data-to-MC \tpp~efficiency ratios between
the two methods in the $\et=$~35--40~GeV bin, which are generally
compatible with zero within less than two standard deviations;
these differences are considered as additional uncorrelated systematic
uncertainties on the primary measurement.

\emph{\Jpsiee~channel:} The baseline sample of \Jpsiee\ events is
similarly modified by using alternative selection criteria to define
the tag electron (additional isolation criteria, tight TRT
requirements) and by enlarging the 2.8--3.3~GeV mass window defining the signal range. The functional fit for the
background from opposite-sign pairs is modified to assess the
uncertainty on the background subtraction (using Chebychev polynomial
functions or exponential fits). The range and the function used for
the pseudo-proper time fit as well as the size of the isolation cone
and its associated threshold are also varied. Both the track-based and
energy-based isolation criteria are investigated. A total of 76 and 52
variations resulting in an S/B ratio distribution RMS of $\sim 30$\%
are used for the lifetime-fit and the short-lifetime methods,
respectively.

The method using the short-lifetime range relies on the non-prompt
fraction, $f_{\mathrm{NP}}$, extracted from the $J/\psi$ differential
cross-section measurement~\cite{jpsimumuPaper}, which is used to combine the MC samples corresponding to prompt and non-prompt \Jpsi\ production. Selections
targeting further suppression of the non-isolated probes decrease
$f_{\mathrm{NP}}$, as expected, and this variation is taken into
account as predicted by simulation. The non-prompt fraction increases
with the probe \et\ and is found to be independent of \eta. It enters
into the computation of the combined MC efficiency prediction with an
uncertainty of 10\%. In contrast, the lifetime-fit method extracts
$f_{\mathrm{NP}}$ from the data, by fitting the lifetime distribution
in the range from --1 to $+3\,\mathrm{ps}$. As in the first method,
this fraction is computed in bins of \et\ only, since no significant
variation was observed as a function of \eta. Systematic uncertainties
on the value of $f_{\mathrm{NP}}$ obtained from data are assessed by
varying the range and the function used in the fit. The results from
the two methods agree reasonably well, within the total uncertainties,
as shown in Figure~\ref{fig:Methodcomp}(b) where the difference of the
data-to-MC \tpp\ efficiency ratios between the two methods is shown
for the bin $\et=$~15--20~GeV. There is an approximate 75\%
statistical overlap between the candidates selected by the two
methods. In the final combination, both the short-lifetime and
lifetime-fit methods are treated as variations of a single
measurement.

\begin{figure}
  \centering
  \includegraphics[width=.5\textwidth] {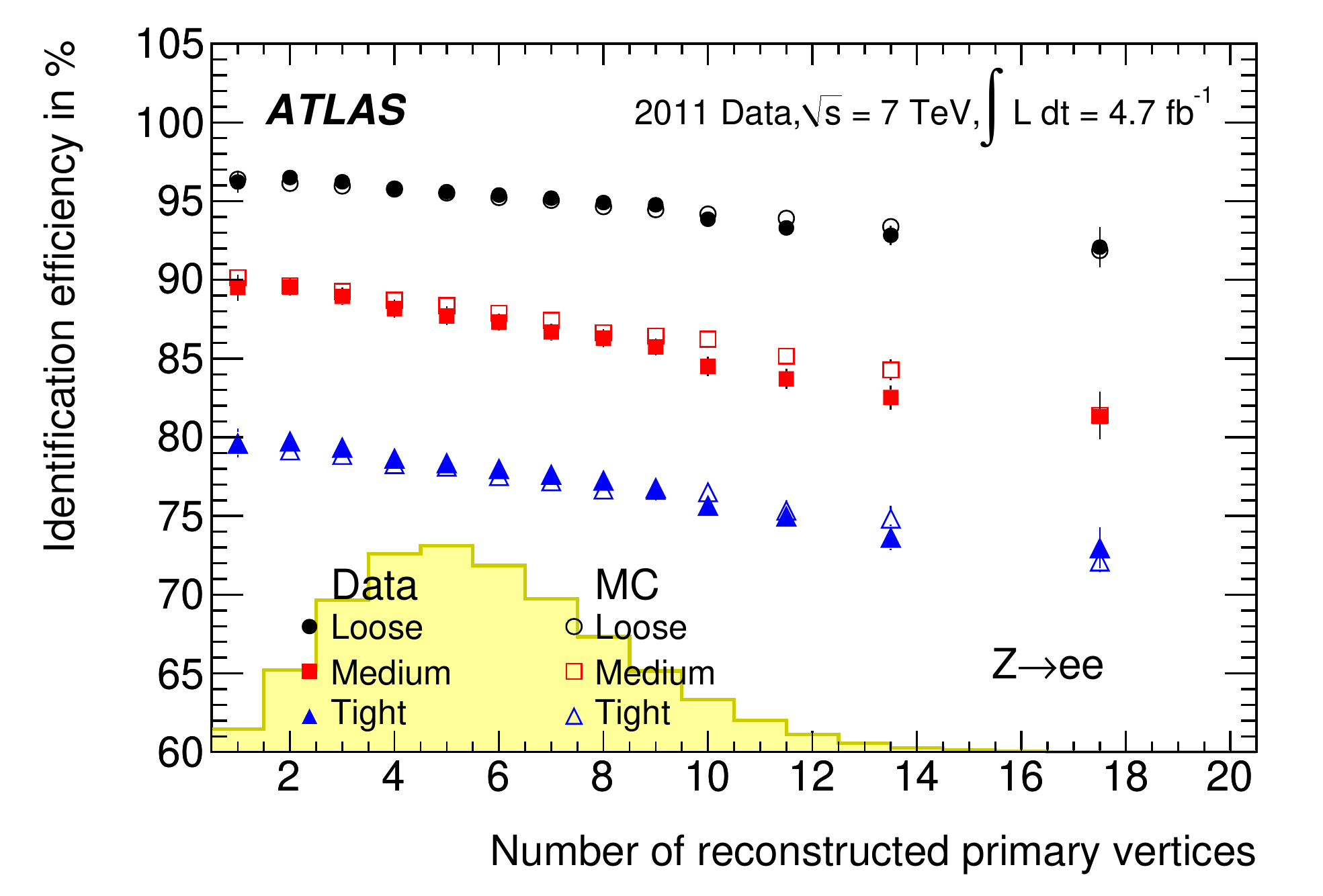}
  \caption{The \lpp, \mpp, and \tpp\ identification efficiencies as a
    function of the number of reconstructed primary vertices in the
    event, for \Zee~events and for central-electron probes in the \et~range 15--50~GeV.
    The quoted error bars correspond to the total uncertainties. The
    observed loss in efficiency is well modelled by the simulation.
    The yellow histogram indicates the $N_{\rm{PV}}$ distribution in
    data. }
  \label{fig:PileupId}
\end{figure}

The steady increase of the instantaneous luminosity during the 2011
period induced pile-up effects that varied proportionally to the
average number of interactions per beam crossing. Increased pile-up
causes higher-energy deposits in the calorimeters and more tracks in
the inner detector, which may impact the electron reconstruction and
identification. These effects are confirmed when measuring the
identification efficiency with \Zee~events as a function of the number
of reconstructed primary vertices in an event (see
Figure~\ref{fig:PileupId}), where the efficiency is seen to drop by up
to 2\% and 5\% for the \lpp\ and \tpp\ criteria, respectively. These
effects are well modelled by simulation with a maximum difference of
approximately two standard deviations observed in the case of
\mpp~criteria. Variations of the pile-up simulation and of the
weighting procedure applied to the simulation to match the pile-up
conditions observed in data impact the efficiency at the per mil
level.

\subsubsection{Combination and results}

\begin{figure*}
  \begin{center}
    \subfigure[]{\includegraphics[width=0.47\textwidth]{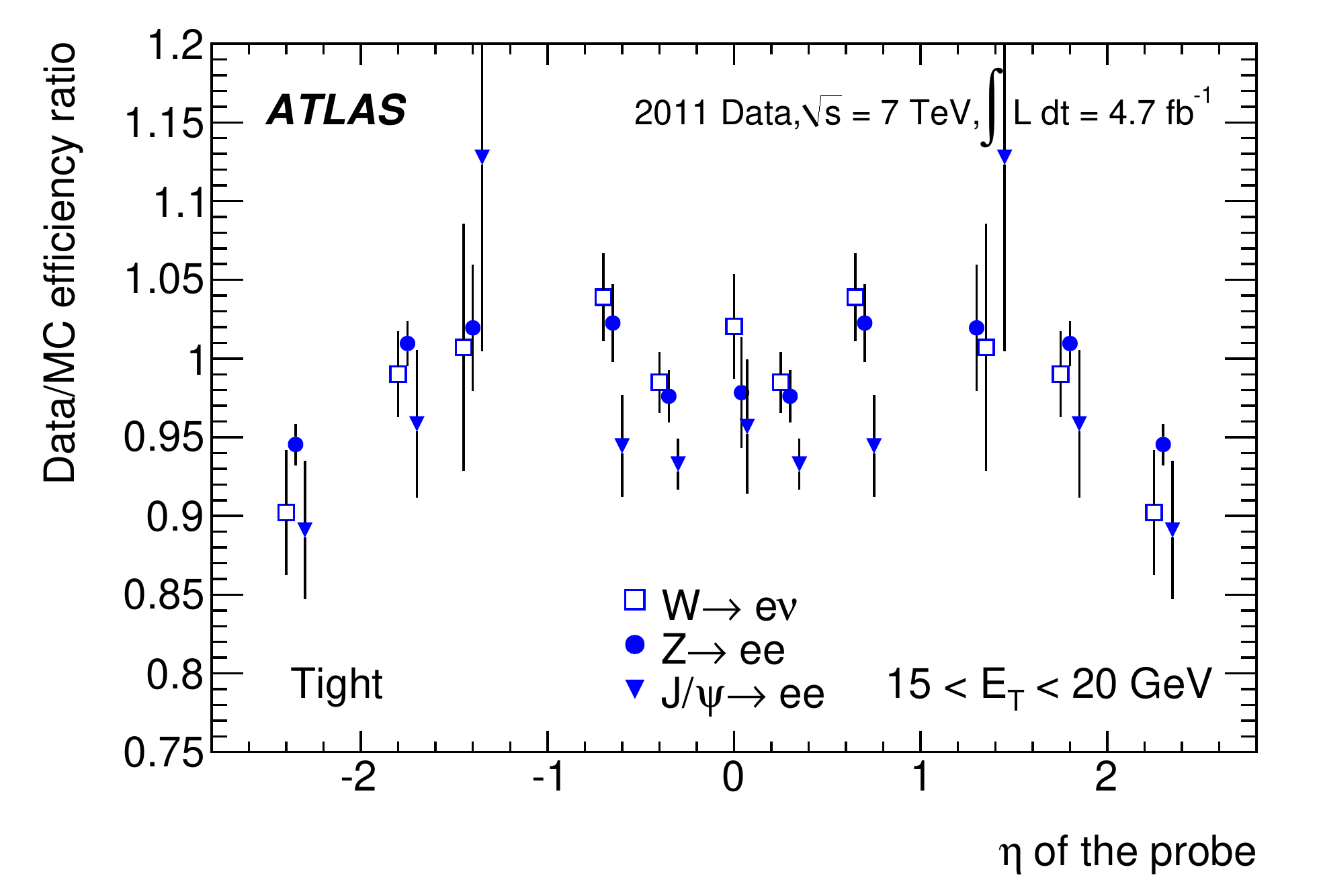}}
    \subfigure[]{\includegraphics[width=0.47\textwidth]{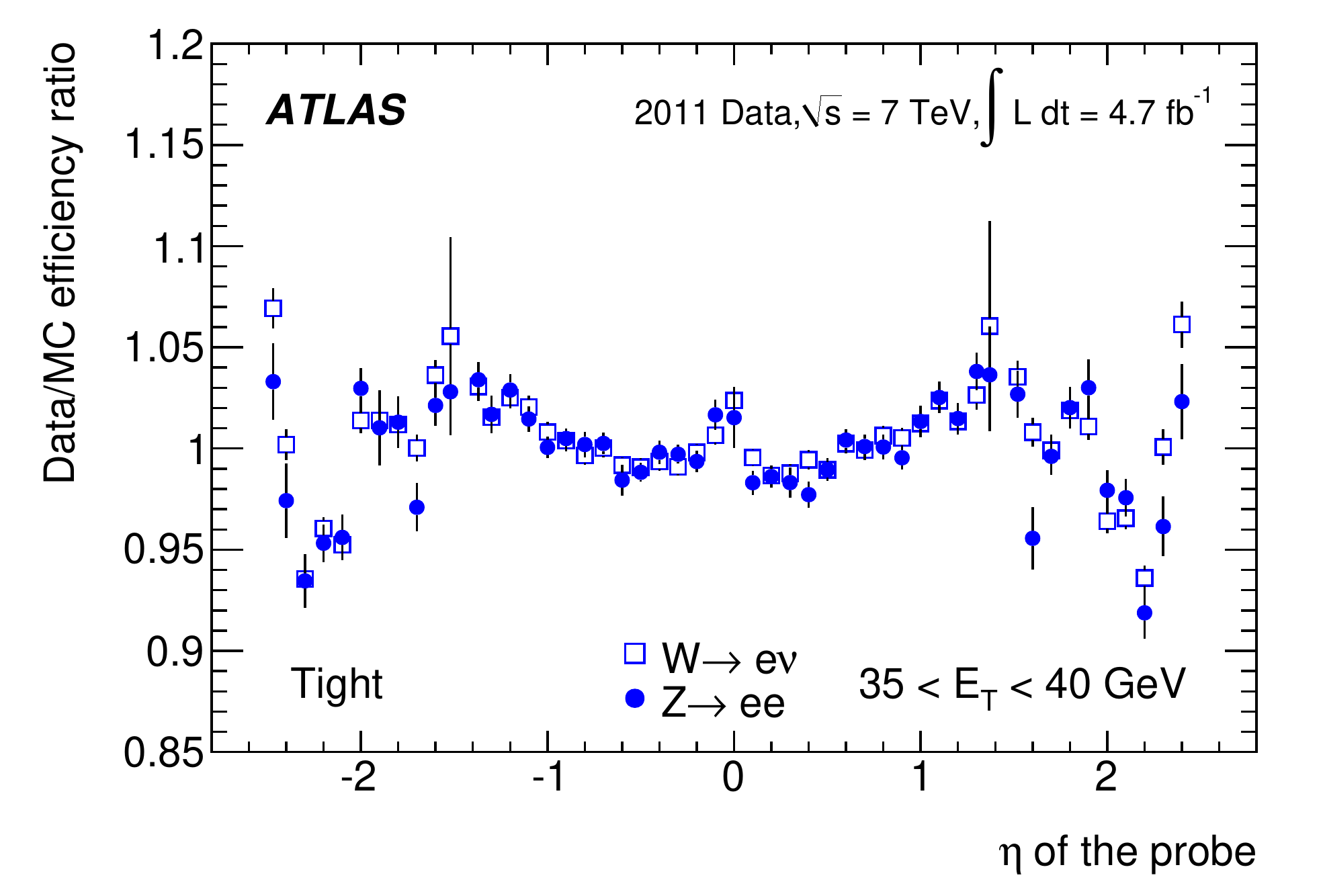}}
  \end{center}
  \caption{Comparison of the scale factors extracted from the various
    channels in two \et~bins, shown as a function of the \tpp\
    probe-electron pseudorapidity. In (a), scale factors from \Zee,
    \Wen, and \Jpsiee\ are compared in the \et~range 15--20~GeV. In
    (b), scale factors from \Zee\ and \Wen\ are shown in the \et~range
    35--40~GeV. The error bars correspond to the total uncertainties
    in each channel. Some points are slightly shifted horizontally
    within the \eta~bin for better visibility.}
  \label{fig:compSF}
\end{figure*}

The \Zee, \Wen, and \Jpsiee\ channels are statistically independent
and so are combined to increase the precision of the identification
efficiency measurements. Although the efficiencies in a given
(\et,\eta) bin may be slightly different in each channel due to
effects related to e.g. resolution and migration effects or the
influence of the trigger, these differences are expected largely to cancel
when taking the data-to-MC efficiency ratios, referred to as
\emph{scale factors} (\emph{\SF}). The combination of the three
channels is therefore performed by first calculating the corresponding
scale factors in a double-differential binning in electron \et\ and
\eta. As examples, the scale factors of the different channels are
shown for two illustrative \et~bins in Figure~\ref{fig:compSF}. The
agreement among the channels is in general fair, with the most notable
discrepancy observed in the \et~range of 15--20~GeV where the
\Jpsiee~results in the barrel region are lower than for \Zee~and
\Wen~with a significance of approximately two standard deviations.

A global $\chi^{2}$ minimisation~\cite{Aaron:2009bp} is used to compute an average value of the $\SF^{i}$ in each bin $i$ common to all channels:
\begin{align*}
\chi^2 = &\sum_{i,k} \frac{\left[\mu^{i,k} - \SF^i - \sum_{j}\gamma^{i,k}_{j}\SF^i b_j\right]^2}
{\left(\delta^{i,k}_\mathrm{sta}\right)^2\mu^{i,k} \SF^i\left(1 - \sum_{j}\gamma^{i,k}_{j}b_j\right) +
\left(\delta^{i,k}_\mathrm{unc}\SF^i\right)^2} 
\\+ &\sum_{j}b^2_j ~~, \label{eq:chidef}
\end{align*}
where $i$, $k$, and $j$ indices run over the (\et,\eta) bins, the
three channels, and the correlated systematics, respectively. The
latter are extracted from the systematic variations used to compute
the scale factor $\mu^i_k$ in each channel. The variables
$\delta^{i,k}_\mathrm{sta}$, $\delta^{i,k}_\mathrm{unc}$, and
$\gamma^{i,k}_{j}$ represent the relative statistical, uncorrelated, and correlated systematic uncertainties,
respectively. The nuisance parameters $b_{j}$ are related to
correlated uncertainties, which are dominated by the background
subtraction uncertainties. The combined scale factors are given by
$\SF^i$.

During the minimisation procedure, the central values of the scale
factors may be shifted by an amount which is a fraction of the
correlated uncertainties, such that the minimal $\chi^{2}$ 
is reached. In 0.5\% of all bins, the absolute value of the
pull~\footnote{The pull gives the deviation from the average value of
  a measurement in units of standard deviation.} is larger than two.
To be conservative, the uncorrelated uncertainties are in this case
inflated by the pull divided by $\sqrt{2}$ and the global minimisation
is performed once again. The combination of independent measurements
constrains the bin-to-bin correlated uncertainties and reduces their
size by up to about 30\%, thereby reducing the total uncertainty. This
reduction is most significant in the range $\et=$~25--40~GeV, where
the \Zee\ and \Wen\ measurements have the highest statistical
precision.

\emph{High-\et\ measurements:} $\et>20$~GeV. In this region, copious
statistics from the low background \Zee\ and \Wen\ channels are
available and so the measurement is performed in all three \eta\
granularities (\emph{coarse, middle, fine}). The total uncertainty in
this region is at most 1--2\% for \tpp~electrons. In general, the
precision reaches the few per mil level at 35~GeV and is statistically
limited.

\emph{Low-\et\ measurements:} $7<\et<20$~GeV. In this region, the
measurement is driven by the \Jpsiee\ sample, although in the
15--20~GeV bin results from both \Wen\ and \Zee\ are also used in the
combination. In this range, only the coarse \eta\ binning is used due
to the statistics available for the measurements. The measurement is
limited by the statistical precision and the total uncertainty varies
from 3\% in the calorimeter barrel regions to 7\% in the endcap
regions.

\begin{figure*}
  \begin{center}
\subfigure[]{    \includegraphics[width=0.47\textwidth]{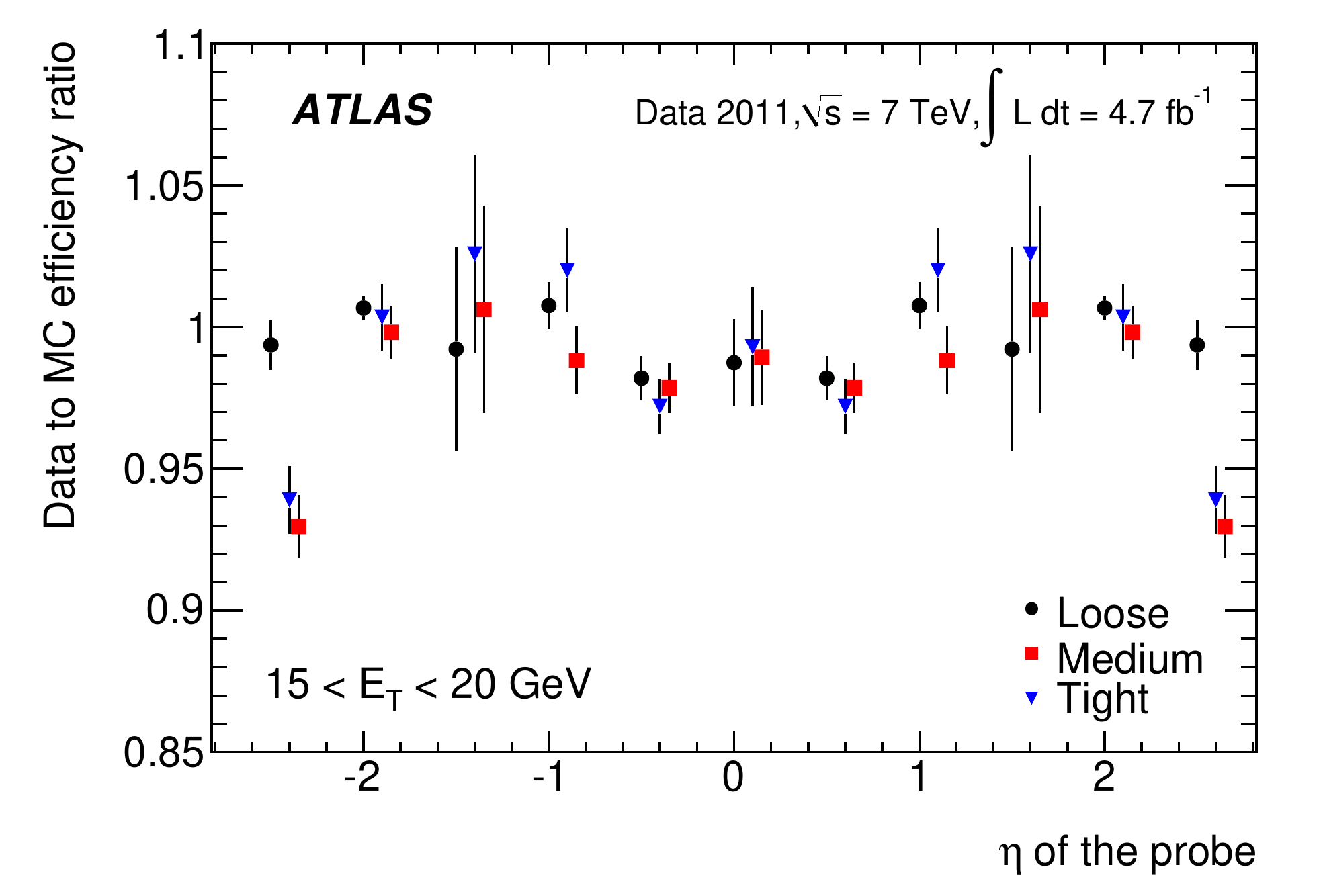}}
\subfigure[]{    \includegraphics[width=0.47\textwidth]{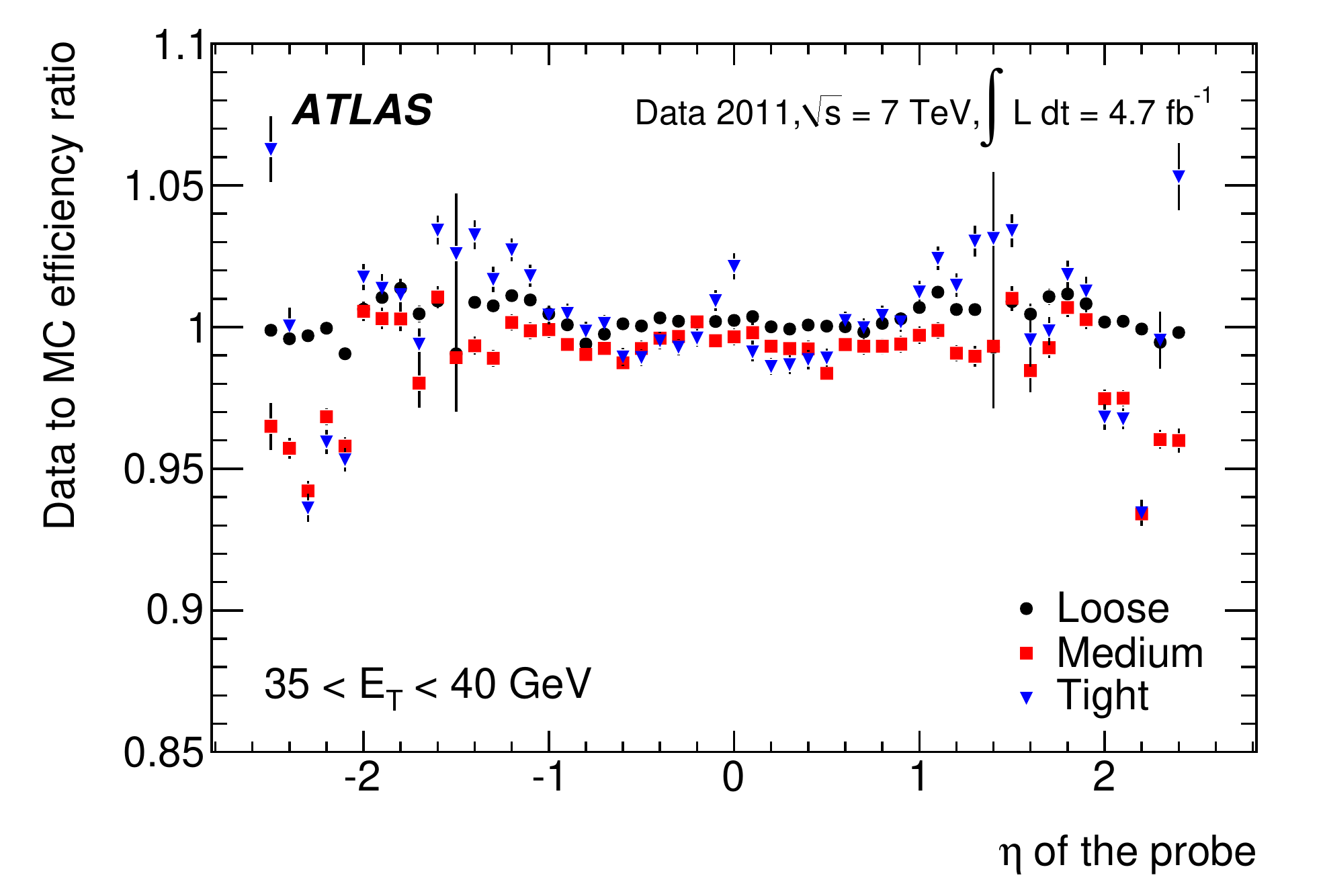}}
  \end{center}
  \caption{Examples of combined scale factors for the three
    identification criteria (\lpp, \mpp, \tpp) as a function of the
    probe-electron pseudorapidity. Results are shown for 15--20~GeV
    in (a) and 35--40~GeV in (b) probes. In each \eta~bin, the points
    for \lpp, \mpp~and \tpp~criteria are slightly shifted horizontally
    for better visibility. The error bars indicate the total
    uncertainties. }
  \label{fig:SFex}
\end{figure*}

Figure \ref{fig:SFex} illustrates some of the combined scale factors
at low and high probe-electron \et\ resulting from this minimisation
procedure. These scale factors are used in all analyses involving
electrons, to correct for residual differences between data and
simulation that are mainly due to the modelling of the shower shapes in the
calorimeter and to the TRT detector calibration in the region
$1<\abseta<2$. These corrections are usually no more than a few
percent.

\begin{figure*}
  \begin{center}
    
    \includegraphics[width=0.32\textwidth]{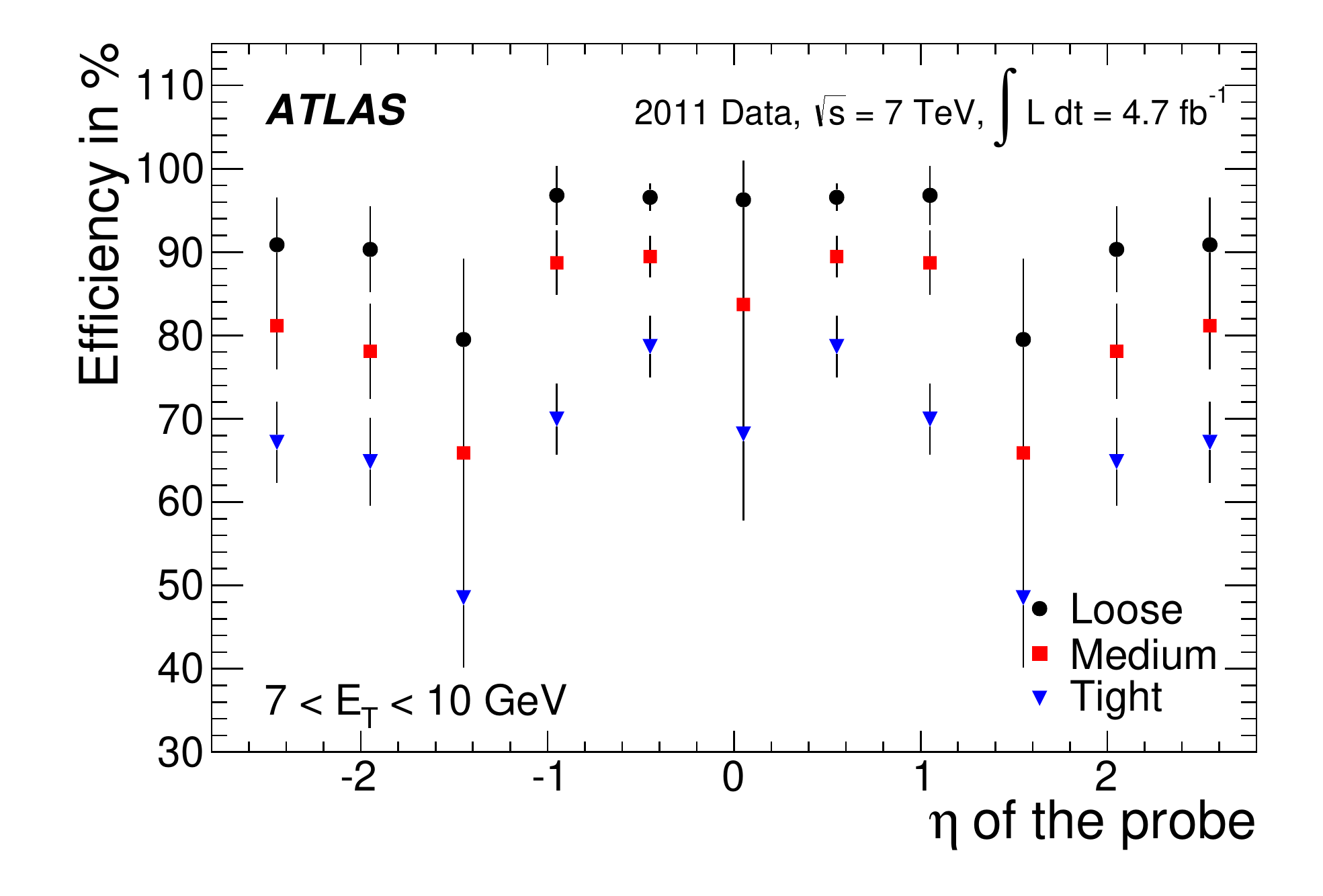}
    \includegraphics[width=0.32\textwidth]{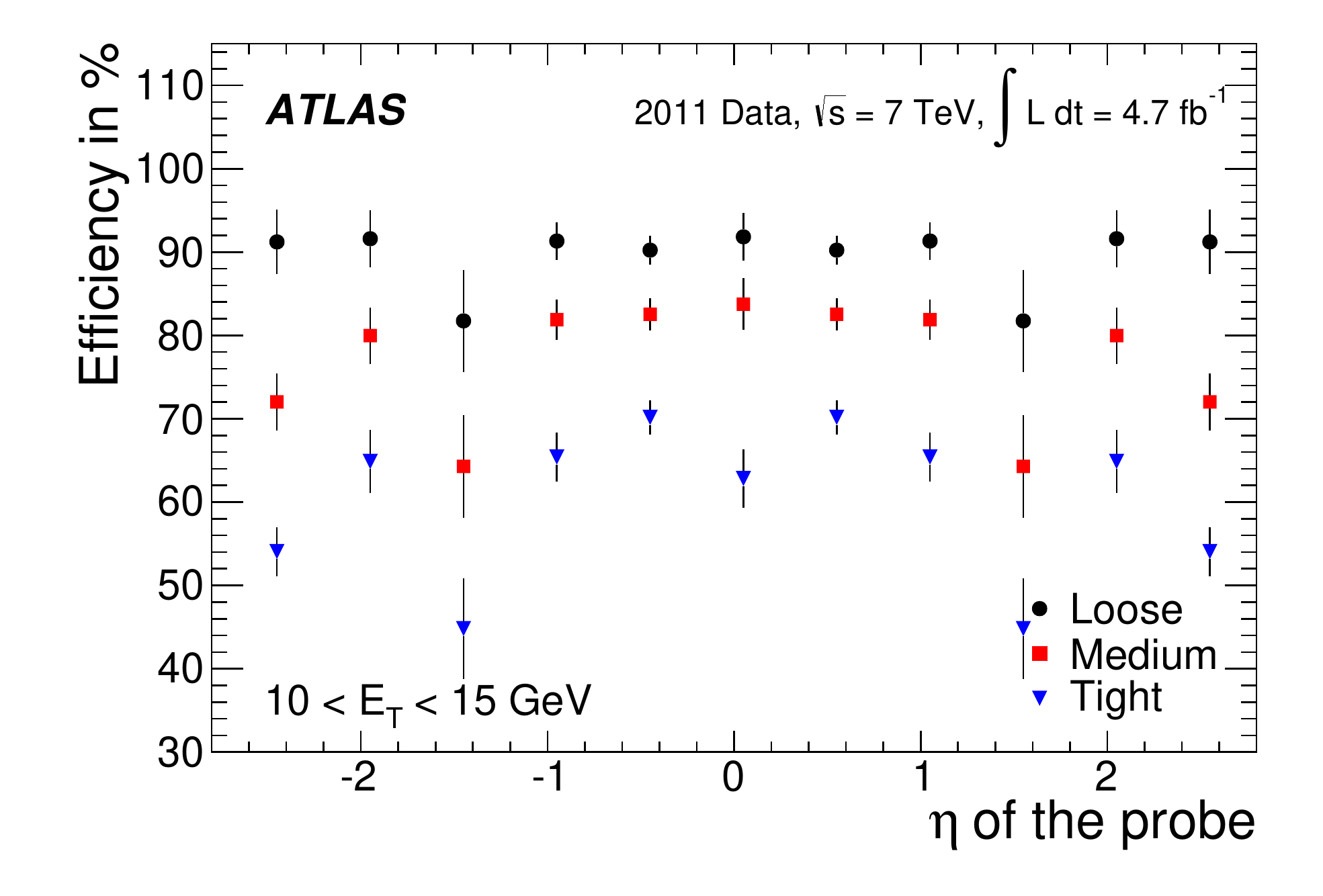}
    \includegraphics[width=0.32\textwidth]{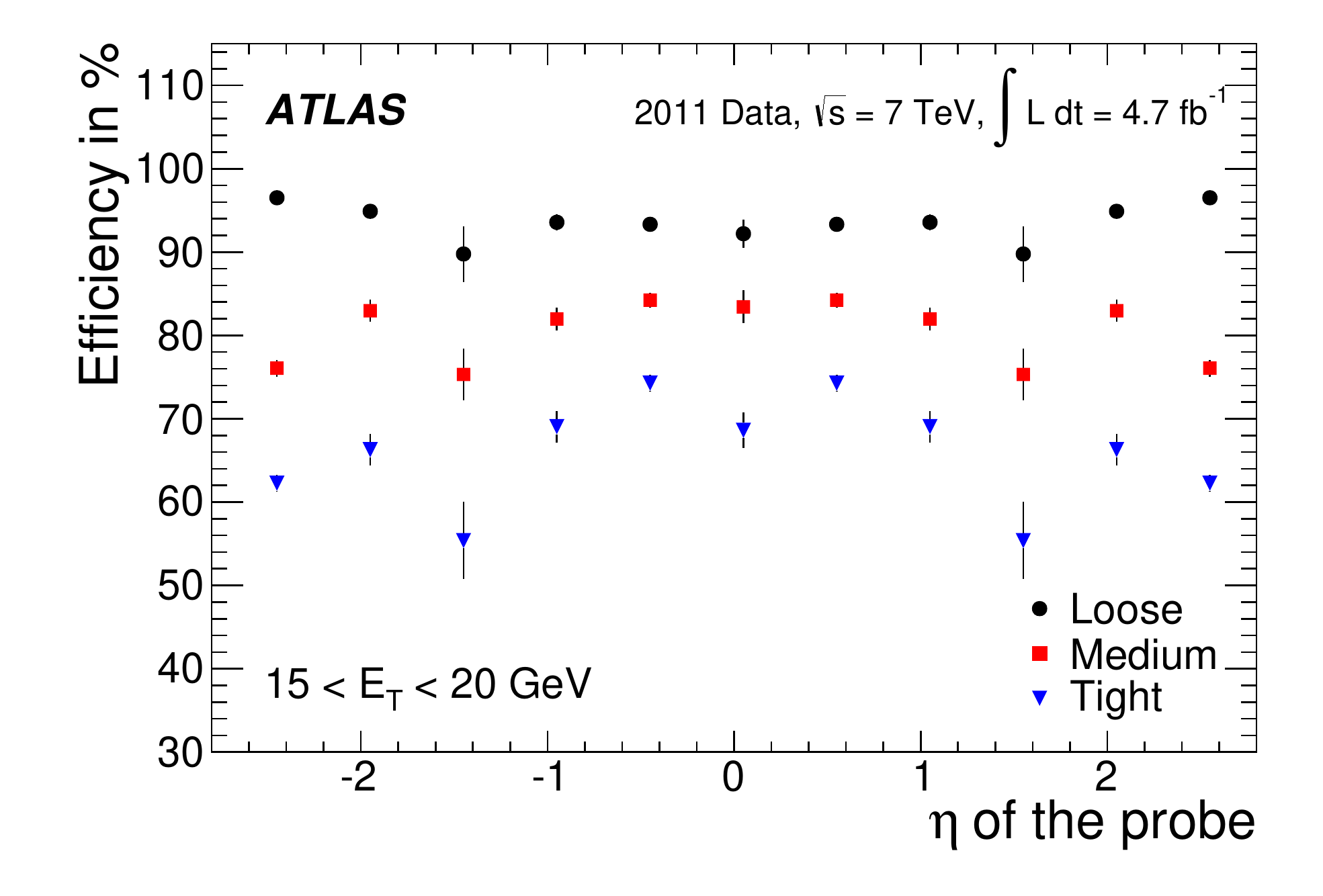}
    \includegraphics[width=0.32\textwidth]{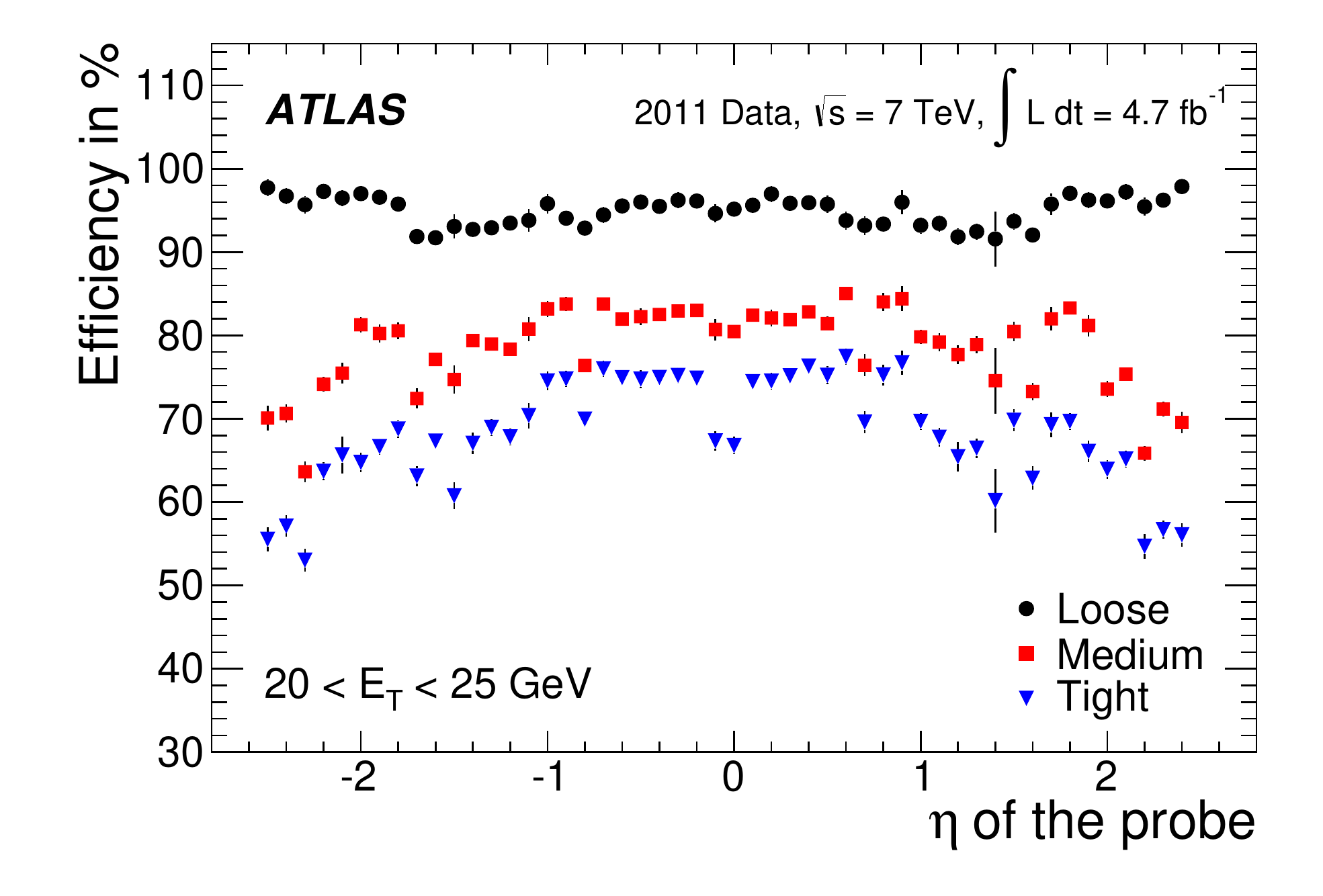}
    \includegraphics[width=0.32\textwidth]{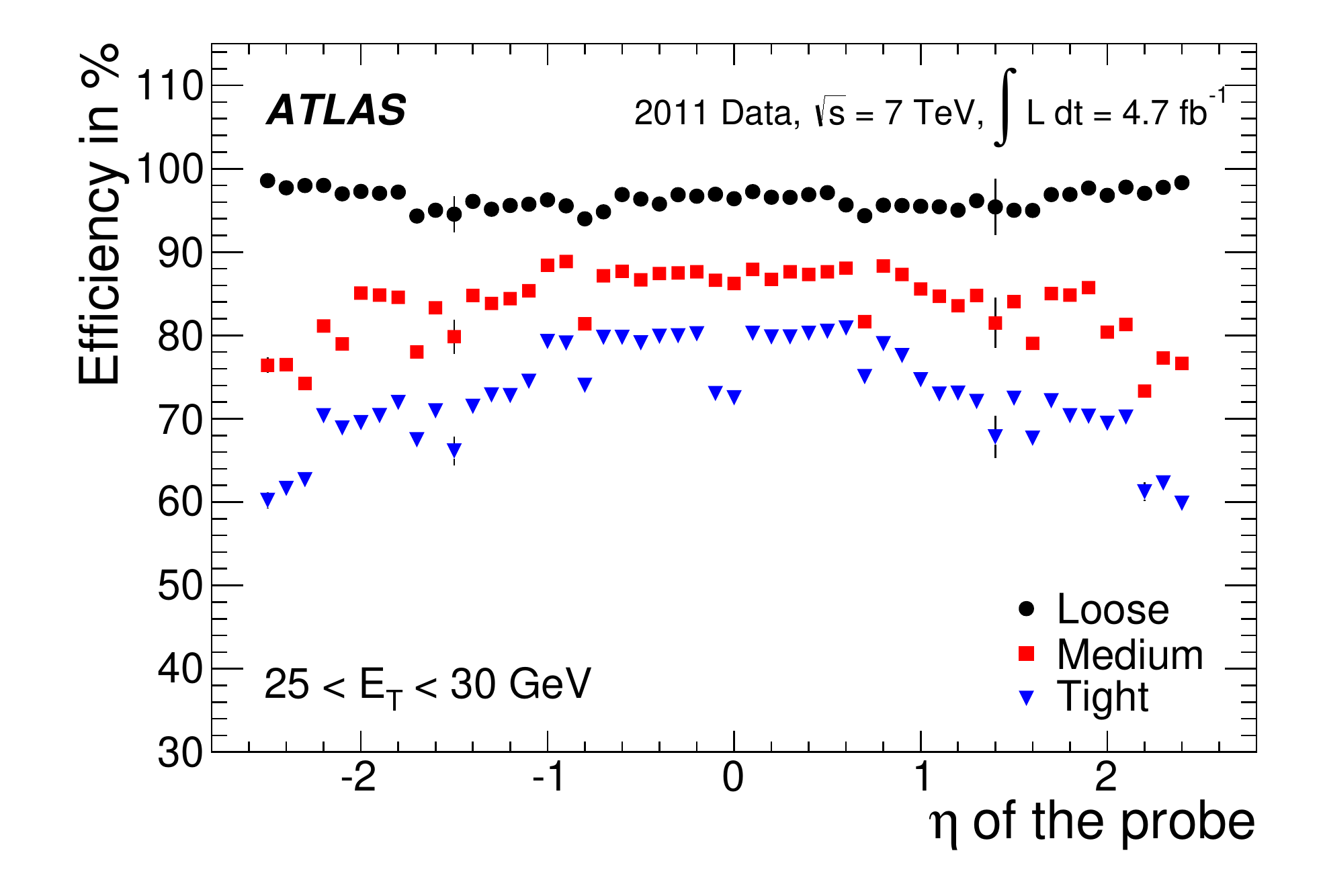}
    \includegraphics[width=0.32\textwidth]{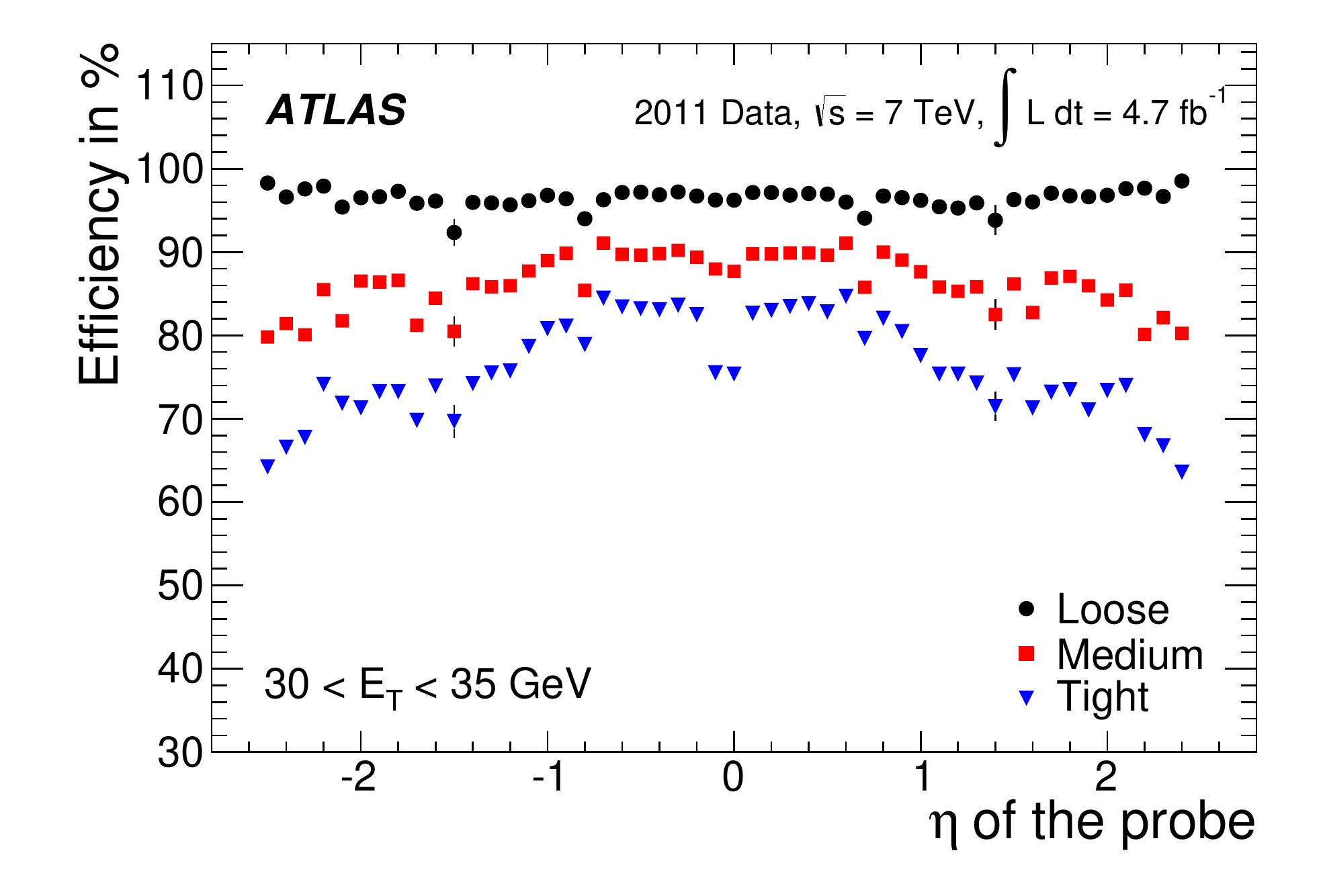}
    \includegraphics[width=0.32\textwidth]{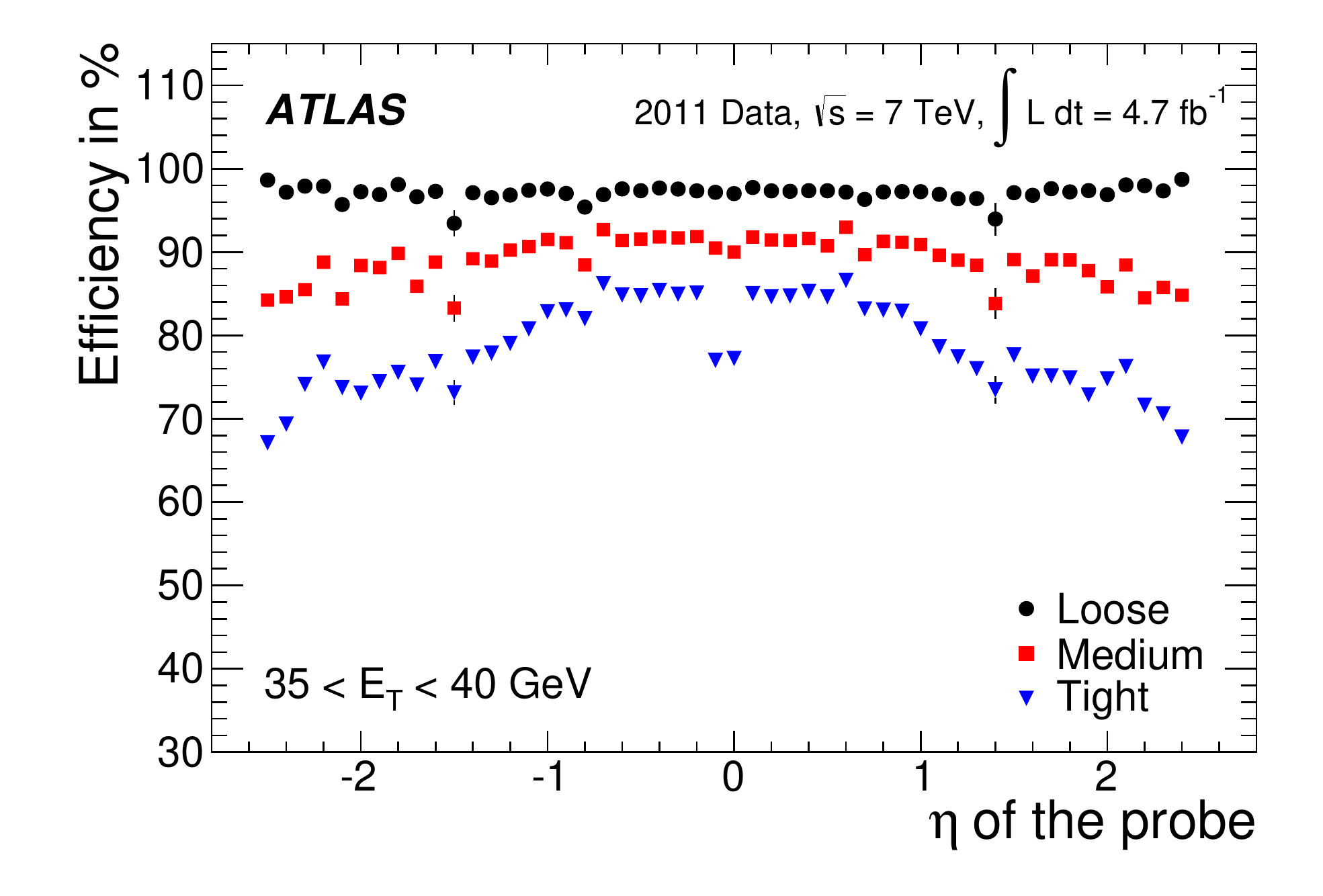}
    \includegraphics[width=0.32\textwidth]{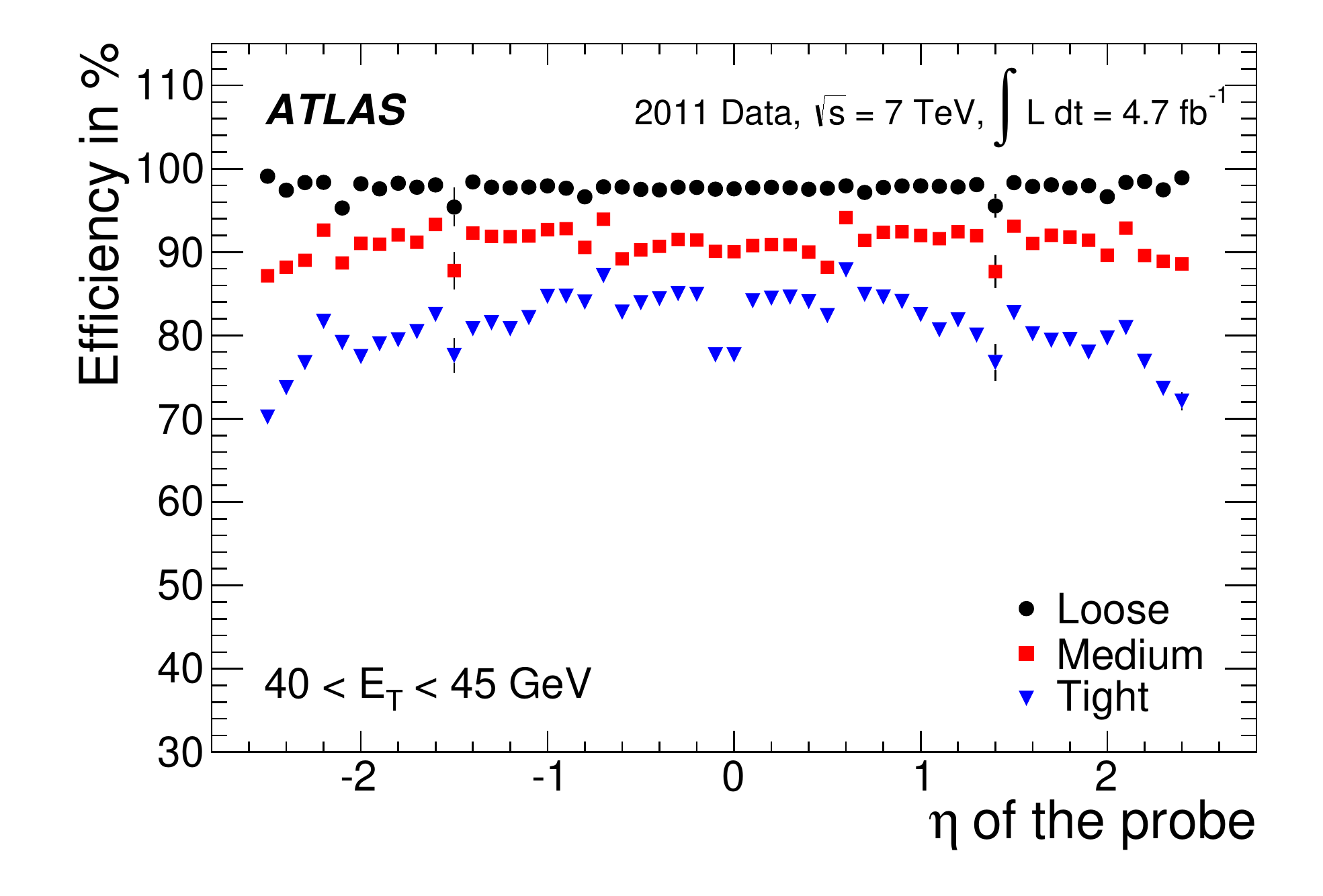}
    \includegraphics[width=0.32\textwidth]{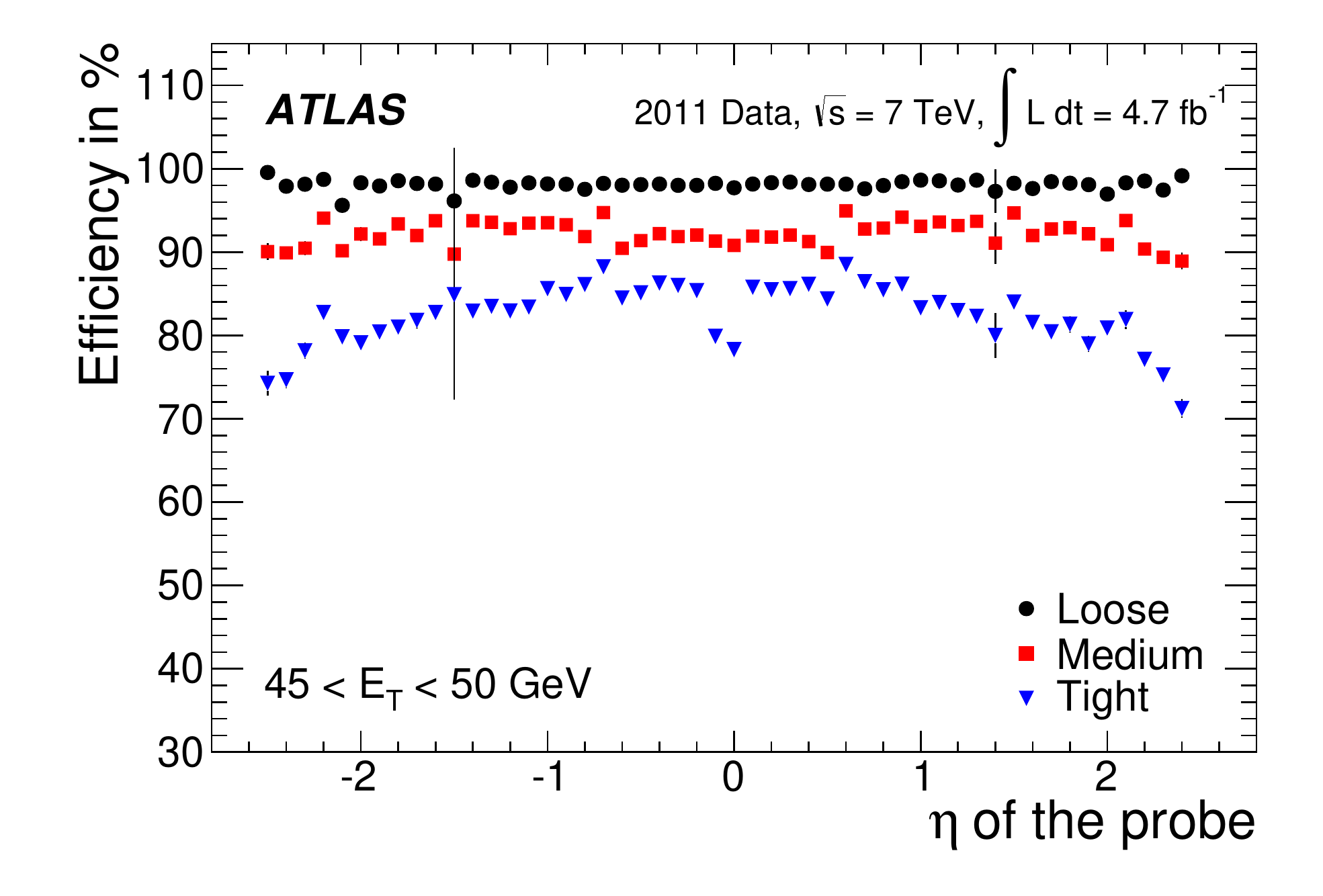}
  \end{center}
  \caption{Electron identification efficiencies, extracted by
    multiplying the combined scale factors evaluated from the \Zee,
    \Wen, and \Jpsiee\ channels by efficiencies computed from a \Zee\
    Monte Carlo simulation, as a function of the \eta\ value of the
    probe for nine \et\ bins, from 7--10~GeV (top) to 45--50~GeV
    (bottom). The three colours correspond to the three identification
    criteria (\lpp, \mpp, \tpp). For $\et<20$~GeV, the coarse binning
    is used and the efficiencies are plotted symmetrically for both
    the positive and negative \eta\ bins. For $\et>20$~GeV, the
    efficiencies are shown in the 50 \eta\ bins available using the
    fine granularity. The error bars indicate the total uncertainties.
  }
  \label{fig:EffPlot}
\end{figure*}

\begin{figure*}
  \begin{center}
    \includegraphics[width=0.32\textwidth]{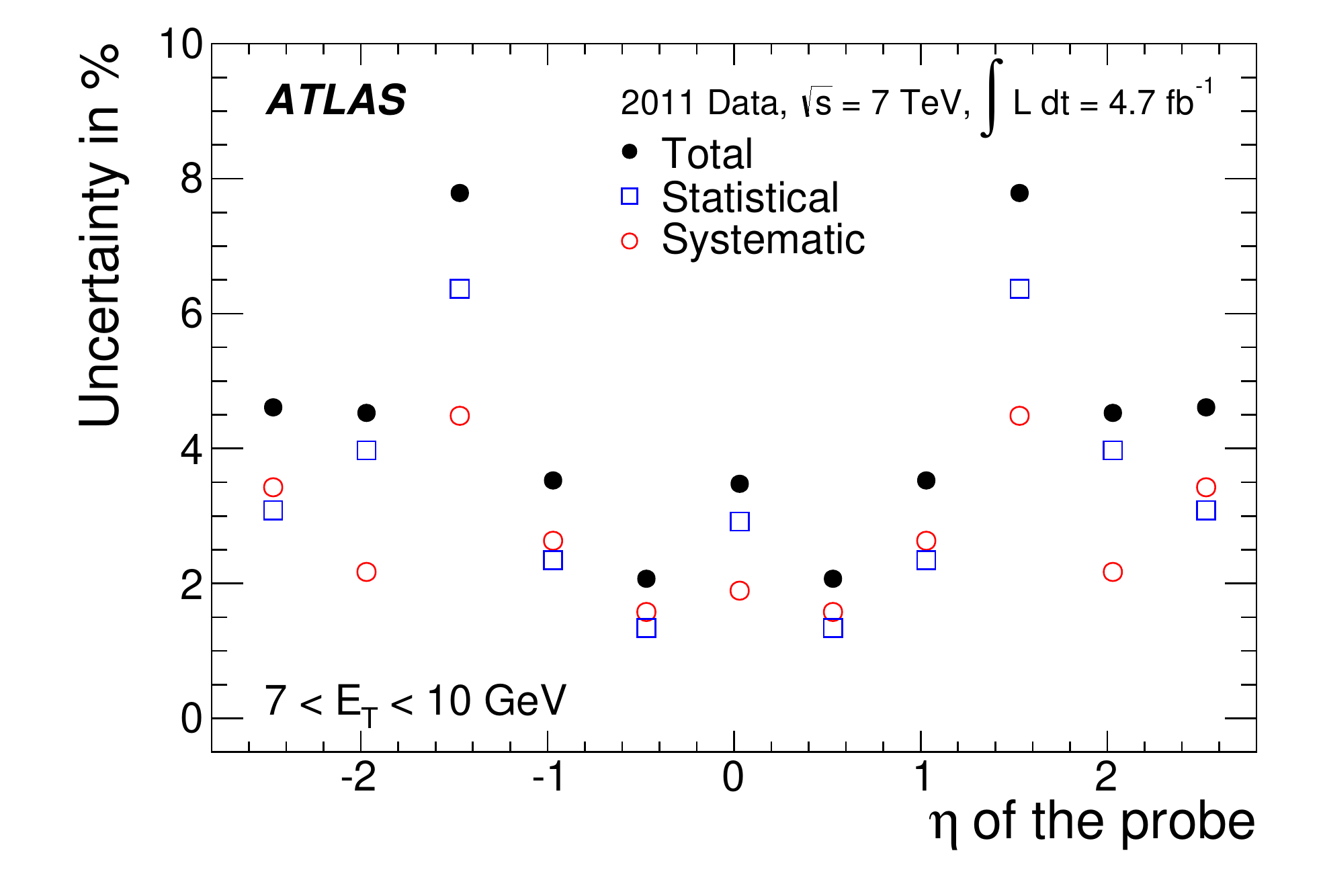}
    \includegraphics[width=0.32\textwidth]{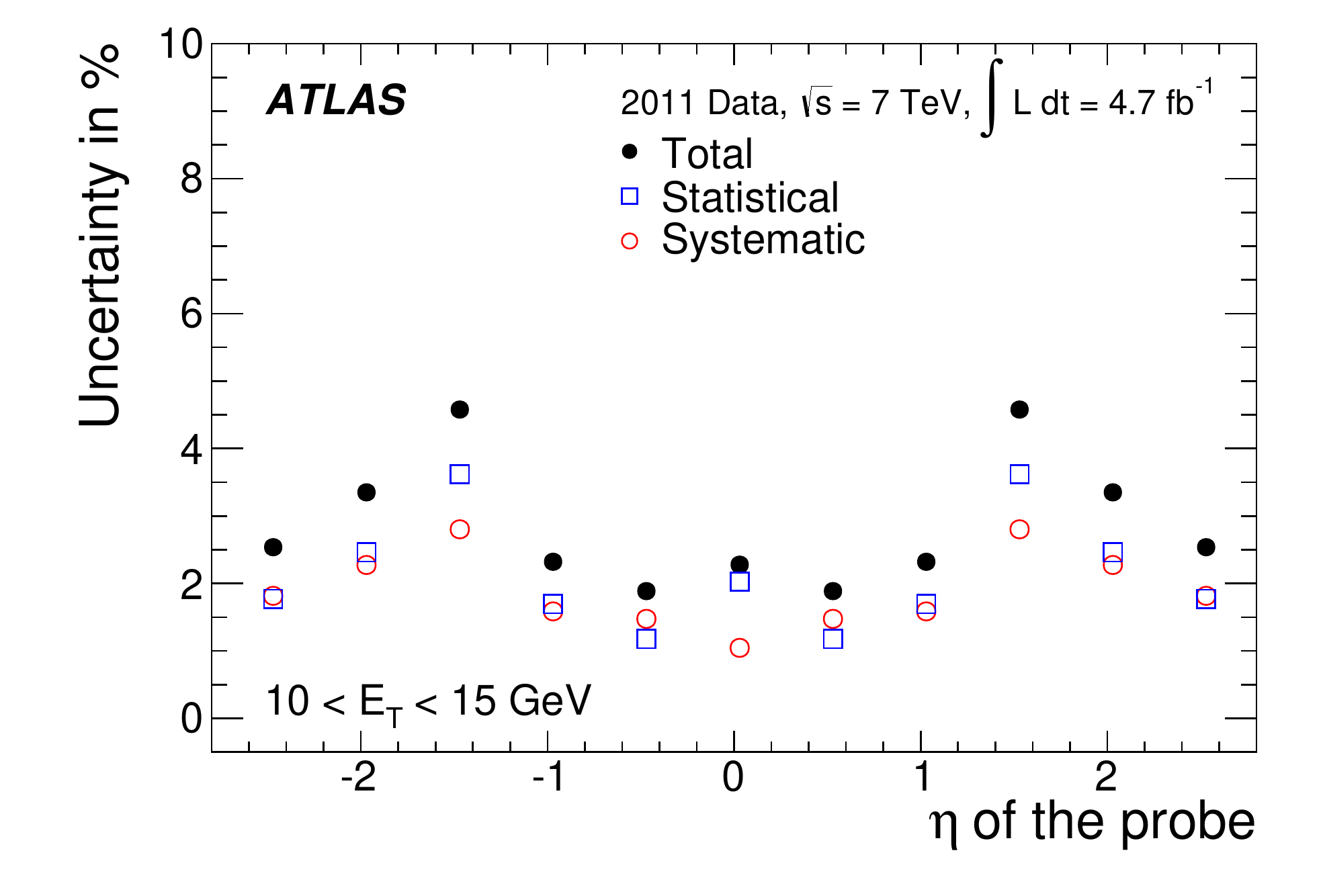}
    \includegraphics[width=0.32\textwidth]{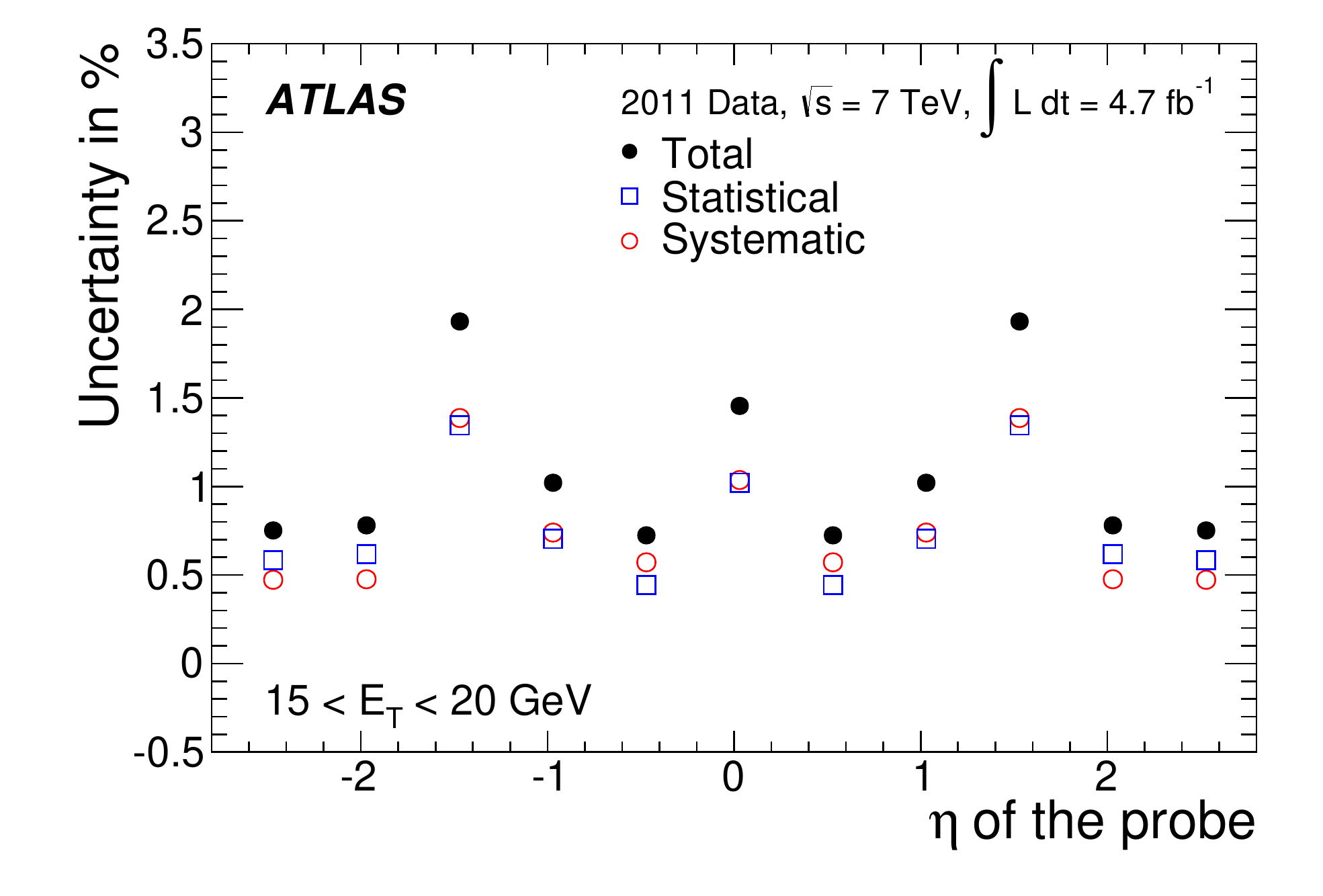}
    \includegraphics[width=0.32\textwidth]{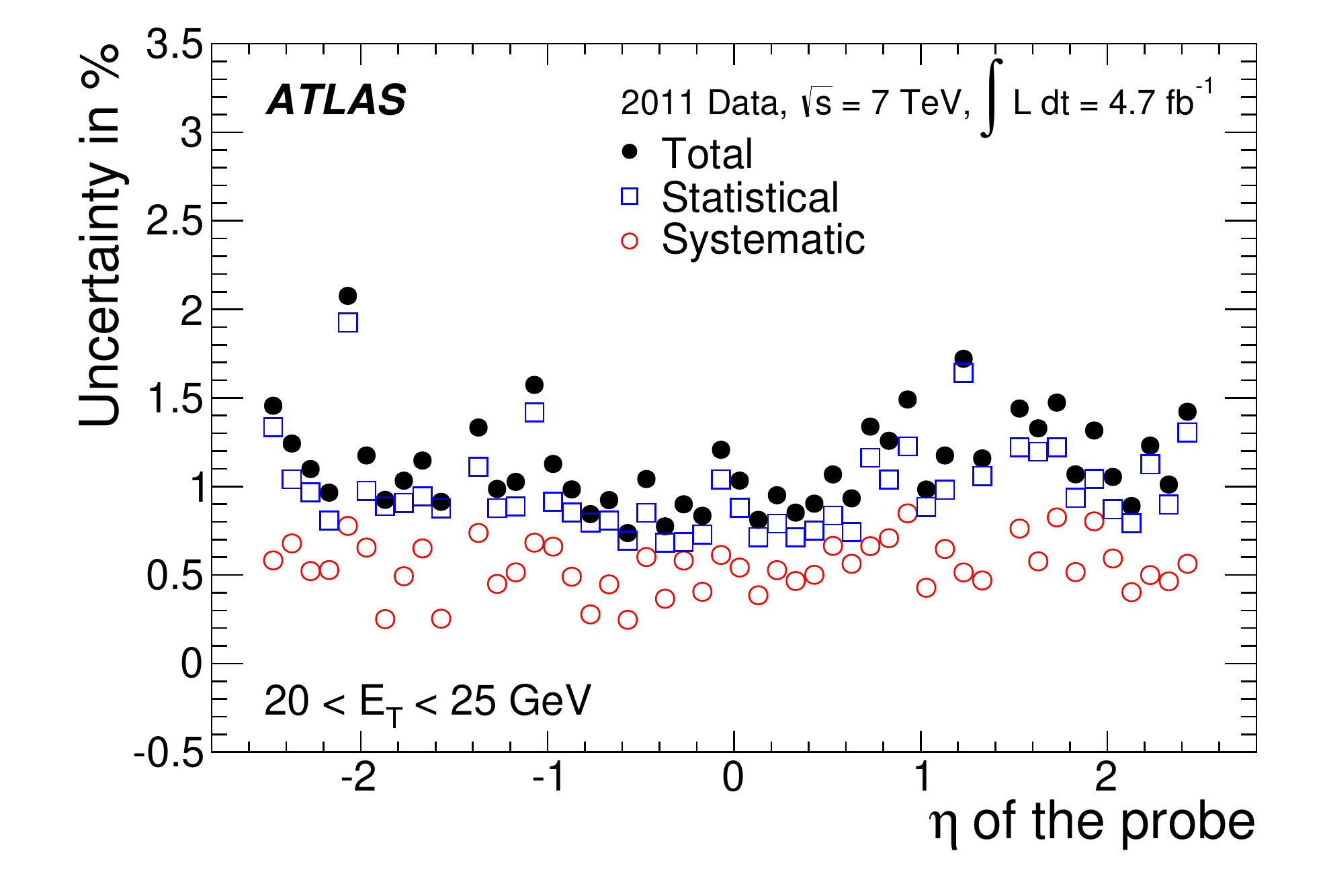}
    \includegraphics[width=0.32\textwidth]{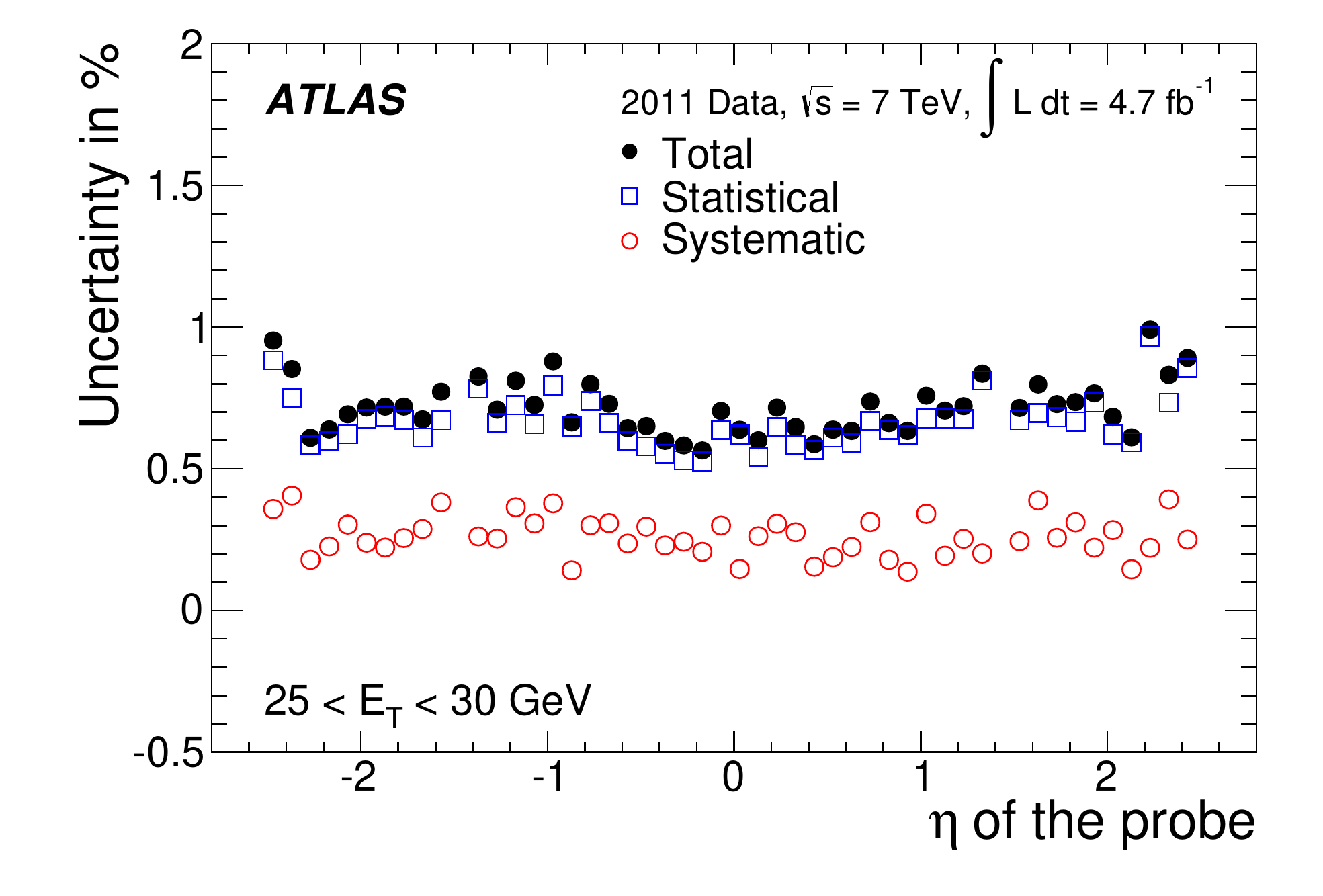}
    \includegraphics[width=0.32\textwidth]{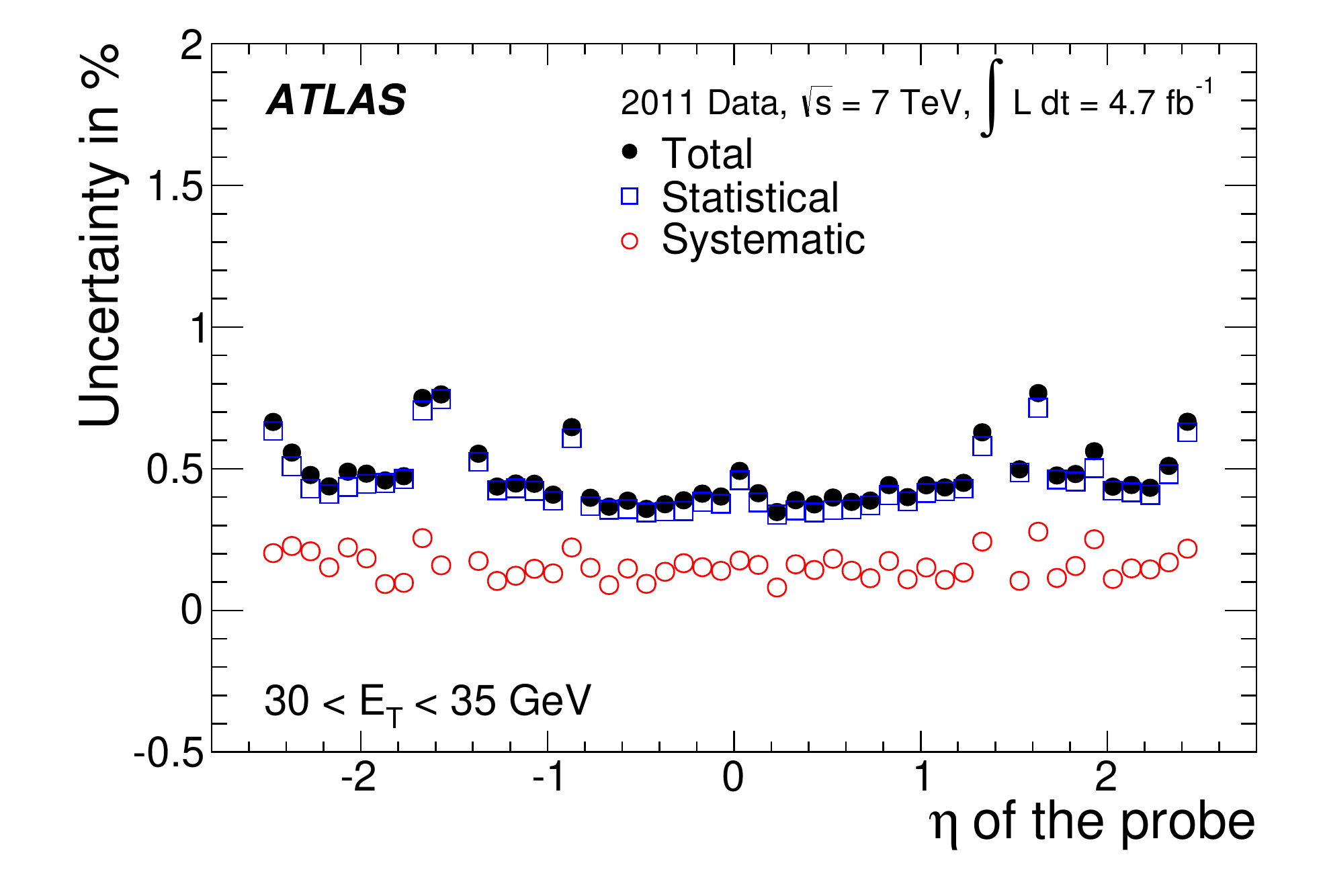}
    \includegraphics[width=0.32\textwidth]{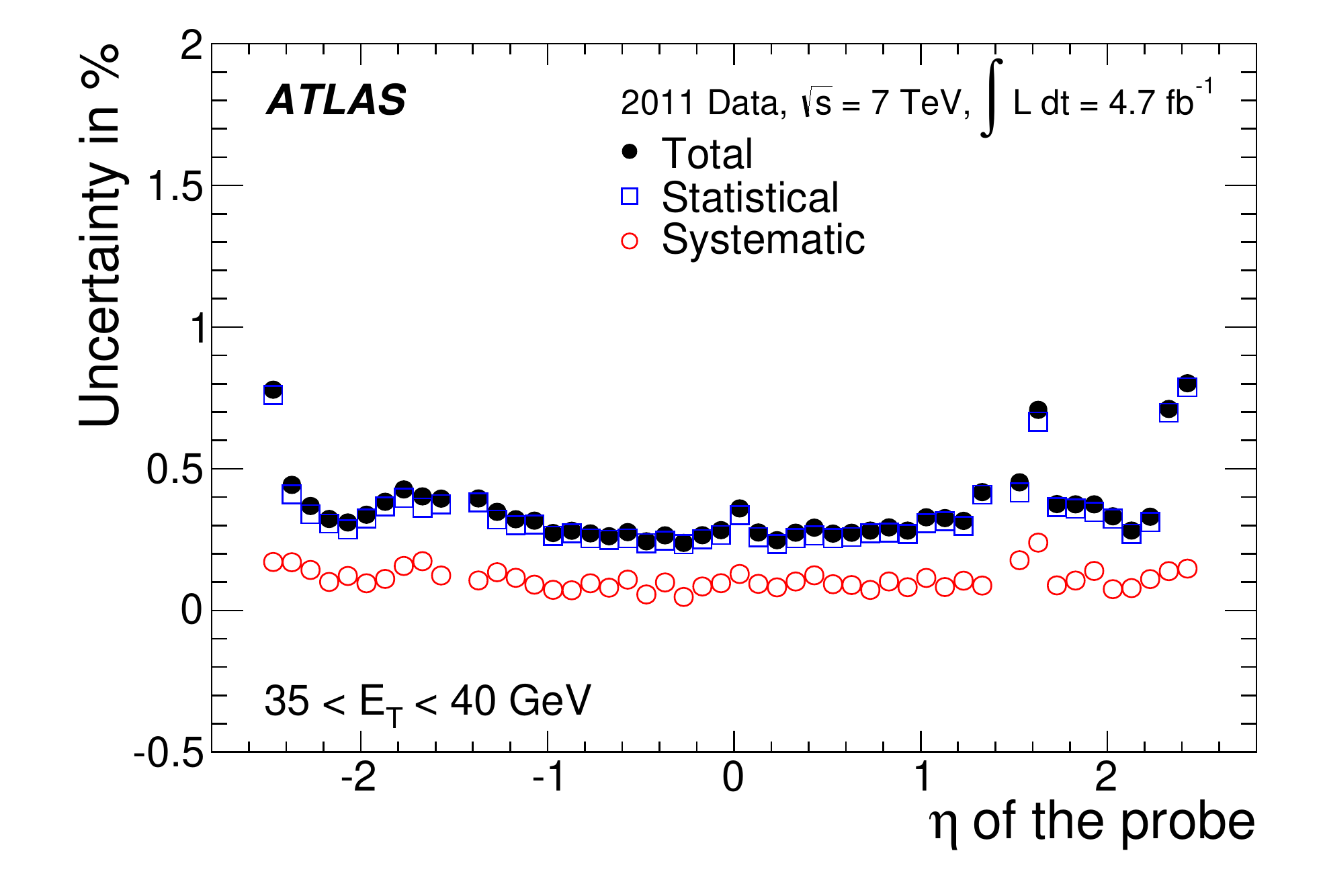}
    \includegraphics[width=0.32\textwidth]{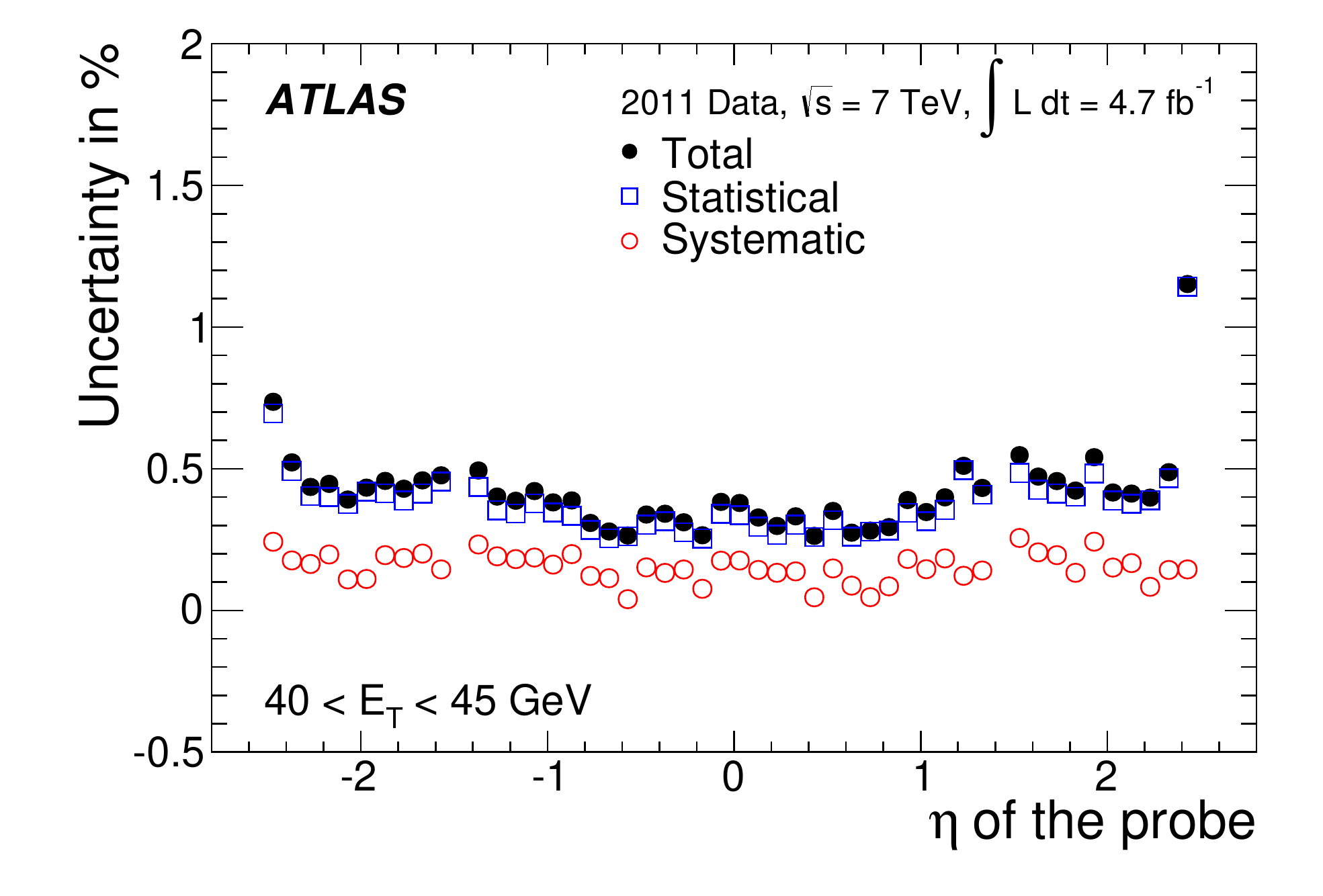}
    \includegraphics[width=0.32\textwidth]{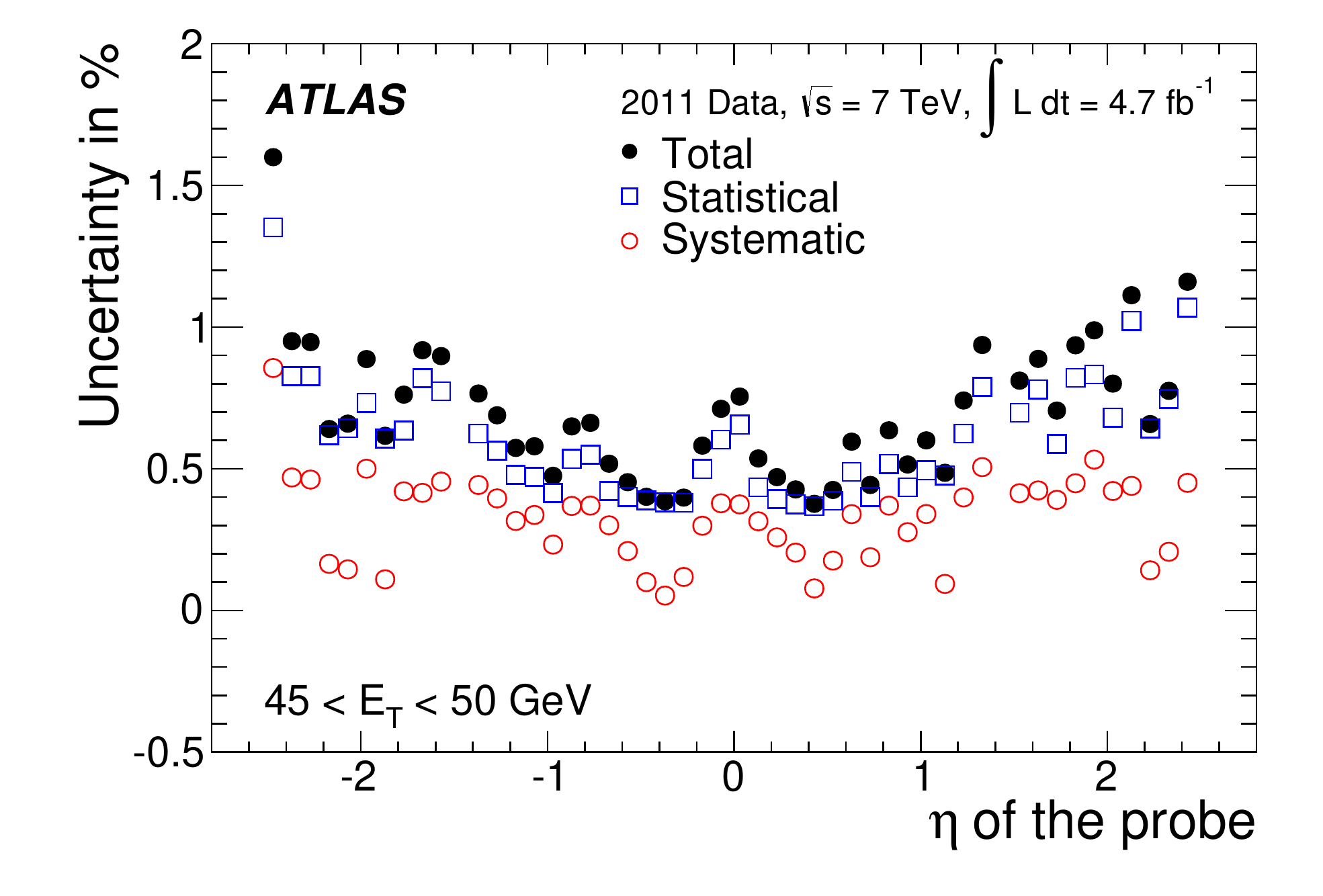}

  \end{center}
  \caption{Total, systematic, and statistical uncertainties of the
    \tpp\ efficiency (extracted by multiplying the combined scale
    factors evaluated from the \Zee, \Wen, and \Jpsiee\ channels by
    efficiencies computed from a \Zee\ Monte Carlo simulation) as a
    function of the \eta\ value of the probe for nine \et\ bins, from
    7--10~GeV (top) to 45--50~GeV (bottom). For $\et<20$~GeV, the
    coarse binning is used and the uncertainties are plotted
    symmetrically for both the positive and negative \eta\ bins. For
    $\et>20$~GeV, the uncertainties are shown in the 50 \eta\ bins
    available using the fine granularity. The total uncertainties are
    dominated by the statistical component. The systematic
    uncertainties are dominated by the uncorrelated component, which
    is largely due to the difference of the two \Zee\ methods and thus
    affected by limited statistics of the different data samples
    employed for the background-subtraction procedures.}
  \label{fig:Efferr}
\end{figure*}

In the following, the combined data efficiencies are extracted by
multiplying the combined scale factors by the efficiencies computed
from a \Zee\ Monte Carlo simulation. Figure~\ref{fig:EffPlot} shows,
for the coarse \eta\ granularity in the low-\et\ region and the fine
granularity in the region $\et>20$~GeV, the efficiencies obtained in
this way. The precision of the efficiency measurements is in general
dominated by the statistical component, as shown in
Figure~\ref{fig:Efferr} for the \tpp\ criteria. Possible sources of
systematic uncertainties arising from the choice of MC generator to
derive the scale factors are not accounted for in this analysis but
are expected to have a negligible impact on these results. This is due
to the fact that the final results as shown in
Figures~\ref{fig:EffPlot} and~\ref{fig:Efferr} are obtained from
data-driven efficiency measurements, combined through the use of scale
factors, but then multiplied by a \Zee\ MC sample.

In the case of \lpp\ identification criteria, efficiencies are fairly
uniform with pseudorapidity, while a slight dependence is observed
both for \mpp\ and \tpp. The identification efficiency is sensitive to
the readout granularity of the detectors and to the non-uniformities
of the material along the path of the electron. These variations are
taken into account in the identification criteria by defining
pseudorapidity-dependent thresholds for the selection variables, in
addition to selections dependent on transverse energy. As the tighter
\mpp\ and \tpp\ criteria make use of both calorimetric and track
information, they are more sensitive to such effects. Dependencies are
most notable at $\abseta <0.1$ and in the transition region
$1.37<\abseta<1.52$. In the region $\abseta>2$, where the requirements
on the shower shapes are tightened to preserve the needed rejection in
the absence of the transition radiation information, a degradation of
the efficiencies is observed.

\begin{figure*}
  \begin{center}
\subfigure[]{    \includegraphics[width=0.47\textwidth]{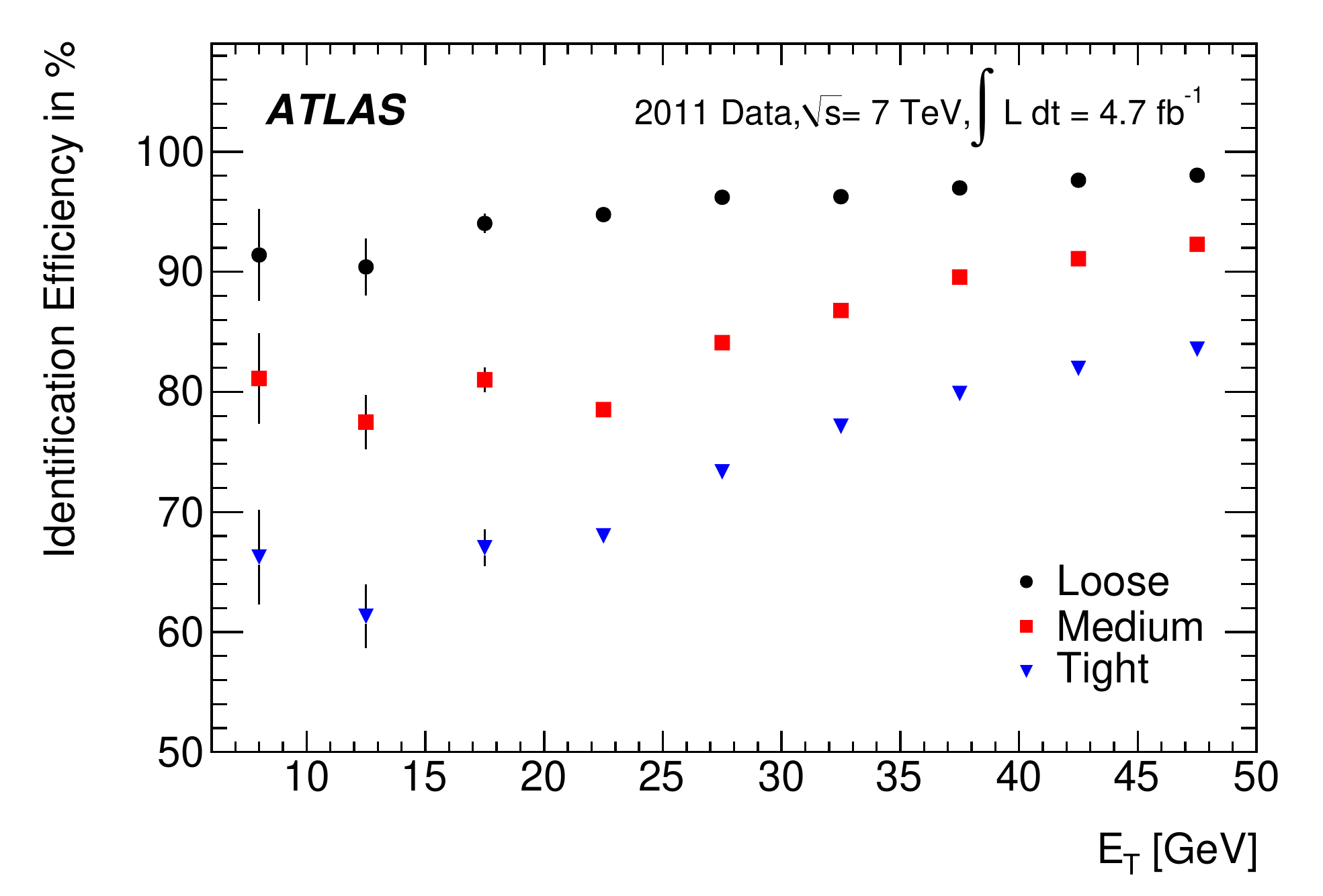}}
\subfigure[]{    \includegraphics[width=0.47\textwidth]{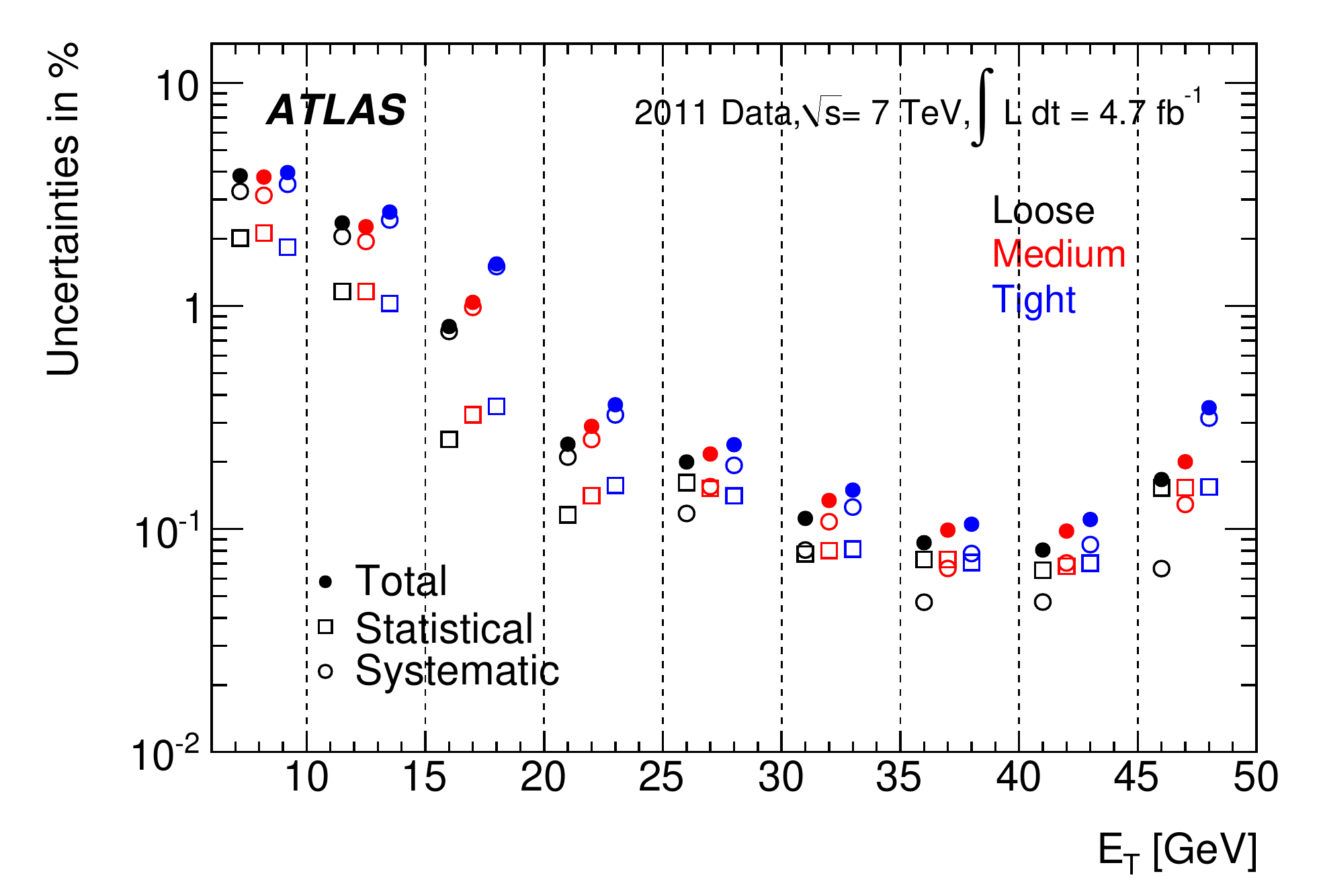}}
  \end{center}
  \caption{(a) Dependence of the combined identification
    efficiencies on the transverse energy of the probe for central
    electrons. Error bars correspond to the total uncertainties.
    (b) Decomposition of the total uncertainty into its statistical
    and systematic components. The three colours correspond to the
    three identification criteria (\lpp, \mpp, \tpp). Some points are
    slightly shifted horizontally within the \et~bin for better
    visibility. In the \et\ region above the Jacobian peak ($\et>$45~GeV), both the statistical and systematic uncertainties increase with respect to the highest precision region (\et~$\sim$~35~GeV), as shown in Figure~\ref{fig:Efferr}.}
  \label{fig:integPt}
\end{figure*}

The dependence of the efficiency on the transverse energy of the
electron is made more explicit when integrating over the whole
pseudorapidity range of the \Zee\ sample, as shown in Figure~\ref{fig:integPt}(a). In
the \et\ range from 7 to 50~GeV, the \lpp\ efficiency varies from
about $90\%$ to $98\%$. The \mpp\ and \tpp\ criteria show a more
significant dependence on energy due to the tighter requirements
applied to provide the desired background rejection. The efficiency
increases from about $80$\% at 7~GeV to $90$\% at 50~GeV for the \mpp\
criteria and from about $65$\% at 7~GeV to $80$\% at 50~GeV for the
\tpp\ criteria. The integration over the pseudorapidity range
decreases the statistical and uncorrelated systematic uncertainties of
the measurement. However, given that almost half of the systematic
uncertainty is correlated amongst all \eta\ bins, the size of
this component does not improve after integration. Thus, the total
systematic uncertainties on the efficiency measurements as a function
of \et\ are dominant over much of the lower \et\ range and of
comparable size to the statistical uncertainties at high \et\, as is
shown in Figure~\ref{fig:integPt}(b).

\subsection{Forward-electron identification efficiency}

In the forward region of the calorimeters, the electron identification
efficiency is measured with a \Zee\ sample where a well-isolated
$\et>25$~GeV tag electron satisfying the \tpp\ requirement is
identified in the central region of the calorimeter and the probe
cluster with $\et>20$~GeV is found in the region $2.5<\abseta<4.9$.
The candidate events are required to have a low missing transverse
momentum, in order to suppress the contributions from \Wen\
background.

\begin{figure*}
  \begin{center}
\subfigure[]{    \includegraphics[width=0.47\textwidth]{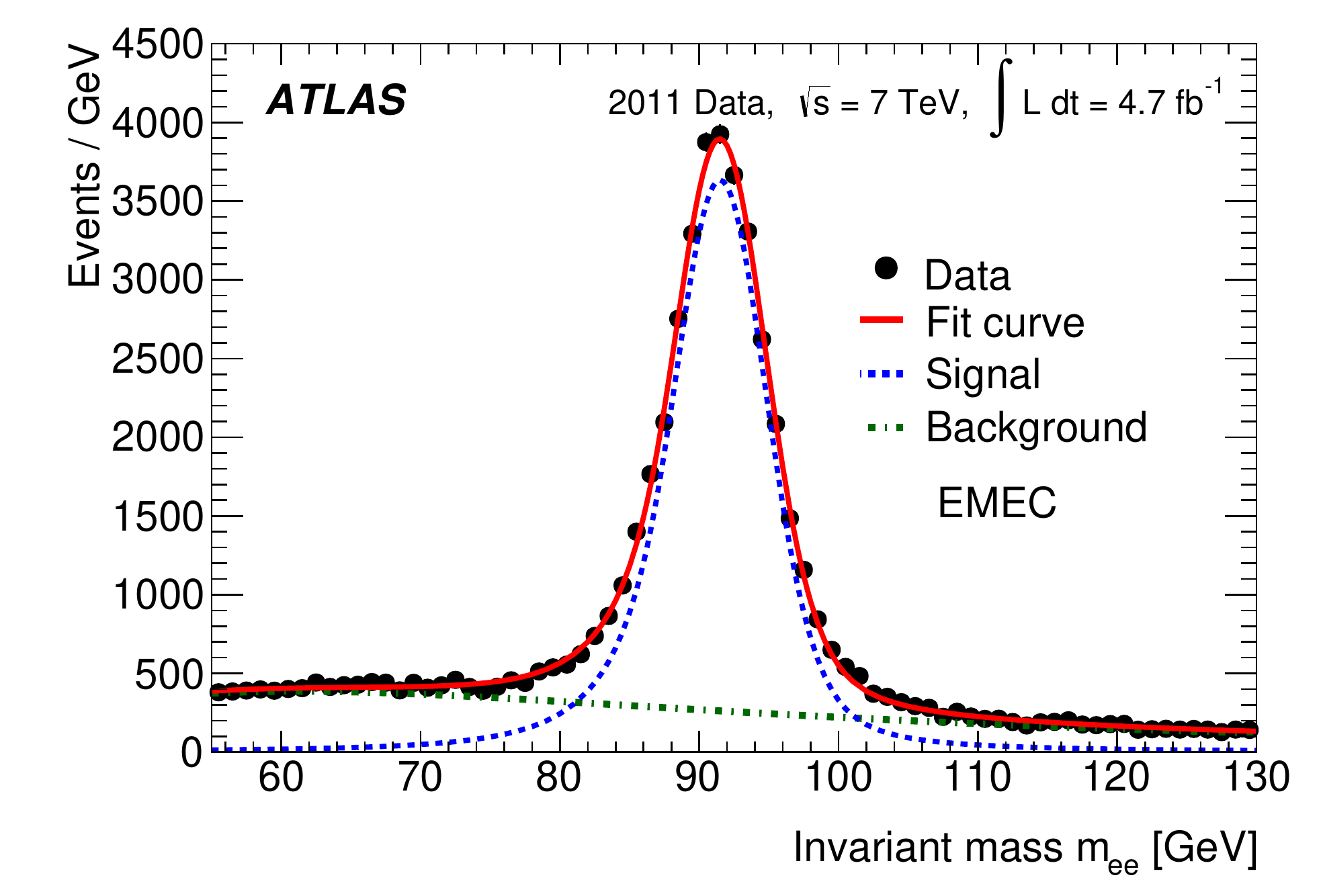}}
\subfigure[]{    \includegraphics[width=0.47\textwidth]{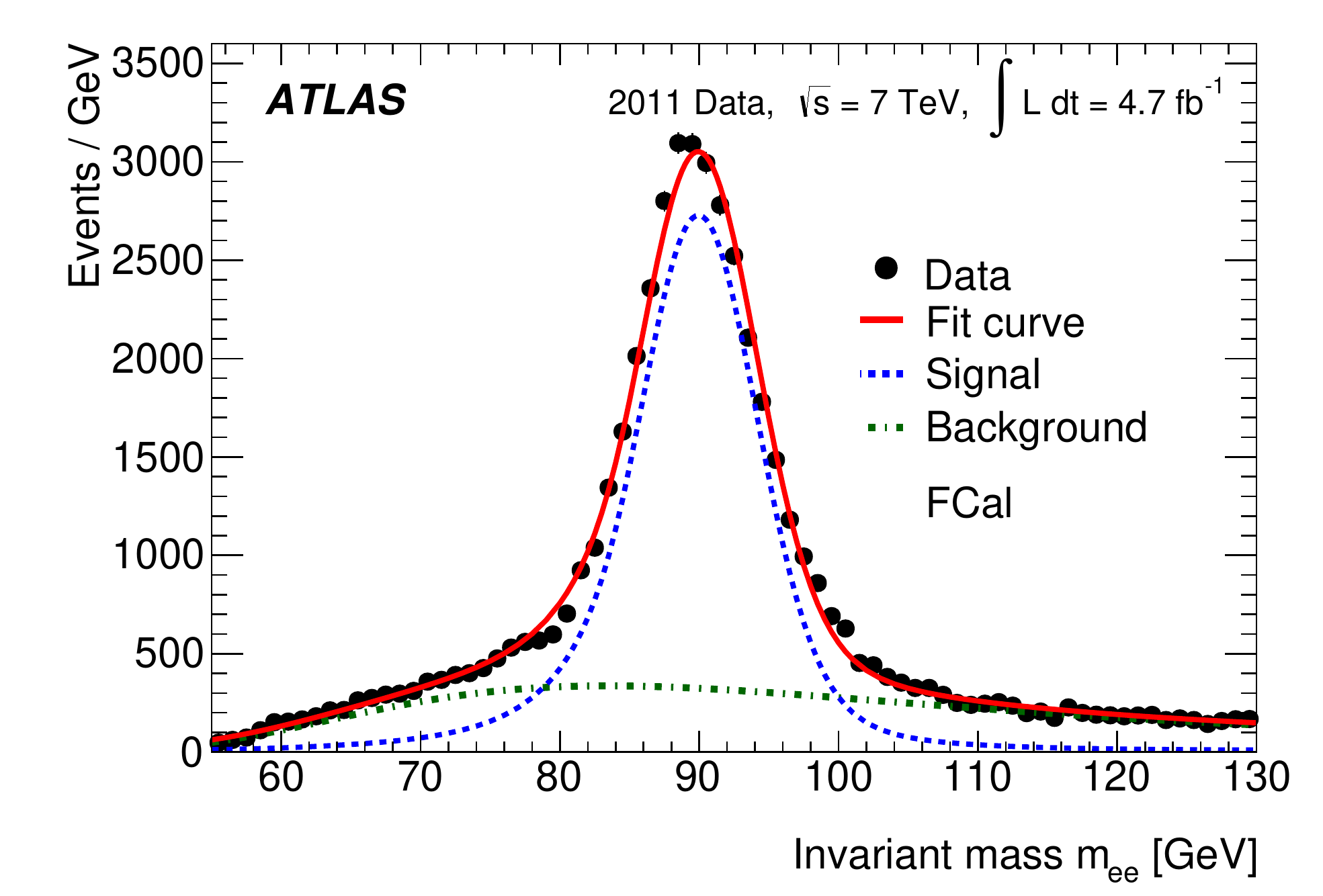}}
  \end{center}
  \caption{Example fits of invariant mass distributions for probes in
    the EMEC in (a) and FCal in (b) regions. }
  \label{fig:forwmass}
\end{figure*}

The invariant mass of the tag--probe system is fitted in each of the 
pseudorapidity bins defined in Section~\ref{sec:fwdid}, in the
range $55<m_{ee}<130$~GeV to a Crystal Ball function convolved with a
non-relativistic Breit--Wigner function with fixed $Z$
width~\cite{Beringer:1900zz} to model the signal, and a Landau
function to model the background. The S/B ratio is $\sim 7$ and $\sim
5$ in the EMEC and the FCal, respectively. After background
subtraction, a total of 192,000 and 76,000 probes remain in the two
regions. Variations of the tag requirements are performed, which
change the S/B ratio by up to a factor of two. In addition,
alternative fit models for signal and background distributions and
different fit ranges are used to assess the systematic uncertainties
on the electron yields. The total systematic uncertainty is computed
by summing in quadrature the effects observed in the individual
variations. The largest contributions are related to the choice of
background model and signal fit range. Examples of invariant mass fits
are shown in Figure~\ref{fig:forwmass}.

\begin{figure*}
  \centering
  \subfigure[]{\includegraphics[width=.47\textwidth]{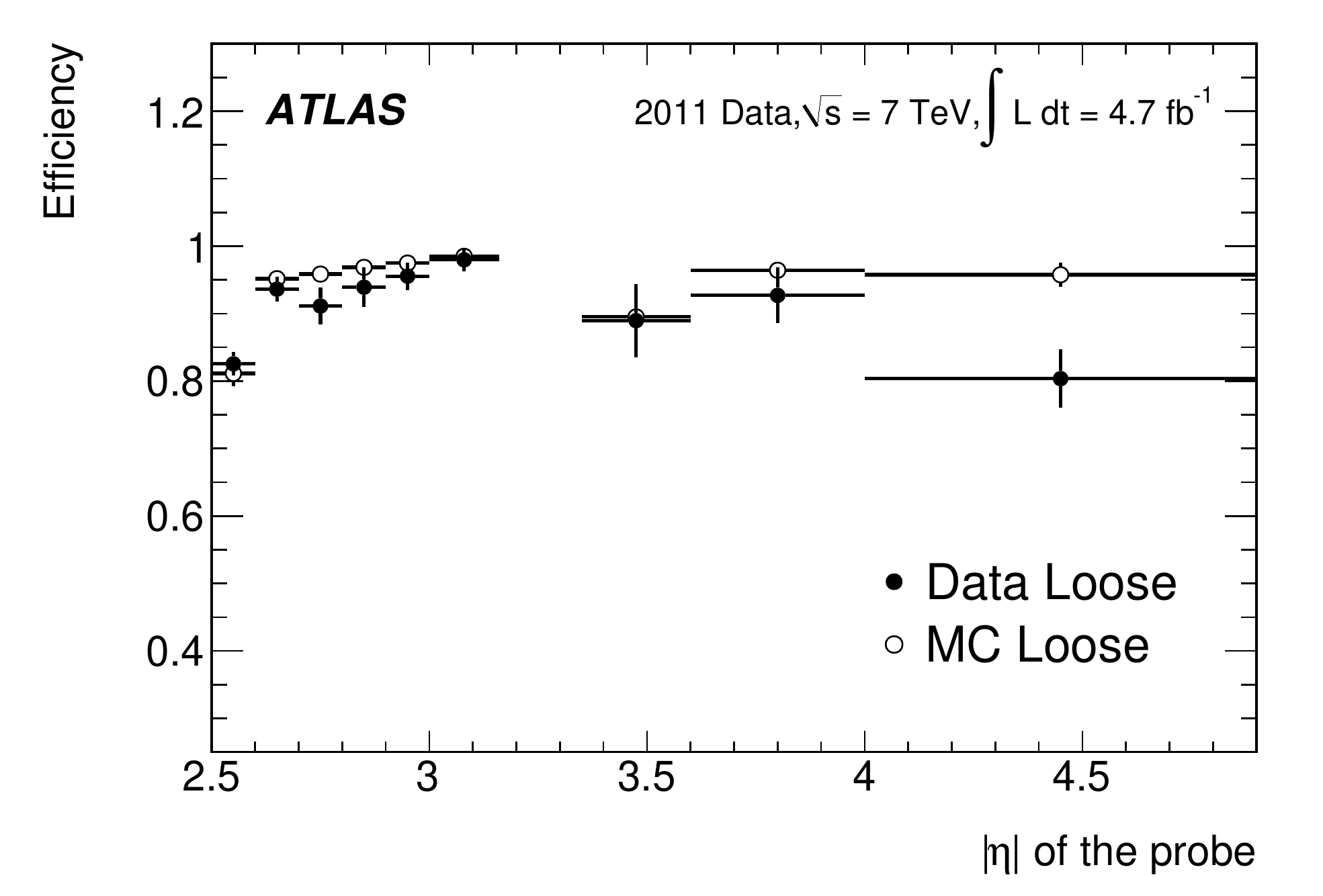}}
  \subfigure[]{\includegraphics[width=.47\textwidth]{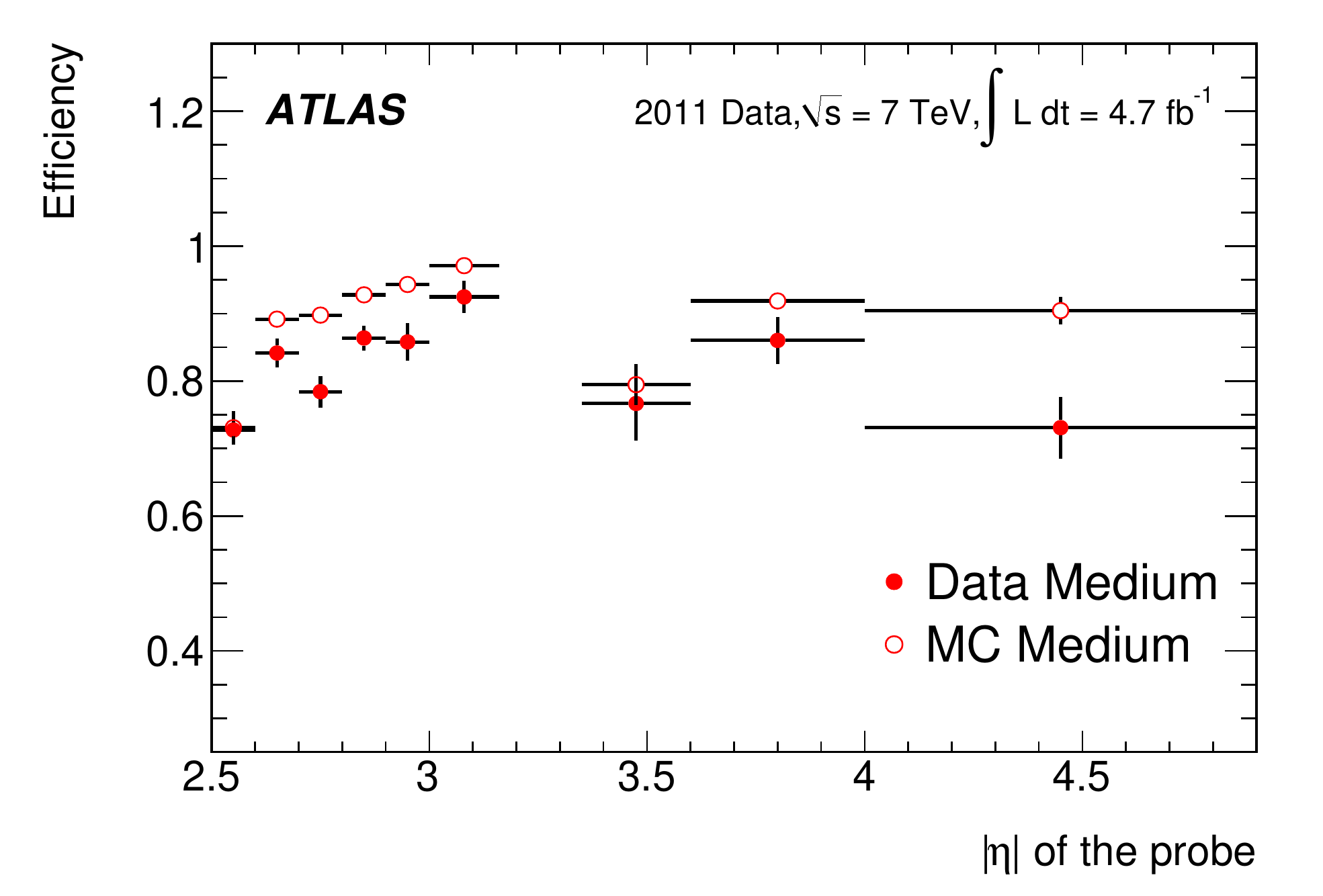}}
  \subfigure[]{\includegraphics[width=.47\textwidth]{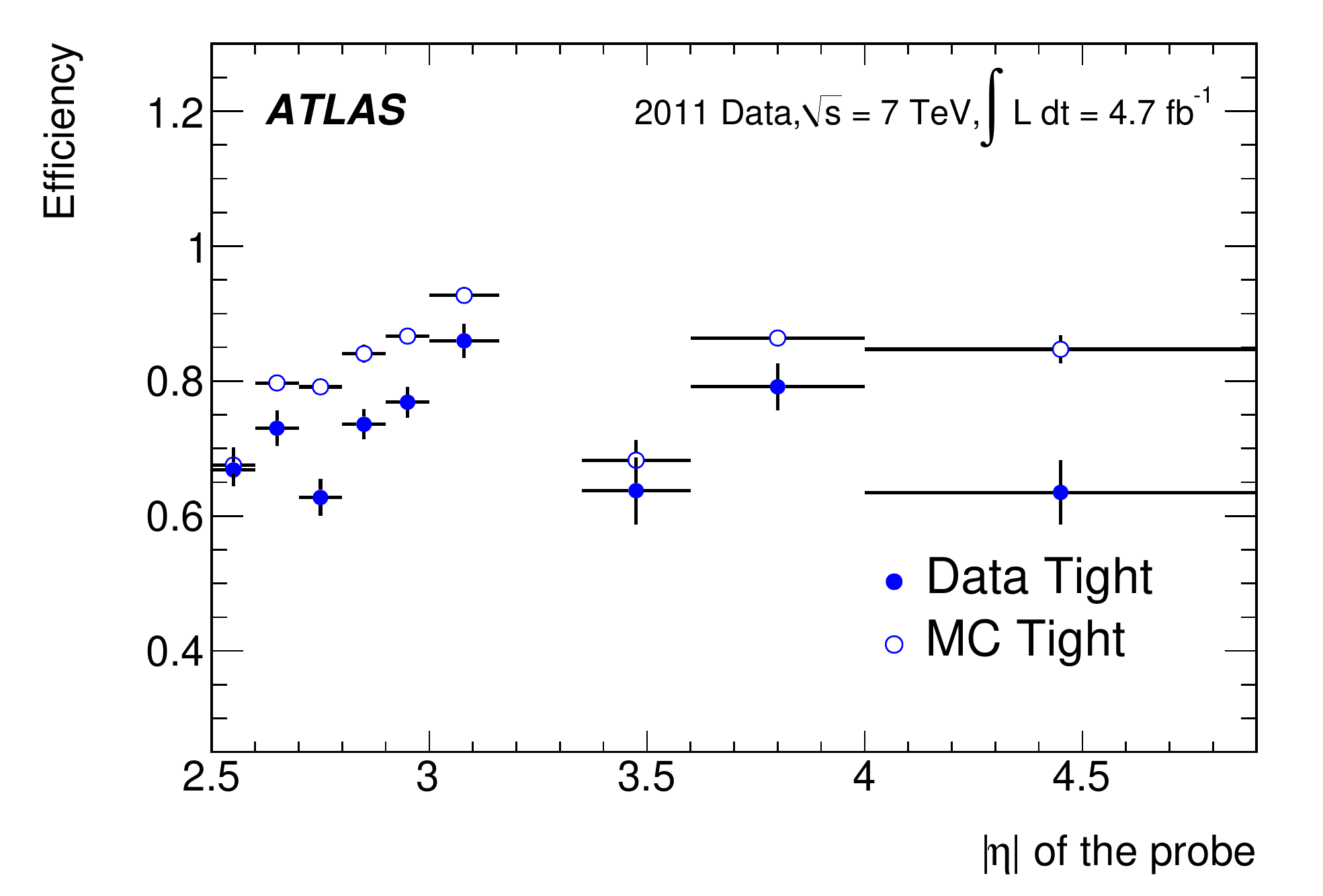}}
  \subfigure[]{\includegraphics[width=.47\textwidth]{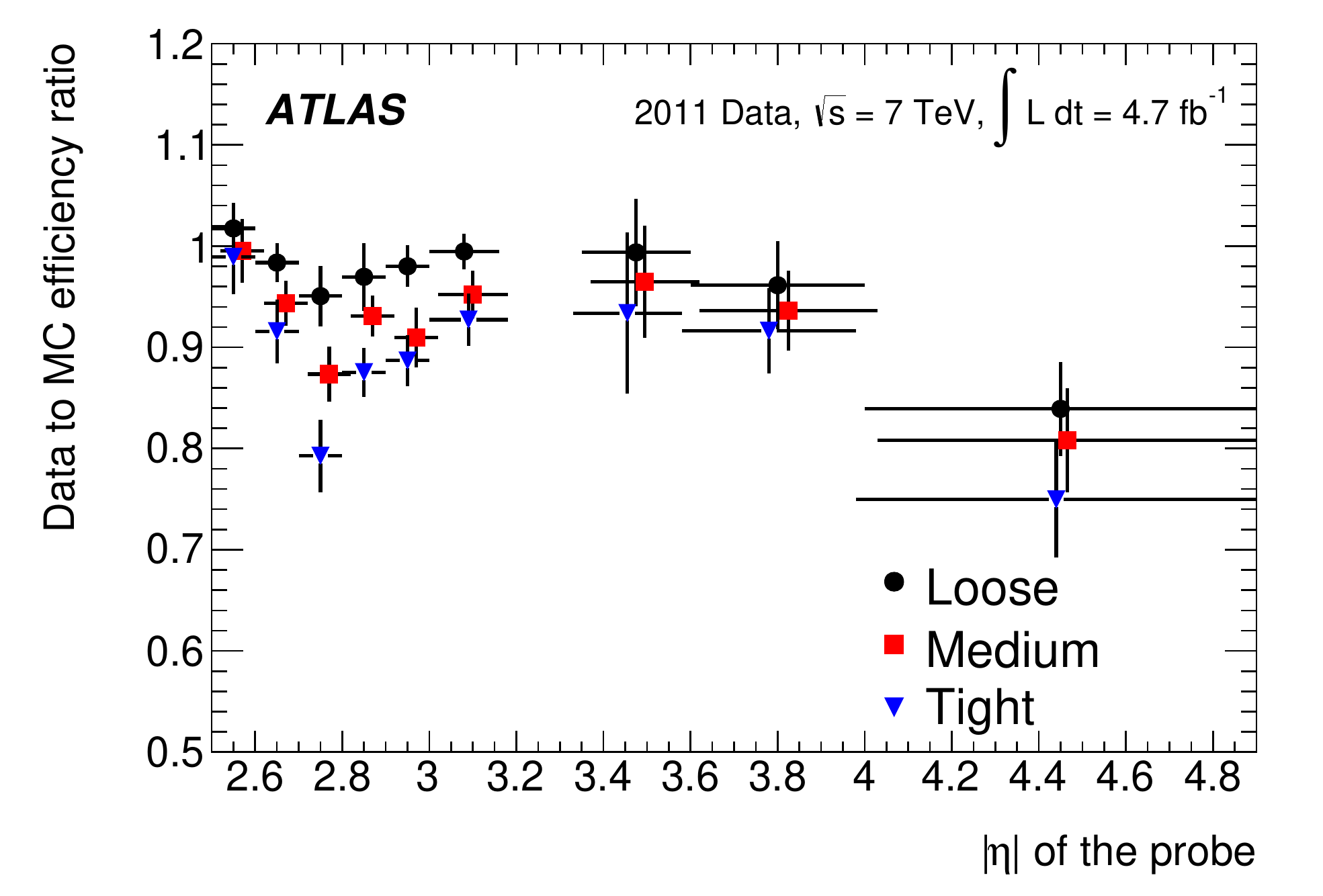}}
  \caption{The (a) \lpp, (b) \mpp, and (c) \tpp\ identification
    efficiencies as a function of the \abseta\ of the probe cluster in
    the forward region of the calorimeters for data and simulation. In
    (d), the data/MC ratio is shown for the three identification
    criteria, slightly shifted for better visibility. All plots are for probe electrons with $\et>20$~GeV. In all four figures, error bars correspond to the total
    uncertainties.}
  \label{fig:forwardID}
\end{figure*}

The electron identification efficiencies measured in data remain
stable with increasing pile-up but vary with \et\ and \abseta. The
simulation models well the measured efficiency shape as a function of
pile-up and of \et. However, it does not describe adequately the
efficiency measurements as a function of \abseta, as shown in
Figure~\ref{fig:forwardID}. This discrepancy is due to a mismodelling
of the shower shapes in the calorimeter and increases
with the tightness of applied identification criteria.
Data-to-simulation scale factors are computed in each \abseta\ bin to
correct for these differences (see Figure~\ref{fig:forwardID}(d)). The
resulting total uncertainty is 2--4\% and 4--8\% in the EMEC and FCal
regions, respectively, and it is dominated by the systematic
component.

\section{Reconstruction efficiency measurement}
\label{sec:RecoEff}

The EM cluster reconstruction efficiency $\epscluster$, for both the
central and forward electrons is determined from simulation of
\Zee~decays. From Section~\ref{sec:RecoIDDescription}, the
forward EM cluster reconstruction efficiency is better than 99\% for $\et>20$~GeV and
the central EM cluster reconstruction efficiency is 97\% and 99\% at 7
and 15~GeV, respectively. It then follows that the central-electron
reconstruction efficiency as measured in data reflects the performance
of the track reconstruction and the track--cluster matching procedure.
Efficiency values are measured for three event samples:
\begin{itemize}
\item{all reconstructed electron candidates;}
\item{electron candidates satisfying in addition a requirement on the quality of the matching track; this is to match the probe definition of the \Jpsiee\ selection used in the electron identification efficiency measurement;}
\item{electron candidates with a good track and satisfying in addition the requirement on the hadronic leakage $R_{\rm had}$ defined by the \lpp\ identification criteria; this is to match the probe definition of the \Wen\ and \Zee\ selections.}
\end{itemize}

For this measurement, only one of the channels available to the
identification efficiency measurement, a \Zee\ sample, can be used.
The \Zee\ event selection follows closely that used for the
measurement of the identification efficiency as described in
Section~\ref{sec:eventSel}, with the two exceptions noted in
Section~\ref{sec:method} related to the inclusion of photons in the
denominator of the efficiency definition and lack of charge
requirements on the tag and probe pairs due to the presence of probe
clusters without a matched track.

\begin{figure*}
  \centering
  \subfigure[]{\includegraphics[width=0.47\textwidth]{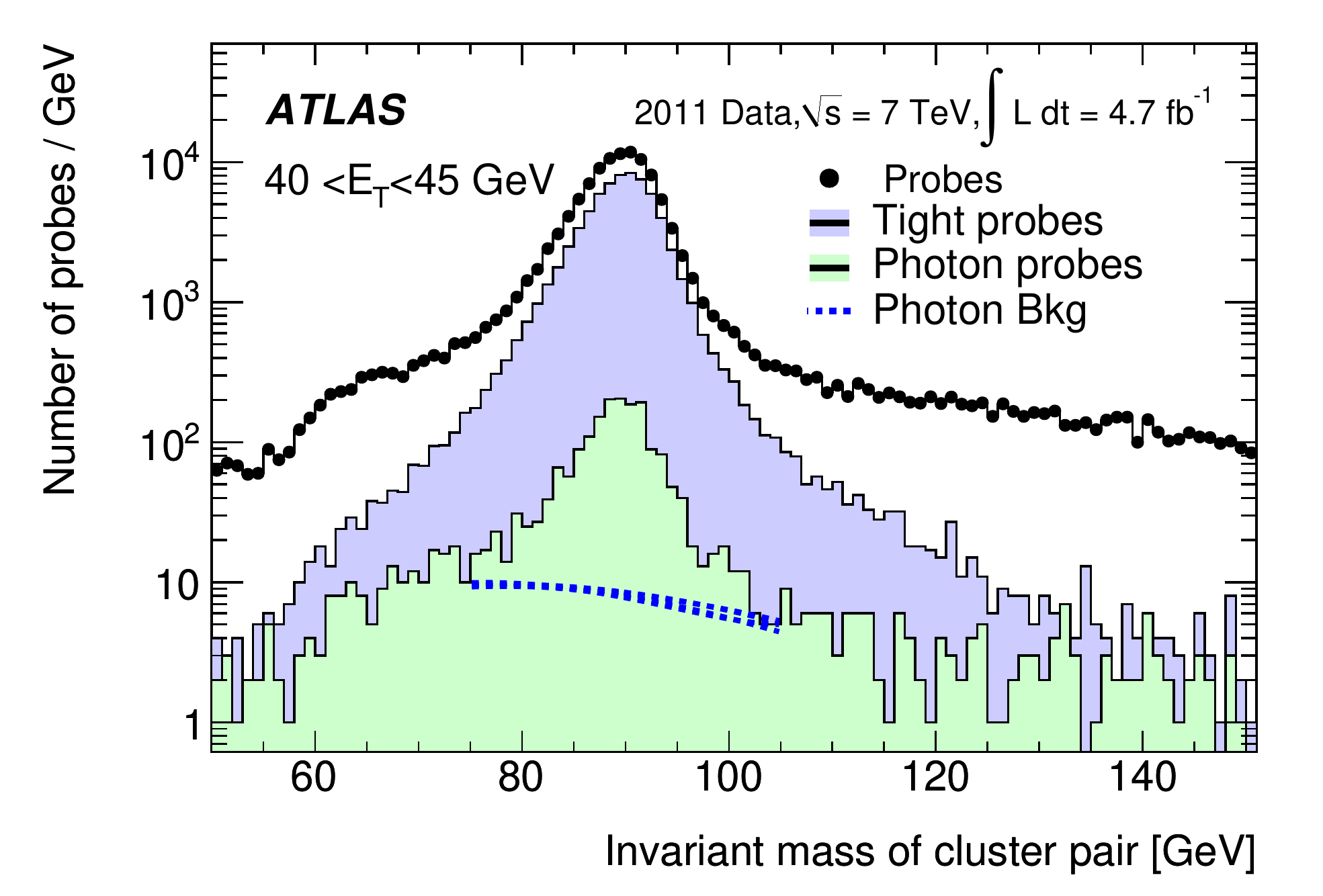} }            
  \subfigure[]{\includegraphics[width=0.47\textwidth]{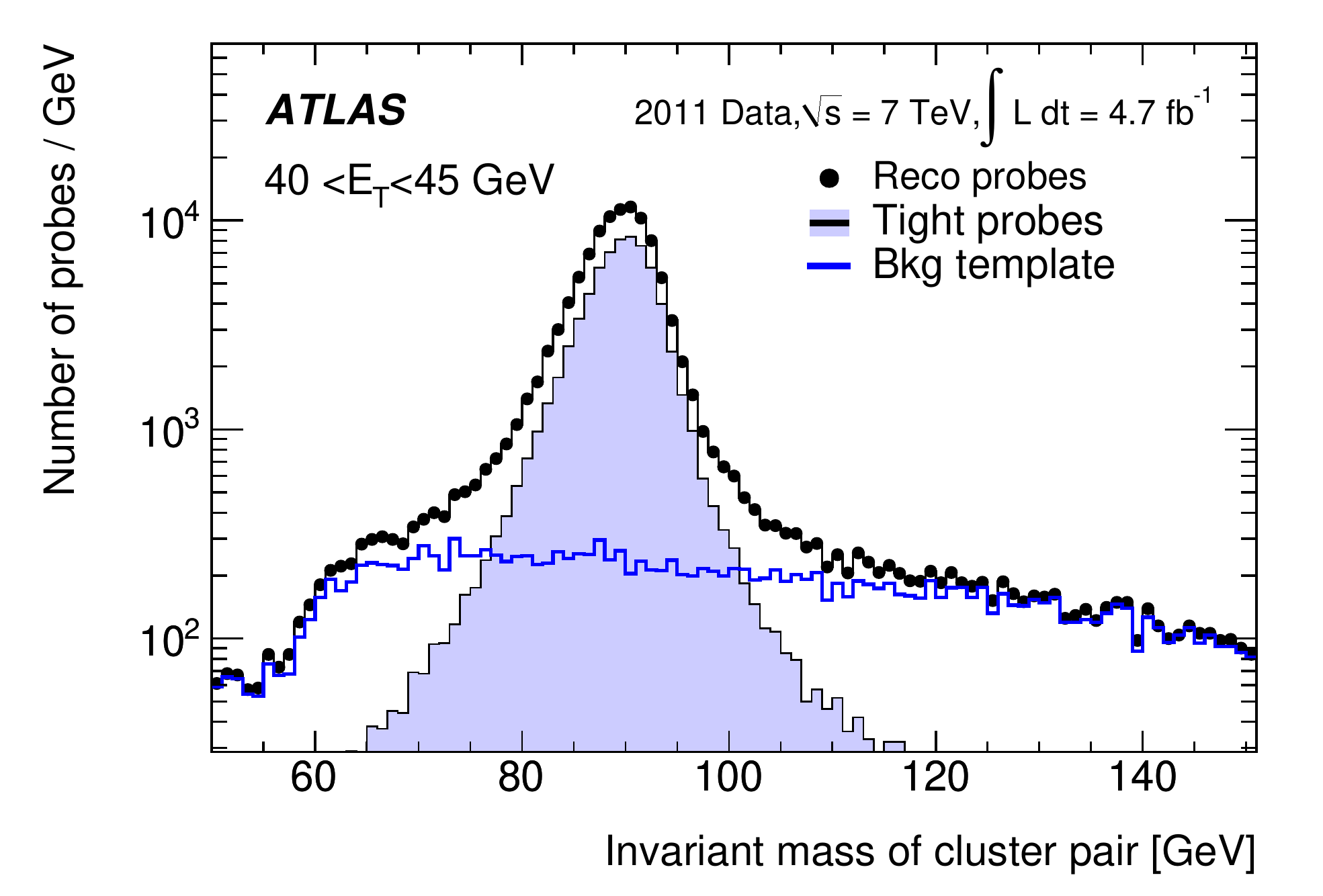}}
  \subfigure[]{\includegraphics[width=0.47\textwidth]{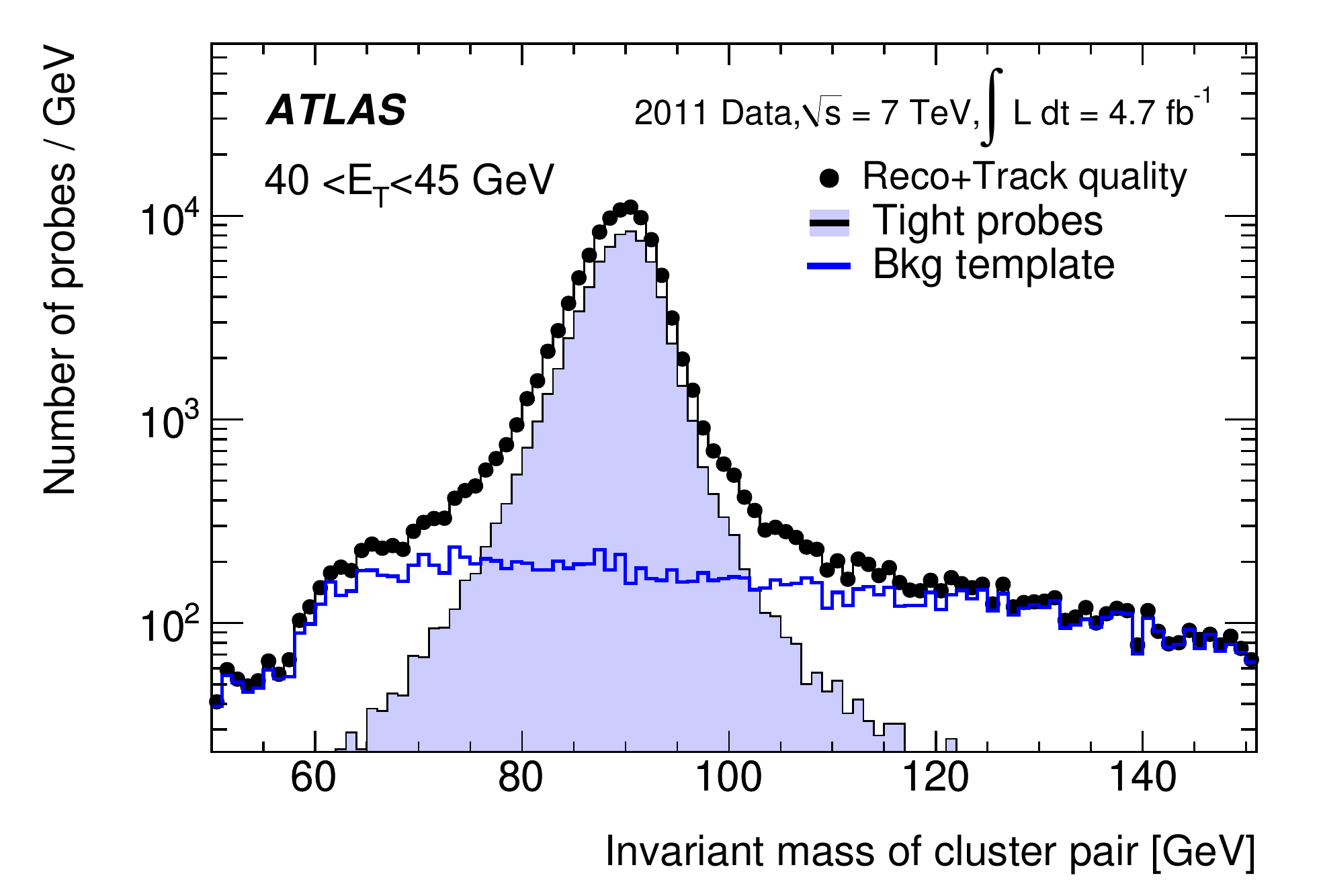} }
  \subfigure[]{\includegraphics[width=0.47\textwidth]{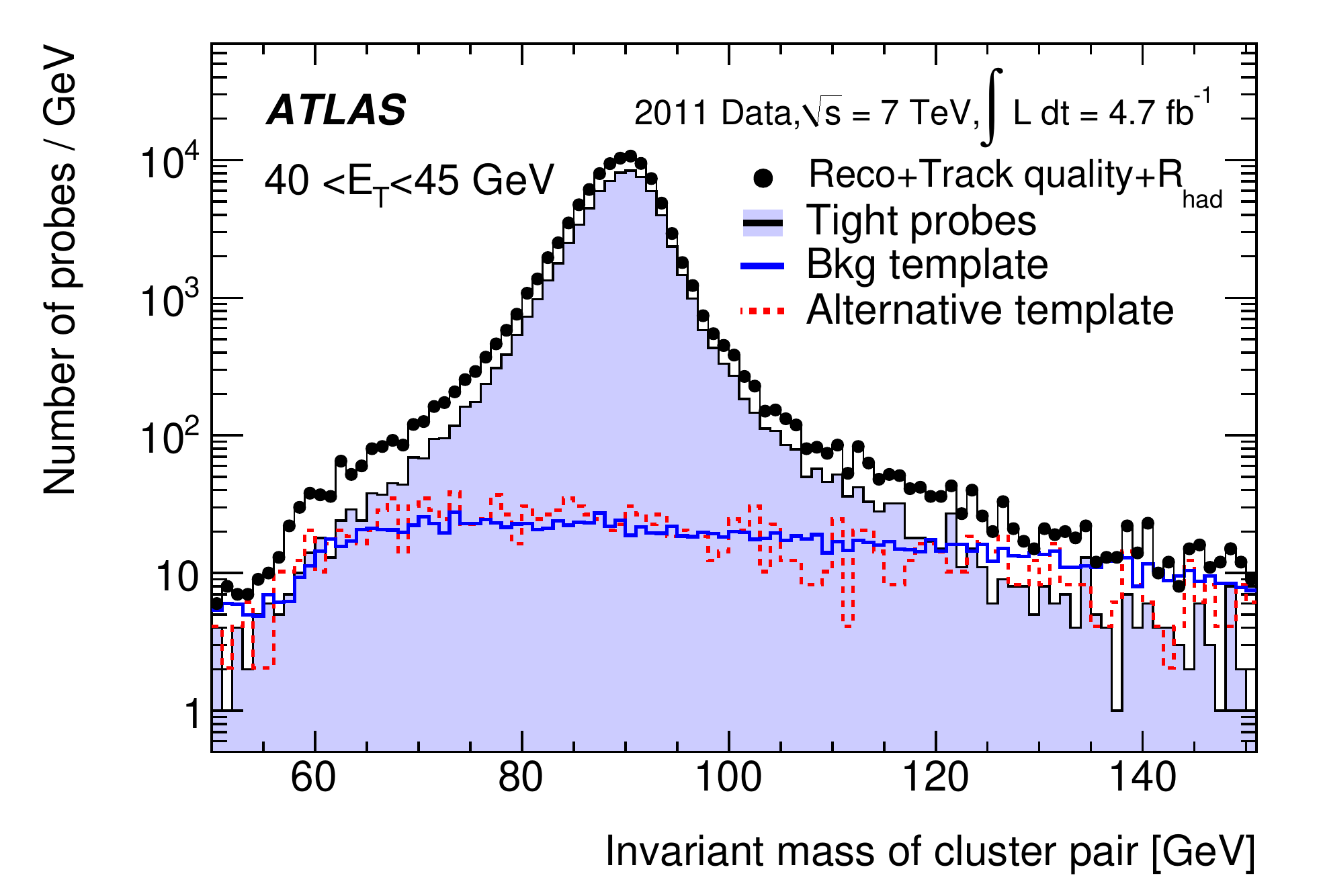} }
  \caption{ 
    Examples of cluster-pair invariant mass distributions at different
    levels of the probe selection, in the bin $40\leq\et<45\GeV$.
    (a) All reconstructed clusters associated with electrons and
    photons, used in the denominator of $\epsreco$; the
    photon background estimation used to evaluate
    the corresponding systematic uncertainty is shown.
    (b) Reconstructed electrons used in the numerator of $\epsreco$.
    (c) Reconstructed electrons passing the track quality requirement.
    (d) Reconstructed electrons passing the track quality and
    $R_{\rm had}$ requirements; two of the different background templates used to
    estimate the associated systematic uncertainty are shown.
    In all cases, the shaded histograms show the distributions
    obtained with probes after \tpp\ identification, to give an
    indication of the expected signal shape. }
  \label{fig:ztp_reco_invm_highpt}
\end{figure*}

Figure~\ref{fig:ztp_reco_invm_highpt} shows typical examples of the
cluster-pair invariant mass distributions used to evaluate $\epsreco$.
A total of 2.2 million probes were used to perform this measurement.
The backgrounds entering the numerator and denominator of $\epsreco$ are evaluated differently due to the inclusion of clusters associated with reconstructed photons in the denominator.

\subsection{Background evaluation}

The electron background contribution is estimated using a methodology
similar to that employed for the identification efficiency
measurements discussed in Section~\ref{sec:backgComp}, based on the
electron--positron invariant mass. The background cluster--pair
invariant mass template is obtained from data by reversing
identification requirements on the probe object and then normalising
the distribution to the background-dominated cluster--pair invariant
mass distribution in the 110--250~GeV region. When measuring the
reconstruction efficiency alone, the probe electrons of the background
template are required to fail at least two of the requirements
defining the \lpp\ identification, with the exception of those
associated with the track quality, and to satisfy the anti-isolation
requirement $\etconqu > 0.05$ (see
Figure~\ref{fig:ztp_reco_invm_highpt}(b)). When measuring the
reconstruction efficiency for electrons passing the track quality
requirement, the background templates are obtained from objects either
passing or failing this extra requirement. The background templates
used for the measurement employing the additional requirement on the
hadronic leakage $R_{\rm had}$ are built in a similar fashion (see
Figures~\ref{fig:ztp_reco_invm_highpt}(c) and (d)).

The background from real photons is estimated using the invariant mass
distribution of pairs composed of an electron tag and a probe
reconstructed only as a photon,
$m_{e\gamma}$. The sideband regions above and below the $Z$-boson
resonance mainly contain background events. These regions, corrected
for the expected number of genuine electron--positron pairs as
estimated from simulation are fit to a third-order polynomial
function. The number of background events associated with photons is
then obtained by integrating this fit function in the signal region
(see Figure~\ref{fig:ztp_reco_invm_highpt}(a)).

\subsection{Reconstruction efficiency and systematics}

\begin{figure*}[p]
  \centering
  \includegraphics[width=0.47\textwidth]{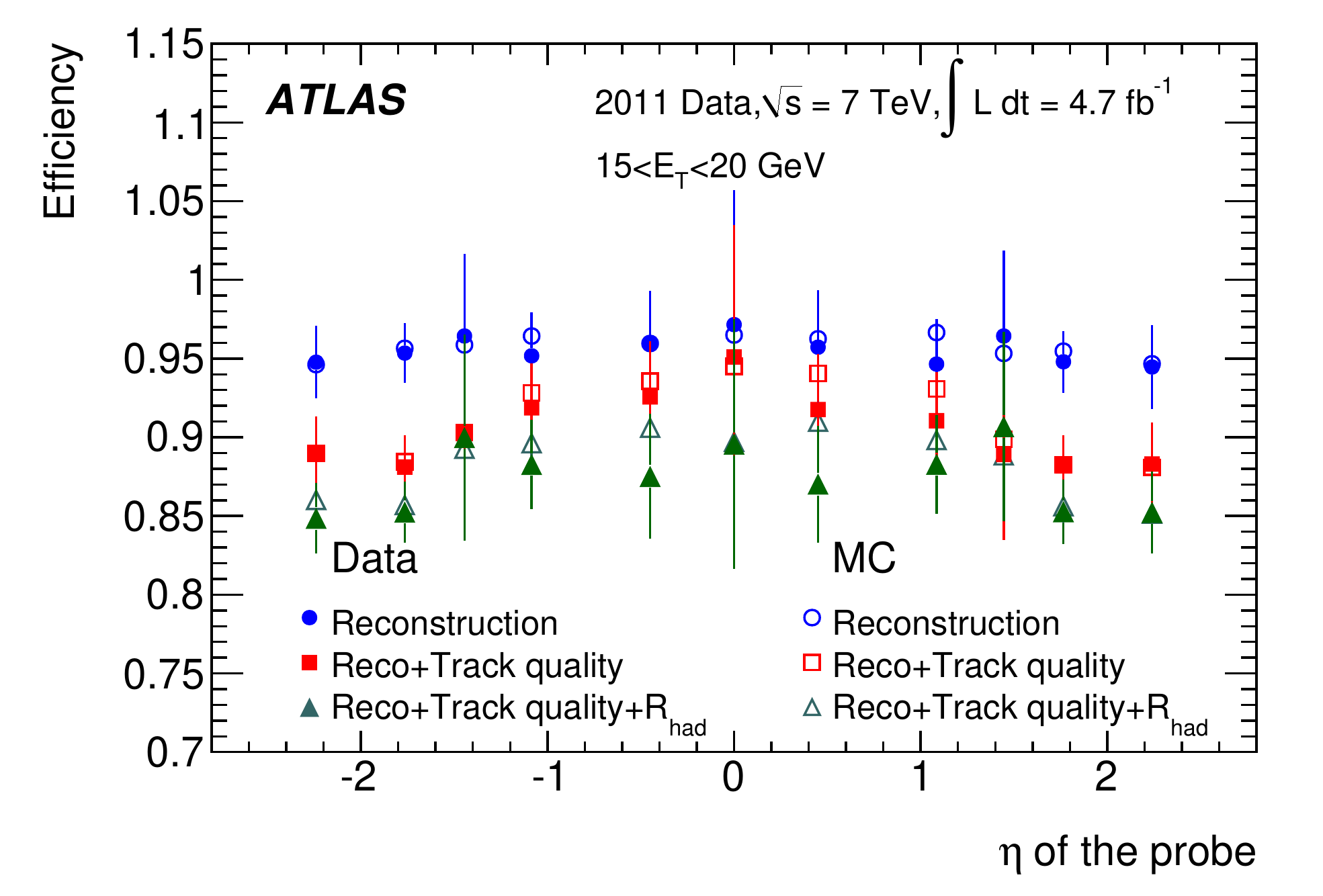}
  \includegraphics[width=0.47\textwidth]{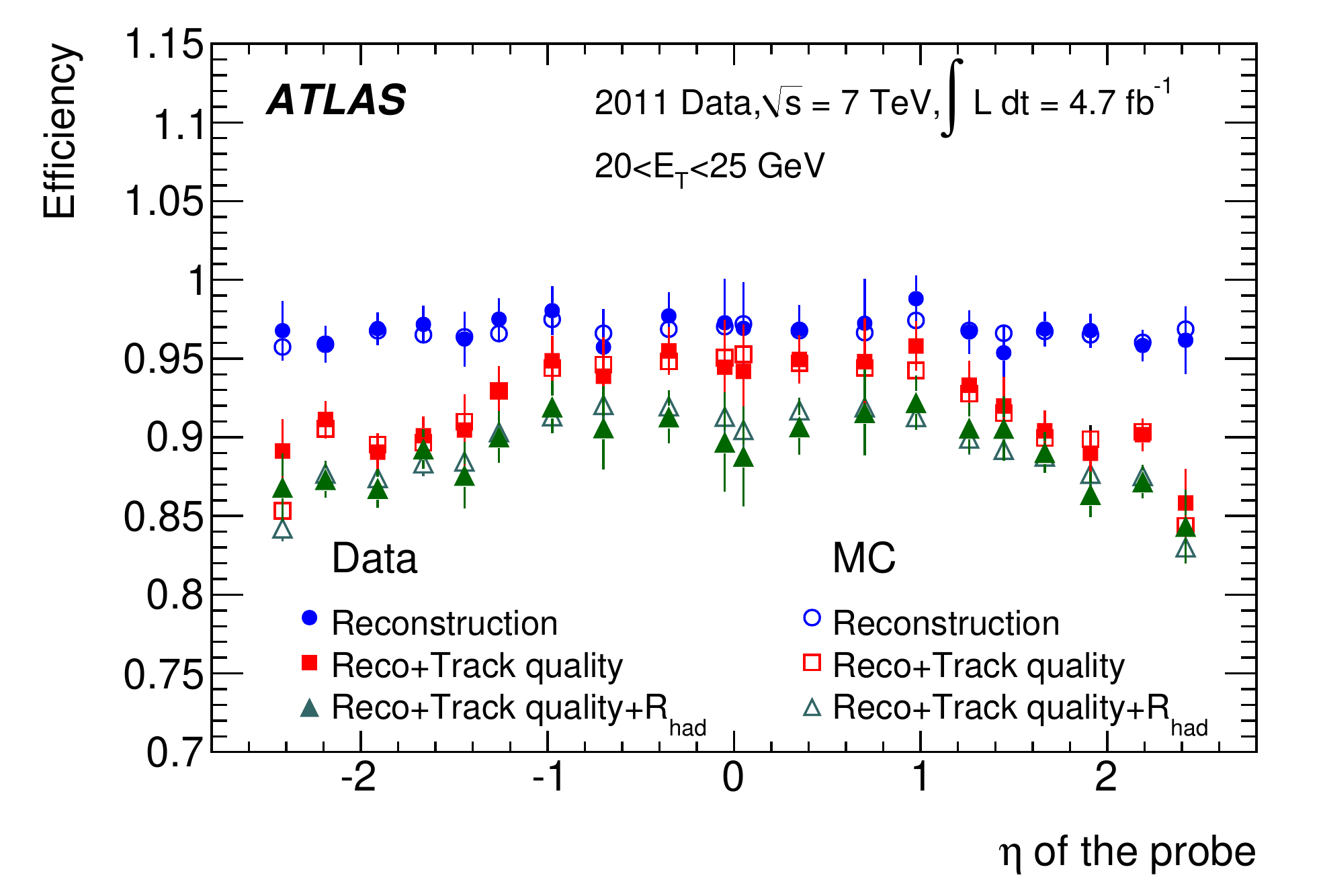}
  \includegraphics[width=0.47\textwidth]{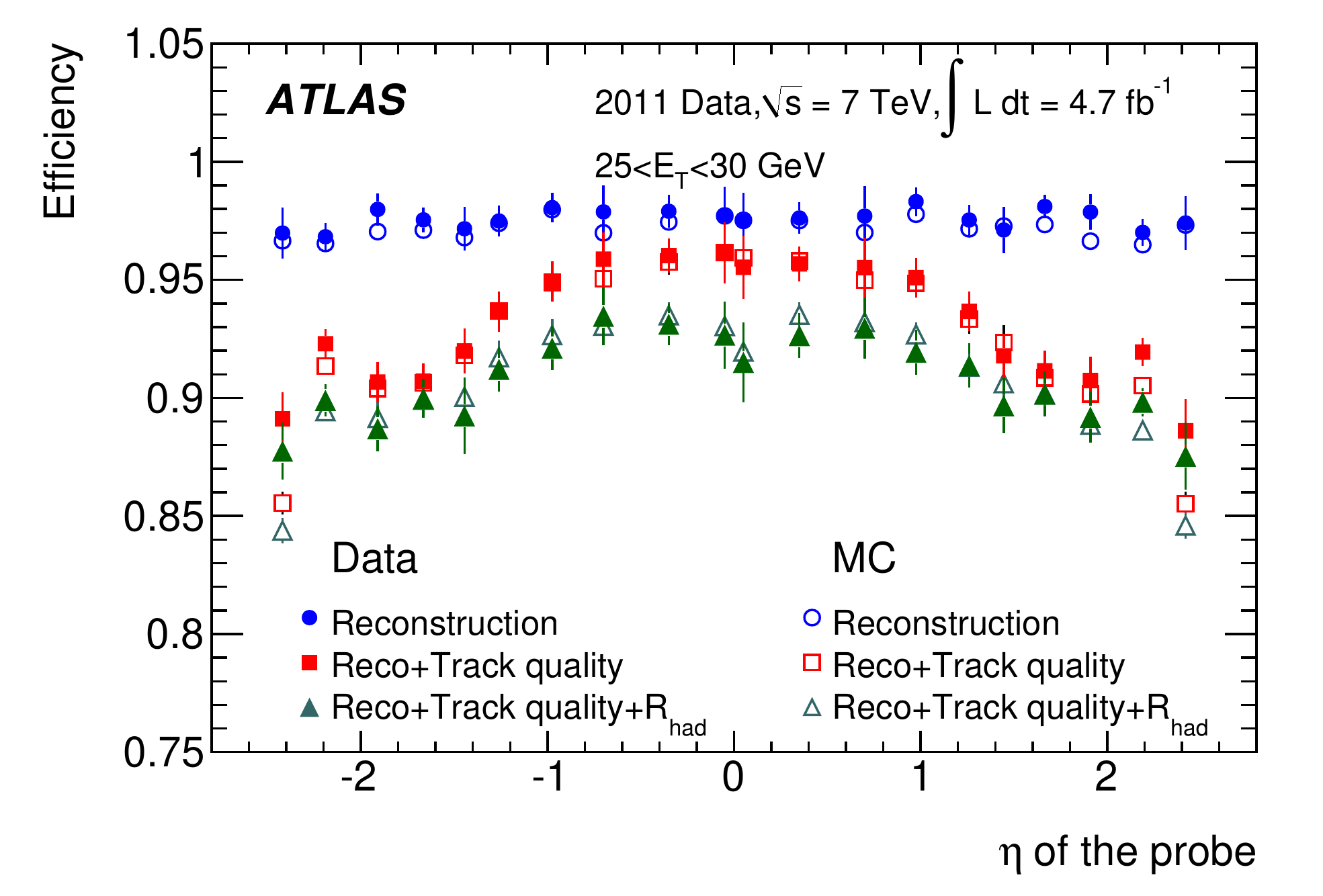}
  \includegraphics[width=0.47\textwidth]{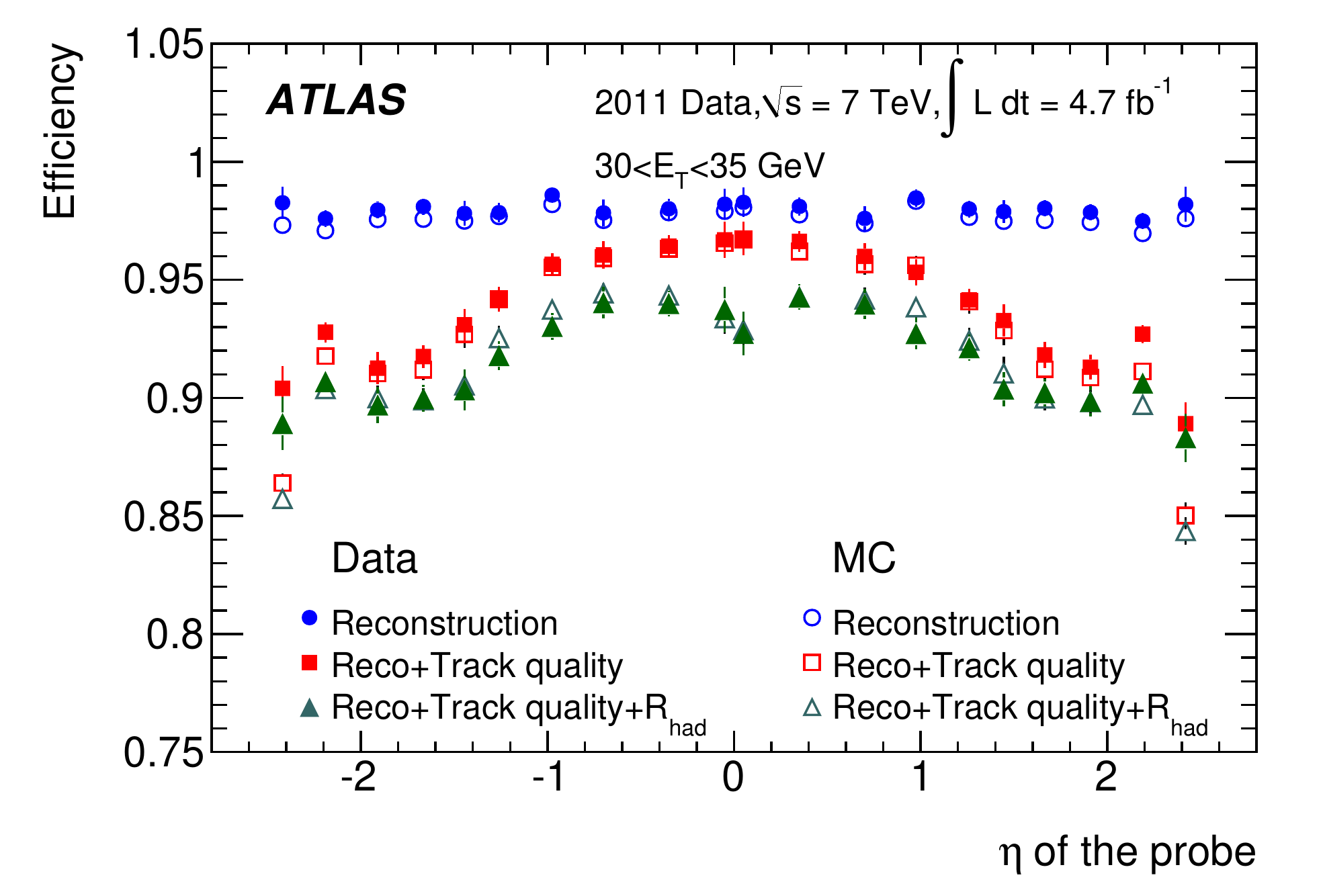}
  \includegraphics[width=0.47\textwidth]{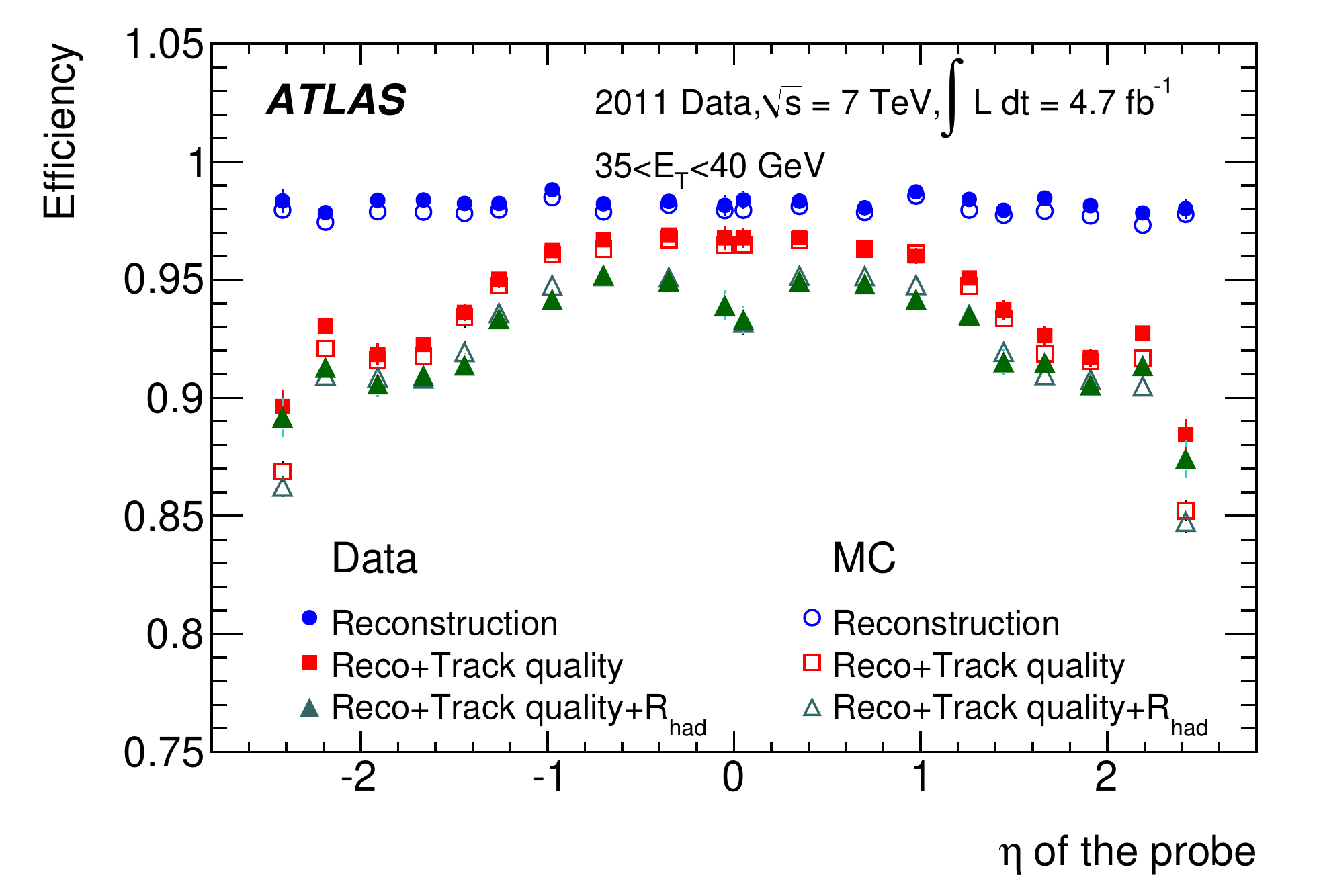}
  \includegraphics[width=0.47\textwidth]{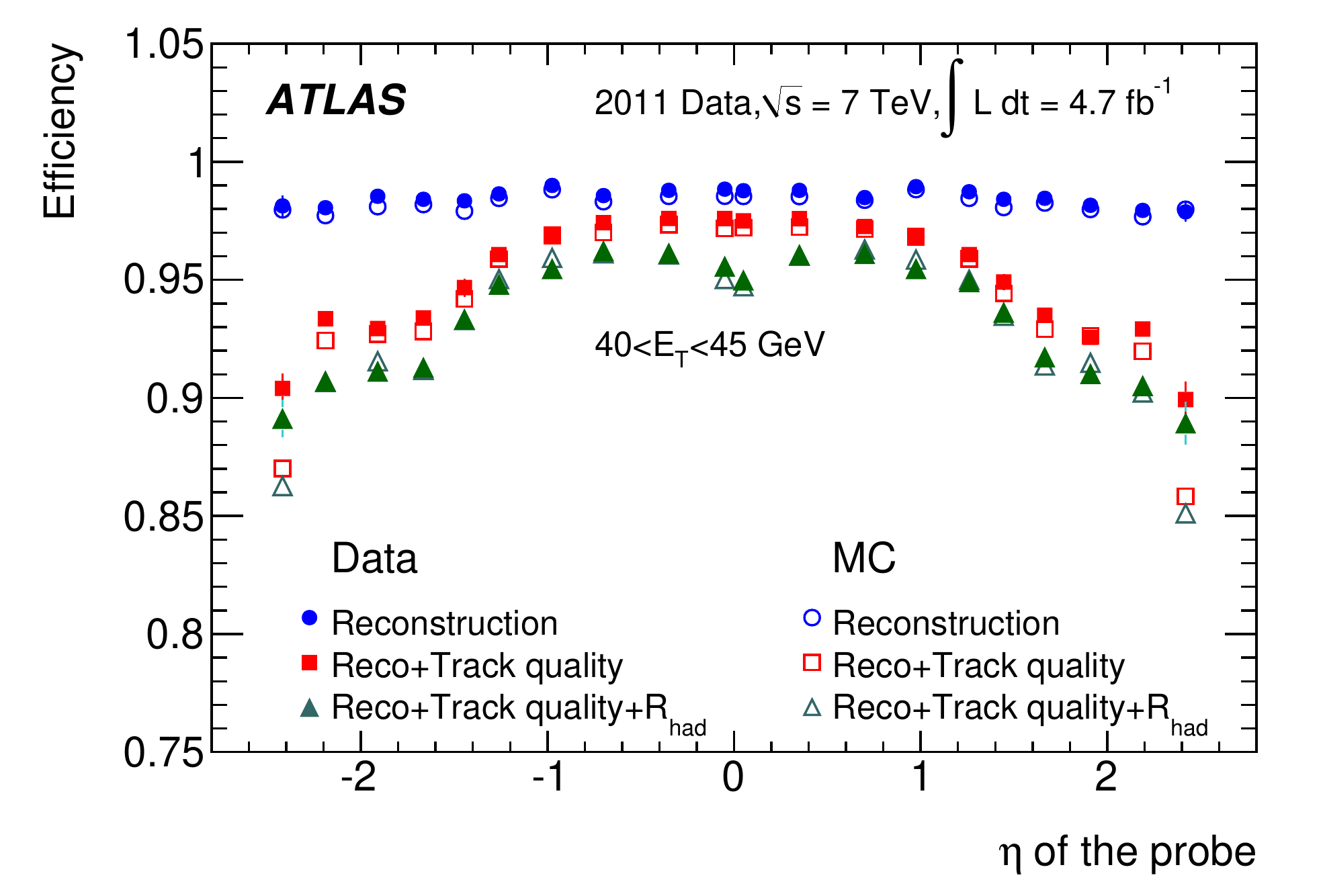}
  \includegraphics[width=0.47\textwidth]{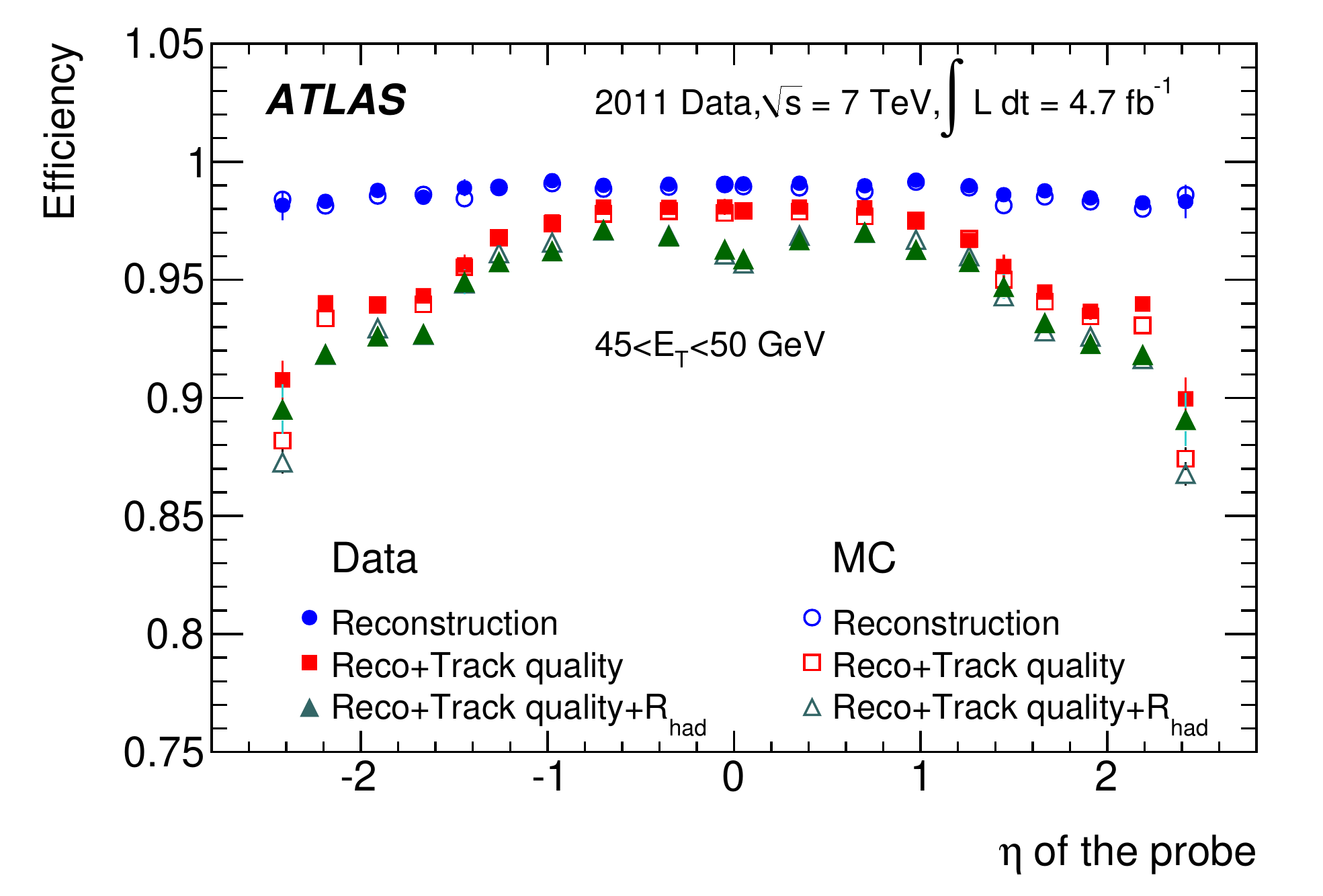}
  \includegraphics[width=0.47\textwidth]{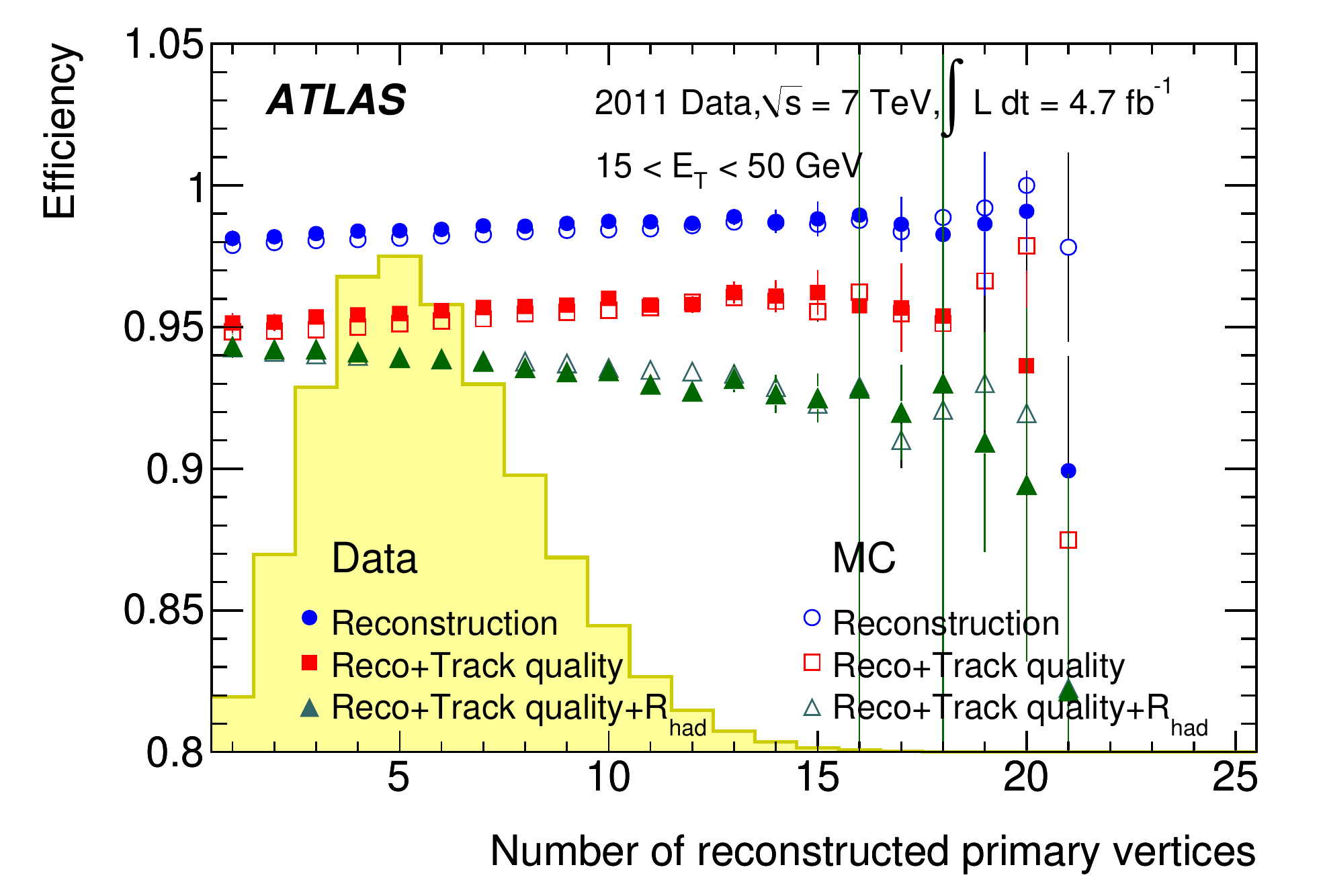}

  \caption{The three types of reconstruction efficiencies, with their total
    uncertainties, as measured in data and simulation in bins of probe \et\
    from $15<\et<20$~GeV to $45<\et<50$~GeV. The final plot on the bottom right shows the efficiency as a function of the number of reconstructed primary vertices in the event. The solid yellow histogram indicates the $N_{\mathrm{PV}}$ distribution in the data. The error bars correspond to the total uncertainties.}
\label{fig:ztp_reco_eff}
\end{figure*}

The reconstruction efficiency as a function of pseudorapidity for all
three event selections is shown in Figure~\ref{fig:ztp_reco_eff} in
\et\ bins ranging from 15 to 50~GeV. In the lowest \et\ bin, a coarser
$\eta$ binning is used to cope with the smaller data sample, and still
ensure that the total uncertainty is equally shared between
statistical and systematic sources.

The systematic uncertainties are assessed by varying parameters in the
fitting procedure and measuring the global systematic uncertainty as
the RMS of the distribution of the results obtained with each
configuration. These variations include identification quality of the
tag electron, the invariant-mass range used to select the signal
events, the template shape used for electrons, and the sideband fit
range for the photon background evaluation. For this latter
uncertainty, the systematic uncertainty associated with the estimate
of the genuine electron--positron events in the sideband region is
evaluated by varying this number by $\pm30$\% in each $m_{e\gamma}$
bin, assuming this variation is fully correlated between bins. The
30\% variation is conservatively estimated from the largest observed
difference between data and simulation for the probability with which
electrons are misidentified as photons. The signal contamination in the template and
in the normalisation region is taken into account by varying the
amount of signal leaking into the cluster-pair invariant mass
template, and by estimating the signal contamination in all other
regions from simulation. Similarly to what is done for the photon
background evaluation, the latter prediction is assigned
a conservative 20\% uncertainty.

From Figure~\ref{fig:ztp_reco_eff}, the efficiency to reconstruct an
electron or positron having a track of good quality and matching an
electromagnetic cluster that fulfils the $R_{\rm had}$ requirement
varies for high-\et\ probes from about 96\% in the barrel region of
the calorimeter, to about 90\% in the endcap region for $\et>30\GeV$.
For $\et<25\GeV$, this efficiency drops to about 93\%~(85\%) in the
barrel (endcap) region. For $\et>35\GeV$, the total uncertainty on the
measured reconstruction efficiencies is well below 0.5\%.

The reconstruction efficiency may be affected by the ambient activity
resulting from pile-up interactions. The final plot in
Figure~\ref{fig:ztp_reco_eff} shows the values of the three
reconstruction efficiencies as a function of the number of
reconstructed primary vertices $N_{\mathrm{PV}}$ in the event. The
$R_{\mathrm{had}}$ requirement introduces the largest sensitivity to
pile-up, demonstrated by the few percent efficiency variation as
$N_{\mathrm{PV}}$ varies from 1 to 20; this dependence is well
modelled by the simulation.

\begin{figure*}
  \centering
\subfigure[]{  \includegraphics[width=0.47\textwidth]{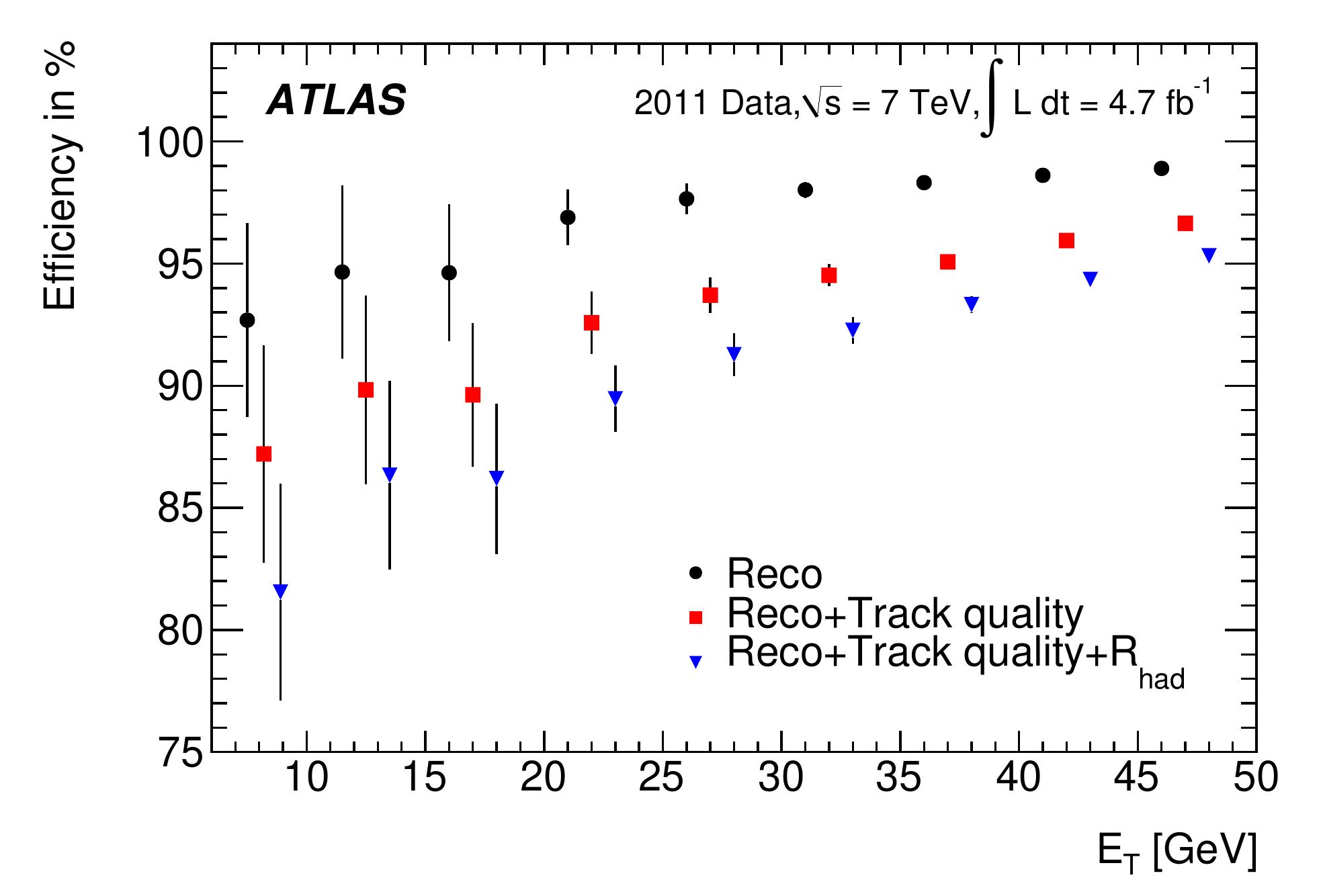}}
\subfigure[]{  \includegraphics[width=0.47\textwidth]{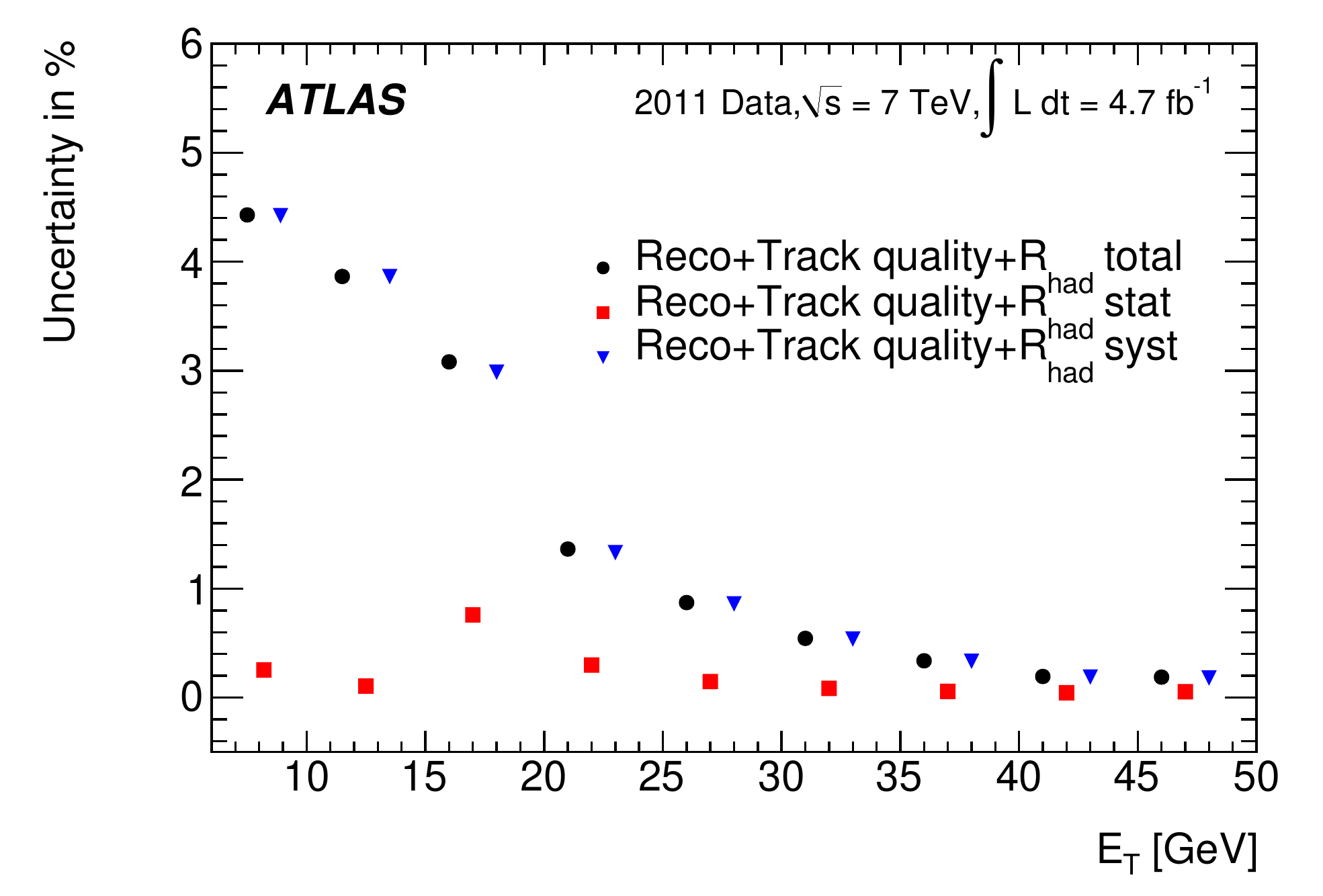}}
  \caption{(a) Reconstruction efficiency as a function of \et~ for central electrons. The
    error bars correspond to the total uncertainties. (b) Composition
    of the total uncertainties is shown as a function of \et . For $\et<15$~GeV no measurement with data was
    possible and the expected efficiencies from the \Zee\ MC sample
    were used directly. In this case, conservative uncertainties of
    2\% and 5\% were assigned for the barrel and the endcap regions,
    respectively. Some points are slightly shifted horizontally within
    the \et~bin for better visibility.}
\label{fig:IntegReco}
\end{figure*}

The significant background contamination and low statistics of probes
at low \et\ does not permit a measurement of the reconstruction
efficiency for $\et<15$~GeV from \Zee\ decays. Furthermore it is not
possible to trigger a sufficiently large sample of \Jpsiee\ or \Wen\
events unbiased with respect to the reconstruction efficiency
measurement. In the region from 7 to 15~GeV, the prediction from
simulation is used instead with fair confidence, based on the observed
good MC modelling in the \et~region beyond 15 GeV. For this
extrapolation, conservative uncertainties of 2\% and 5\% are assigned
in the barrel and endcap regions, respectively.
Figure~\ref{fig:IntegReco} shows the three types of reconstruction
efficiencies as a function of \et. In the two lowest \et\ bins, where
no data measurement exists, the expected efficiencies from a \Zee\ MC
sample were used, assigning the systematic uncertainties quoted above.
The integration of the measurements over the pseudorapidity range
decreases the statistical uncertainty such that the systematic
component dominates overs the entire \et~range.

\subsection{Combined reconstruction and identification efficiency measurement}
 
\begin{figure*}
  \centering
\subfigure[]{  \includegraphics[width=.47\textwidth] {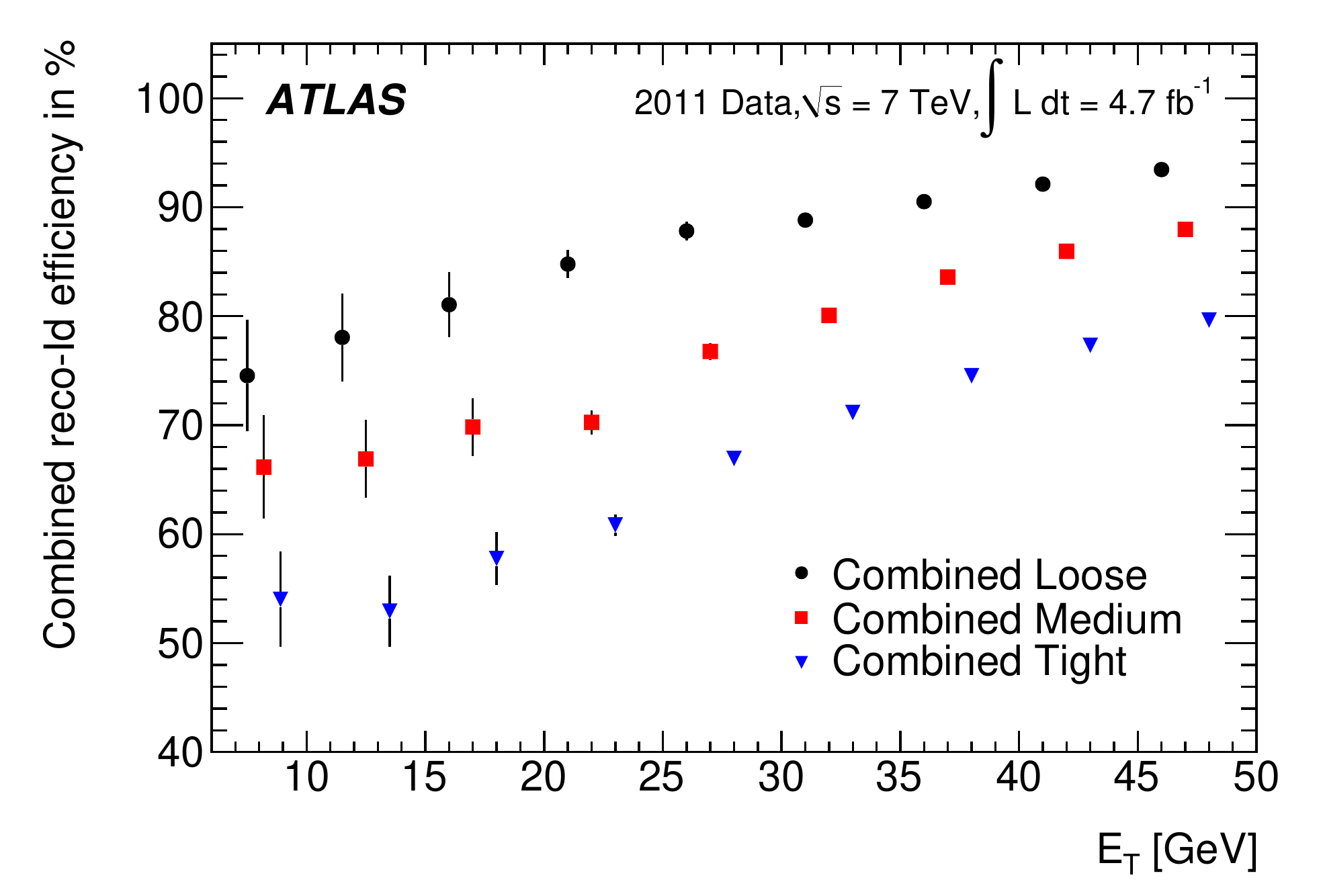}}
\subfigure[]{  \includegraphics[width=.47\textwidth] {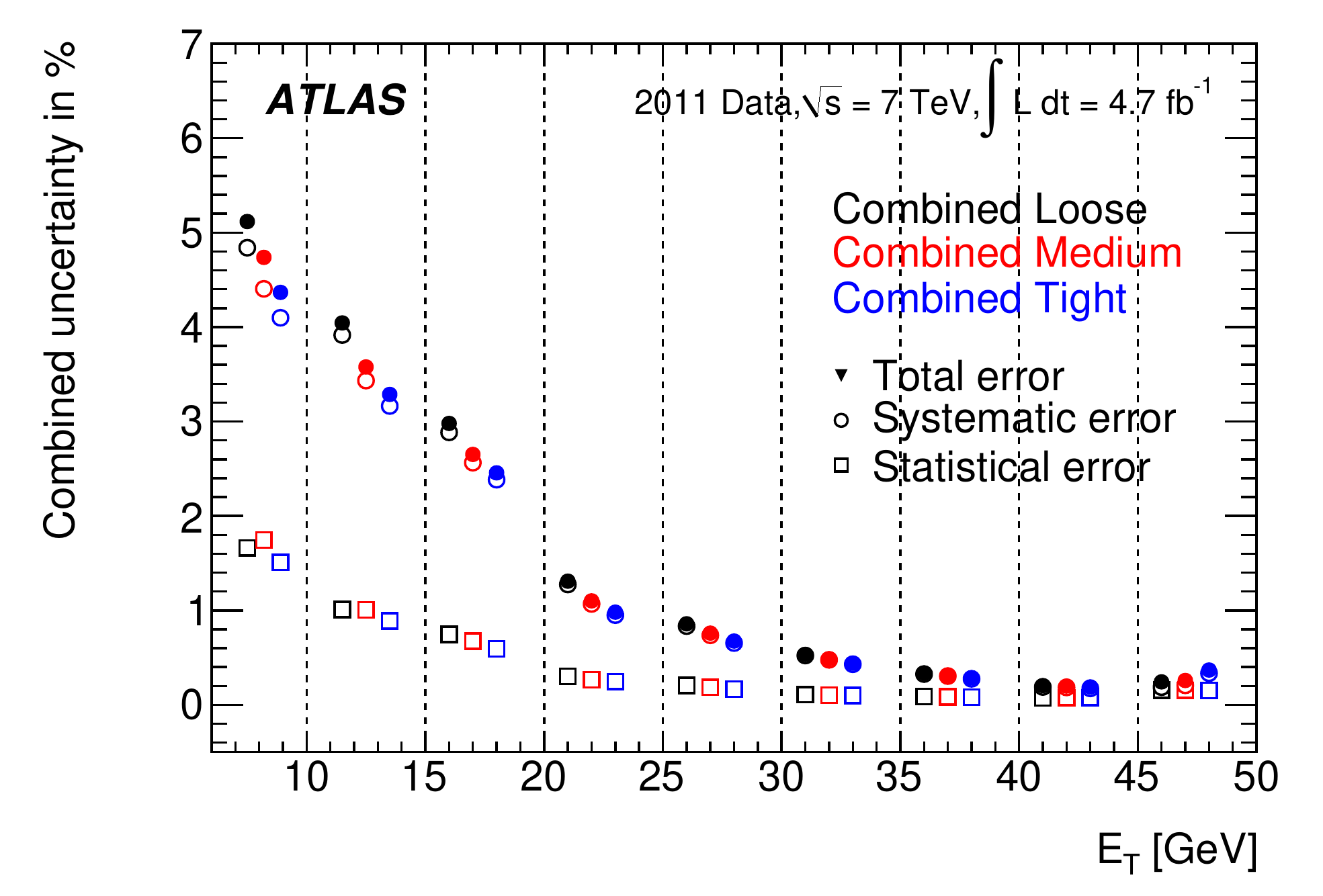}}
  \caption{(a) Central-electron combined reconstruction and identification efficiencies as
    a function of \et~, for the reconstruction plus track
    quality plus hadronic leakage requirements and all three
    identification criteria. The error bars correspond to the total
    uncertainties. (b) Breakdown of the total uncertainty of the
    combined measurement into statistical and systematic components as
    a function of \et. Some points are
    slightly shifted horizontally within the \et~bin for better
    visibility.}
  \label{fig:combinedRID}
\end{figure*}

The reconstruction efficiency presented in this Section and the
identification efficiency in Section~\ref{sec:IDmeas} are combined to
provide the electron reconstruction and identification efficiency
measurement. This combined efficiency, integrated over the range
$\abseta<2.47$, along with the corresponding uncertainty, is presented
as a function of \et\ in Figure~\ref{fig:combinedRID}.

\section{Charge-identification efficiency}
\label{Sec:chargemisID}

The correct identification of the charge of an electron is important
in many analyses, e.g. when exploiting charge correlations of the
final-state particles. Electron charge-misidentification may occur
when electrons radiate early in the detector, such as near the
entrance of the inner tracking detector, and resulting photons
subsequently convert and are reconstructed as high \pt\ tracks. A
particle with reconstructed charge opposite to the parent electron may
then accidentally be associated with the calorimeter cluster. These
effects are expected to follow the distribution of material in the
detector, which Figure~\ref{fig:material} shows to be $|\eta|$
dependent.

The probability to correctly identify the charge of the candidate
electron is evaluated with a \tandp\ analysis employing a \Zee\
sample, considering as probes the ensemble of di-electron pairs
without any requirement on the reconstructed sign of the track. The
tag is required to satisfy \tpp\ identification criteria, to be well
isolated ($\etcontr<0.15$) and to have transverse energy greater than
25~GeV. To ensure a well-measured tag charge, the tag is confined to
the barrel region of the calorimeter ($\abseta <1.37$) where the
charge reconstruction efficiency is observed to be very high. The
probe electron is also required to have $\et>25$~GeV and be anywhere
within the acceptance of the inner detector. No correction is applied
for the misidentification of the tight central tag electron. This
increases the measured charge-identification probability by about
0.2\%.

The invariant mass of the tag--probe system is used as the
discriminating variable to separate signal from background events. The
background template is obtained from events that have the tag
candidate satisfying the \mpp\ criteria and the probe candidate
failing to satisfy the \lpp\ criteria. The invariant mass spectrum as
measured in data is fit in the region $66<\mee < 116$~GeV using the
sum of the \Zee\ signal template from simulation and the background
template from data. The yield in the signal region is counted in the
invariant mass range of 80 to 101~GeV.

\begin{figure*}
  \begin{center}
\subfigure[]{    \includegraphics[width=0.47\textwidth]{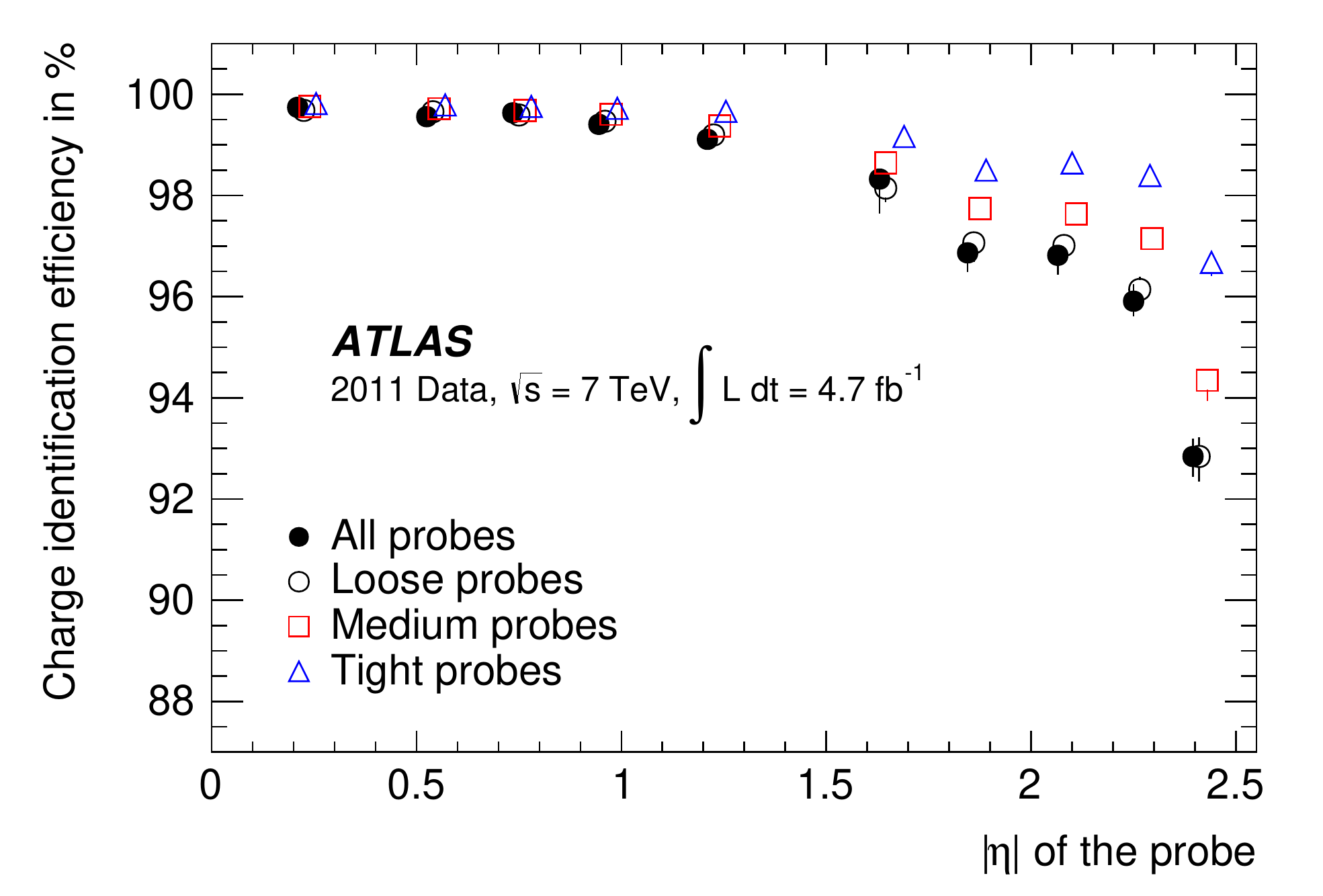}}
\subfigure[]{    \includegraphics[width=0.47\textwidth]{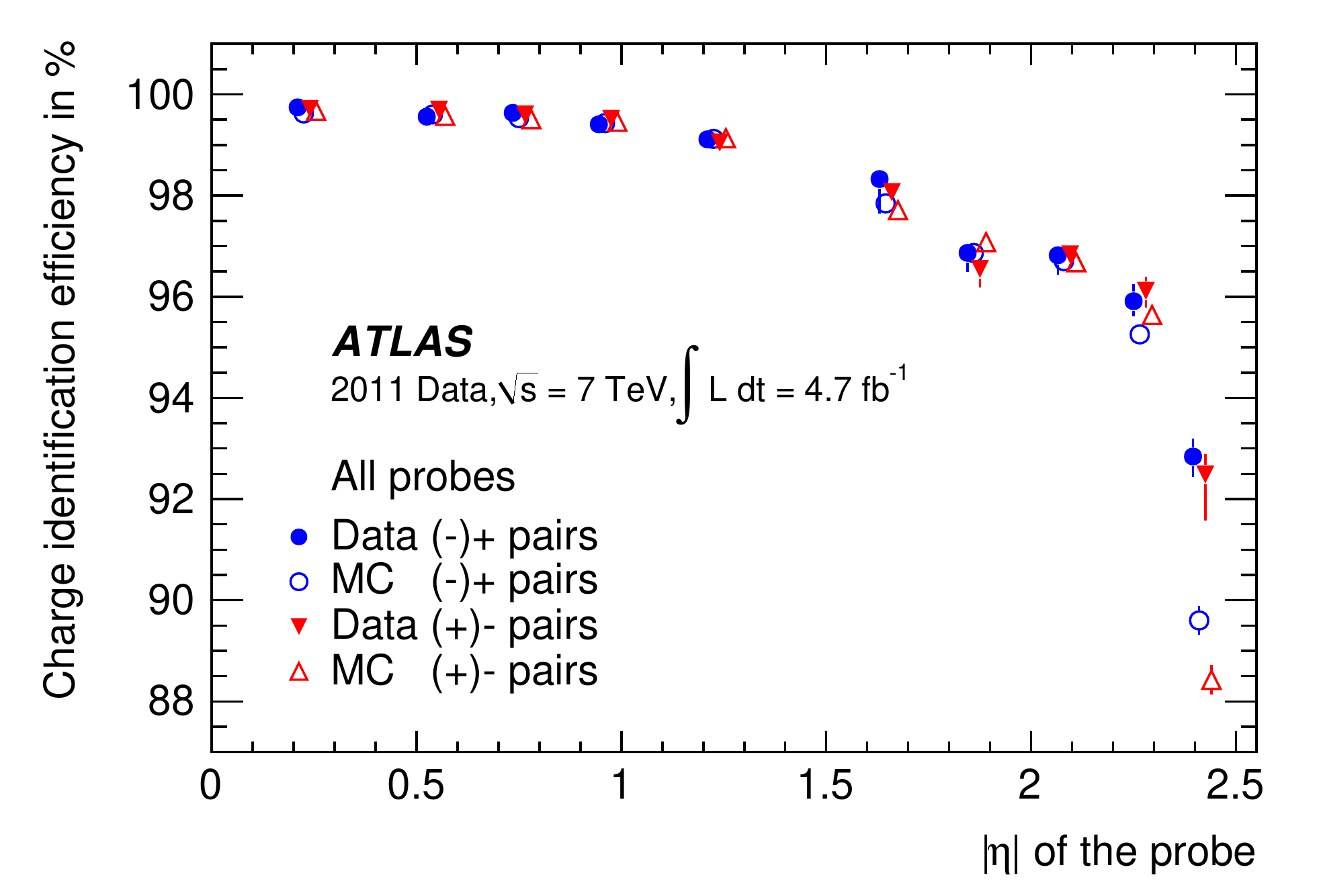}}
  \end{center}
  \caption{Charge-identification efficiency for electrons with $\et>25$~GeV in a \Zee\ sample given as a function of the
    probe \abseta. (a) Measurement of the data charge-identification
    efficiencies for reconstructed, \lpp, \mpp, and \tpp\ probes.
    (b) Comparison of the efficiencies for all electron and
    positron probes as measured in data (closed points) and in
    simulation (open points). The sign within the bracket is the
    charge of the tag while the sign next to it is that of the probe.
    The error bars correspond to the total uncertainties. Some points
    are slightly shifted horizontally within the \eta~bin for better
    visibility.}
  \label{fig:charIDplot}
\end{figure*}

The charge-identification efficiency, extracted by comparing this
yield to the subset of opposite-sign pairs, is measured for all levels
of electron identification: reconstruction, \lpp, \mpp, and \tpp, as
shown in Figure~\ref{fig:charIDplot}(a) and compared with the
equivalent numbers as extracted from simulation in
Figure~\ref{fig:charIDplot}(b). The tightness of the tag selection
and the definition of the signal region are varied to assess the
systematic uncertainty. The charge-identification efficiency is found
to be high ($>99.7$\%) and relatively constant in the barrel region of
the calorimeter decreasing to 93\% in the endcap region. In this
region, the efficiency increases to 97\% when applying the \tpp\
selection to the probes. The agreement between data and simulation is
good for all $\eta$ values except at the outermost edge of the
acceptance where the simulation predicts a higher misidentification
probability. This discrepancy may originate from incorrectly
modelled material in the simulation. The same Figure shows
that the measured efficiencies do not depend on the reconstructed sign
of the probe track.

\section{Summary}
\label{sec:Conclusion}
 
The ATLAS experiment at the LHC recorded approximately 4.7~\ifb\ of proton--proton collision
data in 2011 at a centre-of-mass energy of $\sqrt{s}= 7$~TeV. The
tag-and-probe methods developed to measure the components of the
electron efficiency with these data are described in detail. In
comparison to similar results based on 2010 data~\cite{2011mk}, the
revised analysis methods presented here, in combination with the
higher statistics provided by the 2011 data in the \Zee, \Wen\ and
\Jpsiee\ channels, have enabled precision measurements of electron
efficiency in a finely grained two-dimensional grid of probe electron
(\et,\eta).

The electron reconstruction efficiency, which is related to the
ability to associate a candidate electron track with a corresponding
EM cluster, was extracted from a \Zee\ sample of probe electrons in
the central region of the detector ($\abseta<2.47$) using a fine \eta\
granularity and seven \et\ bins in the range of 15 to 50~GeV. The
statistical precision is the dominant source of uncertainty of the
two-dimensional measurement, with the total uncertainty varying from a
few percent in the lowest \et\ bin to $\sim 0.5$\% at 35~GeV.

The efficiency to identify electrons given the existence of a
reconstructed-electron candidate is assessed in the central region of
the detector ($\abseta<2.47$) for three benchmark selection criteria
called \lpp, \mpp, and \tpp. 
A combination of the data to Monte Carlo efficiency ratios measured from 
\Zee, \Wen, and \Jpsiee\ samples is performed in a fine (\et,\eta) grid, over the
probe \et\ range from 7 to 50~GeV. This results in a typical accuracy on
the efficiency to identify electrons from Z decays of a few per mil at
$\et=35$~GeV and 1--2\% for $\et<20$~GeV and it is dominated by the statistical
uncertainty. As a consequence of improvements in the simulation, the
measured efficiencies demonstrate better agreement with 
expectations compared to the results presented in Ref.~\cite{2011mk}, varying
with \et\ and \eta\ from a few per mil to a few percent. In the forward region
($2.5<\abseta<4.9$), the efficiency of the entirely calorimeter-based
\lpp, \mpp, and \tpp\ criteria was measured in nine \abseta\ bins for
probe $\et>20$~GeV with a total uncertainty of few percent, mostly
arising from systematic effects. In this region, a larger discrepancy
is observed between measured and expected efficiencies.

The efficiency for a correct charge reconstruction for \tpp\ electrons
with $\et>25$~GeV is found from a \Zee\ sample to be $>99.7$\% in the barrel region of the
detector, decreasing to $\sim 97$\% in the endcaps, independent of
lepton charge.

Overall, the work presented in this paper has enabled precision
measurements of two-dimensional efficiencies, improving by
approximately an order of magnitude the uncertainties assigned to the
results presented in Ref.~\cite{2011mk}. These improvements have
greatly benefited the analyses performed by the ATLAS collaboration.


\section{Acknowledgements}

We thank CERN for the very successful operation of the LHC, as well as the
support staff from our institutions without whom ATLAS could not be
operated efficiently.

We acknowledge the support of ANPCyT, Argentina; YerPhI, Armenia; ARC,
Australia; BMWF and FWF, Austria; ANAS, Azerbaijan; SSTC, Belarus; CNPq and FAPESP,
Brazil; NSERC, NRC and CFI, Canada; CERN; CONICYT, Chile; CAS, MOST and NSFC,
China; COLCIENCIAS, Colombia; MSMT CR, MPO CR and VSC CR, Czech Republic;
DNRF, DNSRC and Lundbeck Foundation, Denmark; EPLANET, ERC and NSRF, European Union; IN2P3-CNRS, CEA-DSM/IRFU, France; GNSF, Georgia; BMBF, DFG, HGF, MPG and AvH
Foundation, Germany; GSRT and NSRF, Greece; ISF, MINERVA, GIF, I-CORE and Benoziyo Center, Israel; INFN, Italy; MEXT and JSPS, Japan; CNRST, Morocco; FOM and NWO, Netherlands; BRF and RCN, Norway; MNiSW and NCN, Poland; GRICES and FCT, Portugal; MNE/IFA, Romania; MES of Russia and ROSATOM, Russian Federation; JINR; MSTD,
Serbia; MSSR, Slovakia; ARRS and MIZ\v{S}, Slovenia; DST/NRF, South Africa;
MINECO, Spain; SRC and Wallenberg Foundation, Sweden; SER, SNSF and Cantons of
Bern and Geneva, Switzerland; NSC, Taiwan; TAEK, Turkey; STFC, the Royal
Society and Leverhulme Trust, United Kingdom; DOE and NSF, United States of
America.

The crucial computing support from all WLCG partners is acknowledged
gratefully, in particular from CERN and the ATLAS Tier-1 facilities at
TRIUMF (Canada), NDGF (Denmark, Norway, Sweden), CC-IN2P3 (France),
KIT/GridKA (Germany), INFN-CNAF (Italy), NL-T1 (Netherlands), PIC (Spain),
ASGC (Taiwan), RAL (UK) and BNL (USA) and in the Tier-2 facilities
worldwide.


\bibliographystyle{atlasnote}
\bibliography{effNote}
\newpage

\onecolumn
\clearpage
\input{atlas_authlist}

\end{document}

%% file: atlas_authlist.tex
\begin{flushleft}
{\Large The ATLAS Collaboration}

\bigskip

G.~Aad$^{\rm 84}$,
T.~Abajyan$^{\rm 21}$,
B.~Abbott$^{\rm 112}$,
J.~Abdallah$^{\rm 152}$,
S.~Abdel~Khalek$^{\rm 116}$,
O.~Abdinov$^{\rm 11}$,
R.~Aben$^{\rm 106}$,
B.~Abi$^{\rm 113}$,
M.~Abolins$^{\rm 89}$,
O.S.~AbouZeid$^{\rm 159}$
% ...
\bigskip
\\
% ...
$^{11}$ Institute of Physics, Azerbaijan Academy of Sciences, Baku, Azerbaijan\\
% ...
$^{21}$ Physikalisches Institut, University of Bonn, Bonn, Germany\\
% ...
$^{84}$ CPPM, Aix-Marseille Universit{\'e} and CNRS/IN2P3, Marseille, France\\
% ...
$^{89}$ Department of Physics and Astronomy, Michigan State University, East Lansing MI, United States of America\\
% ...
$^{106}$ Nikhef National Institute for Subatomic Physics and University of Amsterdam, Amsterdam, Netherlands\\
% ...
$^{112}$ Homer L. Dodge Department of Physics and Astronomy, University of Oklahoma, Norman OK, United States of America\\
$^{113}$ Department of Physics, Oklahoma State University, Stillwater OK, United States of America\\
% ...
$^{116}$ LAL, Universit{\'e} Paris-Sud and CNRS/IN2P3, Orsay, France\\
% ...
$^{152}$ Institute of Physics, Academia Sinica, Taipei, Taiwan\\
% ...
$^{159}$ Department of Physics, University of Toronto, Toronto ON, Canada\\
% ...
\end{flushleft}
